\documentclass{acmart}
\AtBeginDocument{%
  \providecommand\BibTeX{{%
    \normalfont B\kern-0.5em{\scshape i\kern-0.25em b}\kern-0.8em\TeX}}}

\usepackage{verbatim} 
\usepackage{tabulary} 
\usepackage[inline]{enumitem} 
\usepackage{xhfill}
\usepackage{makecell} 
\usepackage{tabularx}
\usepackage{array}
\usepackage{caption}
\usepackage{subcaption}
\usepackage{lscape}
\usepackage{pifont}
\usepackage{adjustbox}
\usepackage{afterpage}
\mathchardef\mhyphen="2D 
\newcommand{\cmark}{\ding{51}}
\newcommand{\PreserveBackslash}[1]{\let\temp=\\#1\let\\=\temp}
\newcolumntype{C}[1]{>{\PreserveBackslash\centering}p{#1}}
\newcolumntype{R}[1]{>{\PreserveBackslash\raggedleft}p{#1}}
\newcolumntype{L}[1]{>{\PreserveBackslash\raggedright}p{#1}}

\setcopyright{none}
\copyrightyear{}
\acmYear{}

\begin{document}

\title{FairSNA: Algorithmic Fairness in Social Network Analysis}


\author{Akrati Saxena}
\email{a.saxena@tue.nl}
\affiliation{%
  \institution{Eindhoven University of Technology}
  \city{Eindhoven}
  \country{The Netherlands}
}

\author{George Fletcher}
\email{g.h.l.fletcher@tue.nl}
\affiliation{%
  \institution{Eindhoven University of Technology}
  \city{Eindhoven}
  \country{The Netherlands}
}

\author{Mykola Pechenizkiy}
\email{m.pechenizkiy@tue.nl}
\affiliation{%
 \institution{Eindhoven University of Technology}
 \city{Eindhoven}
 \country{The Netherlands}}

\renewcommand{\shortauthors}{Saxena et al.}

\begin{abstract}
In recent years, designing fairness-aware methods has received much attention in various domains, including machine learning, natural language processing, and information retrieval. However, understanding structural bias and inequalities in social networks and designing fairness-aware methods for various research problems in social network analysis (SNA) have not received much attention. In this work, we highlight how the structural bias of social networks impacts the fairness of different SNA methods. We further discuss fairness aspects that should be considered while proposing network structure-based solutions for different SNA problems, such as link prediction, influence maximization, centrality ranking, and community detection. This paper clearly highlights that very few works have considered fairness and bias while proposing solutions; even these works are mainly focused on some research topics, such as link prediction, influence maximization, and PageRank. However, fairness has not yet been addressed for other research topics, such as influence blocking and community detection. We review state-of-the-art for different research topics in SNA, including the considered fairness constraints, their limitations, and our vision. This paper also covers evaluation metrics, available datasets, and synthetic network generating models used in such studies. Finally, we highlight various open research directions that require researchers' attention to bridge the gap between fairness and SNA.
\end{abstract}



\maketitle

\section{Introduction}

In the real world, people connect with each other based on different relationships, such as friendships, acquaintances, kinship, collaboration, or having the same interests, and form different types of complex social networks, such as friendship networks, collaboration networks, or communication networks. All these social networks evolve based on day-to-day human interactions, and social network analysis (SNA) has been used to understand human behavior as an individual and in groups \cite{freeman2004development}. In recent years, due to the vast amount of available data from online social networks (OSNs), social network analysis has been of great interest to understand and model human behavior in several dynamic processes, such as information spreading or opinion formation, at a large-scale. Structural patterns of social networks has been used to answer various interesting questions, such as which users will connect in the future, which users are more important in the network given the application context, which users have high control on the network for a given dynamic phenomenon, or which users are more vulnerable?  

In SNA, the solutions proposed based on network structure and the characteristics of nodes and networks have provided state-of-the-art accuracy for different problems; however, most of them have not accounted for structural bias and inequality of social networks. For example, in health-intervention programs, such as awareness for HIV or prevention of suicide, we aim that the useful information should reach as many people in need as possible. However, due to the limited resources, it is not possible to personally reach every user at risk. Therefore, we target influential community leaders who can spread the vital information to other members of the community or detect suspicious cases beforehand and respond appropriately. The success of these intervention programs highly relies on the underlying social networks. \cite{tsang2019group, stoica2020seeding} highlighted the bias in fairness-oblivious social intervention methods \cite{kempe2003maximizing, li2018influence} that aim to maximize the outreach by focusing on major communities and are not fair for minor communities at risk. They might disproportionately exclude the users from racial minorities or LGBTQ communities from the benefits of the intervention. Therefore, it is important to ensure that the resource allocation should respect the diverse composition of communities, and each community should receive a fair proportion. In recent era, SNA has been used in many different applications, including studying marketing strategies \cite{bolotaeva2010marketing, doyle2007role}, identifying scientific leaders \cite{bordons2015relationship}, news and rumor propagation \cite{saxena2022fakeprop}, opinion and innovation diffusion \cite{cho2012identification}, job hiring \cite{tassier2008social}, disseminating terrorist networks \cite{roberts2011strategies, miller2018discovering}, and identifying gang leaders or predicting criminal activities using criminal networks \cite{carrington2011crime, albanese2007criminal}. 
Several works, including \cite{farnad2020unifying, mehrabi2019debiasing, khajehnejad2021crosswalk, rahman2019fairwalk, masrour2020bursting}, have highlighted unfairness in different SNA algorithms, i.e., mainly occurred because of the structural bias. 

In social networks, due to homophily \cite{mcpherson2001birds}, people prefer to connect with similar kinds of people. Stoica et al. \cite{stoica2018algorithmic} collected an online social networking dataset from Instagram and observed that male users on Instagram exhibit much stronger homophily among them as compared to female users. A similar phenomenon has been observed in other social networks \cite{stoica2018homophily, gallivan2015co, messias2017white, tarbush2012homophily}. The homophily and preferential attachment further give rise to the glass ceiling effect, i.e., an unbreachable barrier and keeps minorities from achieving higher ranks even having comparable characteristics as majorities. They further showed that the bias of recommendation algorithms used for the network evolution in OSNs further worsens these pre-existing inequalities. Avin et al. \cite{avin2017modeling} also highlighted power inequality and the glass-ceiling effect in different social networks and co-authorship networks. Besides homophilic connectivity, in scale-free networks, communities' size follows power-law distribution \cite{arenas2004community}. Therefore, the algorithms might provide higher accuracy by optimizing their results for large-size communities and might not be accurate for small-size communities. All these factors should be considered while designing methods based on network structure so that the proposed methods are fair for all kinds of users and all communities, irrespective of their size or type. Fair algorithms are required to increase diversity in a network, which will further reduce inequalities. 

The inequality of social networks has been an interesting topic for sociologists and economists \cite{finneran2003social, pena2021inequality, dimaggio2012network, bottero2007social, mcguire2002gender, nishi2015inequality}, though this issue is not well addressed by network scientists while designing solutions based on network structure that are applied in real-life. In the past three years, some researchers have proposed fairness-aware methods for some of the SNA problems, such as link prediction, influence maximization, and PageRank \cite{rahman2019fairwalk, masrour2020bursting, tsioutsiouliklis2021fairness, stoica2019fairness}. However, fairness is not yet defined and explored for several other problems, including influence blocking, community detection, different centrality rankings, anomaly detection, and network anonymization. This paper discusses the fairness constraints that should be considered for achieving individual or group fairness \cite{dwork2012fairness} for different research problems, including state-of-the-art literature and future directions. The main aim of this paper is to get the attention of researchers towards this gap. 

\textbf{Related Surveys.} Initial surveys on fairness in machine learning (ML) focused on independent and identically distributed (i.i.d.) data \cite{caton2020fairness, corbett2018measure, du2020fairness, mehrabi2021survey, mitchell2021algorithmic, pessach2020algorithmic, pitoura2021fairness}. These surveys did not mainly focus on relational graph datasets. There are some recent tutorials that provide a good taxonomy of fairness in ML and graph data mining, and might be interesting for readers to explore \cite{kang2021fair, cisse2019fairness, barocas2017fairness, venkatasubramanian2021fairness}. 

We would like to highlight recent surveys that have focused on graph data mining from machine learning and deep learning perspectives \cite{zhang2022fairness, dong2022fairness, choudhary2022survey}. In \cite{dong2022fairness}, the authors focused on fairness in graph data mining algorithms for knowledge graphs, recommender systems, and network embedding. 
Zhang et al. \cite{zhang2022fairness} provided a brief review on quantifying different types of fairness studied for graph structured datasets. 
Choudhary et al. \cite{choudhary2022survey} reviewed fair-ML methods for graph structured relational datasets. This survey considers structural bias and comes close to our work, though they have only focused on ML methods to achieve fairness. Besides this, these surveys are not focused on social networks and have considered only a few downstream tasks in graph mining, mainly node classification, link prediction, and influence maximization. 

An important difference to note is that most of the works in fair graph mining have considered bipartite or heterogeneous graphs; for example, recommending movies on Netflix where nodes are users and movies, recommending items on Amazon where nodes are users and items, or hiring job candidates where nodes are employers and prospective employees. In such cases, it is easy to compute the ranking of items based on their characteristics and the given requirement context, and individual and group fairness are defined based on these rankings. However, computing the ranking of users in the case of social networks is not that straightforward as the ranking depends on the inter-connectivity of users as well as on the application context. Another important point is that in such applications, the data is well-organized, and several attributes of the nodes are known in advance, though that is not the case with OSNs. For example, sensitive attributes, such as gender or demographic location, are not available for OSN users from their profiles. 

To the best of our knowledge, this is the first survey-cum-vision paper that focuses on fairness in a wide range of research problems in Social Network Analysis. One important point is that this survey considers all types of algorithms, including heuristic, probabilistic, machine learning, and deep learning based methods, to achieve fairness. This paper also highlights many open research problems in this area and aims to bridge the gap between fairness and SNA. 

\textbf{Prerequisite.} In the paper, we have briefly explained all required terms to understand the discussion. Still, a basic understanding of (i) network science terminologies \cite{barabasi2014network}, and (ii) fairness constraints \cite{pessach2022review} will be helpful in easily following the paper. 

The rest of the paper is structured as follows. In Section \ref{taxonomy}, we discuss the taxonomy of fairness in SNA. In Section \ref{section2}, we discuss various research topics for which the network structure based solutions have been proposed, and for each topic, we discuss fairness constraints, state-of-the-art, and future directions. In Section \ref{datasets}, we summarize datasets used for such studies. The paper is concluded in section \ref{conclusion}. All abbreviations used in the paper are summarized in Appendix~\ref{appendixabbrv}.

\section{Taxonomy of Fairness-aware SNA (FairSNA)}\label{taxonomy}

In fairSNA, fairness has been defined by extending fairness definitions from ML \cite{gajane2017formalizing, caton2020fairness}, and also proposing some novel definitions to the network context. Here, we discuss a high-level taxonomy of fairSNA that can be looked at from different perspectives, including group vs. individual fairness, feature-aware vs. feature blind fairness, and in/pre/post-processing based fair methods. 

\textbf{Group vs. Individual Fairness:} In recent works on algorithmic fairness, there are two basic frameworks, group fairness and individual fairness \cite{pessach2022review}.  
In SNA, group fairness demands that different communities, irrespective of their size or protected attribute, should be treated equally and should receive equal resources. For example, in the case of influence maximization for a social-awareness cause, the influence should be propagated equally to all the communities, and the number of influenced users in each community should be proportional to its size. In fairSNA, group fairness is defined using \textit{equality}, \textit{equity}, and \textit{statistical parity} fairness constraints \cite{farnad2020unifying, rahman2019fairwalk}. 
Similar to ML, in SNA, individual fairness aims that similar users should receive similar treatment or similar resources. In some SNA problems, such as link prediction and influence maximization, the individual fairness has been modeled using \textit{equality at user level} \cite{rahman2019fairwalk} and \textit{information access gap} \cite{fish2019gaps}, respectively, i.e., different than the individual fairness constraints in ML \cite{caton2020fairness} due to the application settings. Individual fairness for other problems, such as community detection or centrality ranking, is still not defined (further discussed in Subsections \ref{seclp} and \ref{seccr}). 
Besides these, in SNA, fairness has also been incorporated by proposing novel fairness constraints, including \textit{maximin}, \textit{disparity}, and \textit{diversity}, that ensure fairness for small and marginalized groups \cite{farnad2020unifying}. Maximin fairness \cite{farnad2020unifying} aims to improve the fairness for the least benefited community. The diversity constraint \cite{farnad2020unifying} is inspired by the game-theoretic approach and guarantees that each community is at least as well off as if it receives the resources proportional to its size and allocates them internally.

\textbf{Feature-aware vs. Feature-blind:} In ML, the methods that explicitly use protected attributes to achieve fair results are called feature-aware methods, otherwise feature-blind. 
In fairSNA, all proposed methods have used community membership of users or its size so that each community is treated fairly or receives equal resources. The community membership is identified using protected attributes or by applying community identification methods on the network. Therefore, fairSNA methods fall under the feature-aware category. As per the best of our knowledge, no methods provide feature-blind fairness in SNA. The scientific community will appreciate such methods as they will avoid the cumbersome task of collecting ground-truth community data.

\textbf{Pre/In/Post-Processing:} Fair algorithms can be categorized as (i) pre-processing, (ii) in-processing, and (iii) post-processing approaches \cite{caton2020fairness}. 
In SNA, fairness is mainly achieved by using in-processing methods, in which the algorithm learns to provide fair output by considering the fairness constraints that diminish the effects of structural bias in the data. However, Laclau et al. \cite{laclau2021all} proposed a pre-processing method to repair the adjacency matrix of the graph by adding links that will obfuscate dependency on the protected attribute, and the repaired matrix can be used further to generate network embedding that provides individual as well as group fair results in downstream tasks, such as link prediction. One important point to note is that such methods can only be applied if the entire network structure is known in advance. The real-world networks are highly dynamic, and their size is increasing exponentially with time. Therefore, in SNA, researchers have proposed several methods that use partial network information to provide fast and efficient solutions \cite{stein2017heuristic, saxena2018estimating, eshghi2019efficient, tran2021community}. Therefore, the pre-processing steps, such as repairing adjacency networks, will require the entire network information and will not be suitable for large-scale dynamic networks. One can still explore the possibility of efficiently generating repaired adjacency matrix for dynamic and partial networks. 

\cite{saxena2021hm, masrour2020bursting} proposed post-processing-based link prediction methods that achieve fairness by increasing the prediction likelihood of the edges of underrepresented node-pair groups that might be less likely to be predicted using classical link prediction methods \cite{al2011survey}. In SNA, heuristic methods have been proposed for several tasks, such as link prediction \cite{liben2007link, adamic2003friends,zhou2009predicting, jeon2017community}, influence maximization \cite{chen2010scalable,kundu2011new}, which are comparatively faster than probabilistic modeling, machine learning, or deep learning-based methods. The post-processing for such heuristic methods might provide fair and fast solutions.

\section{FairSNA: Fairness Constraints, State-of-the-art, and Future Directions}\label{section2}

In the following subsections, we discuss various research topics in SNA, including the respective fairness constraints, state-of-the-art, challenges, and future directions.\\
\textbf{Notations.} $G(V, E)$ denotes a network, where $V$ is the set of nodes and $E$ is the set of edges. $u, v, w$ are nodes in the network, and $A$ is the sensitive attribute. $C=\{C_1, C_2, \cdots, C_i, \cdots \}$ denotes communities in the network $G$. 

\subsection{Fair Link Prediction}\label{seclp}

Link prediction (LP), also known as link recommendation, has been widely used to predict future or unknown links in the network \cite{al2011survey}. Predicting promising links in online social networks for recommending friends is important so that users will be more loyal to the website. Most of the existing link prediction methods have not considered structural biases \cite{saxena2021hm}.

\begin{figure}[h]
     \centering
     \begin{subfigure}[b]{0.48\textwidth}
         \centering
         \includegraphics[width=\textwidth]{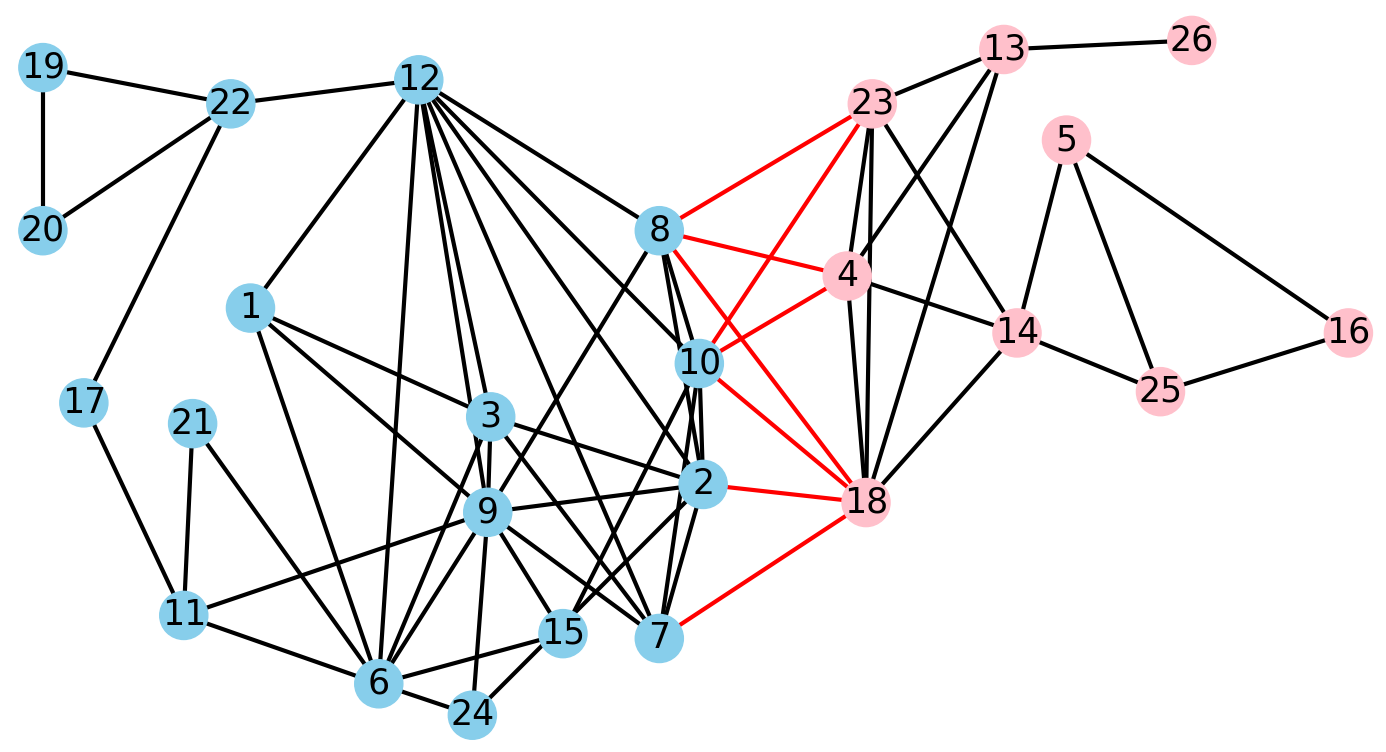}
         \caption{}
         \label{}
     \end{subfigure}
     \hfill
     \begin{subfigure}[b]{0.48\textwidth}
         \centering
         \includegraphics[width=\textwidth]{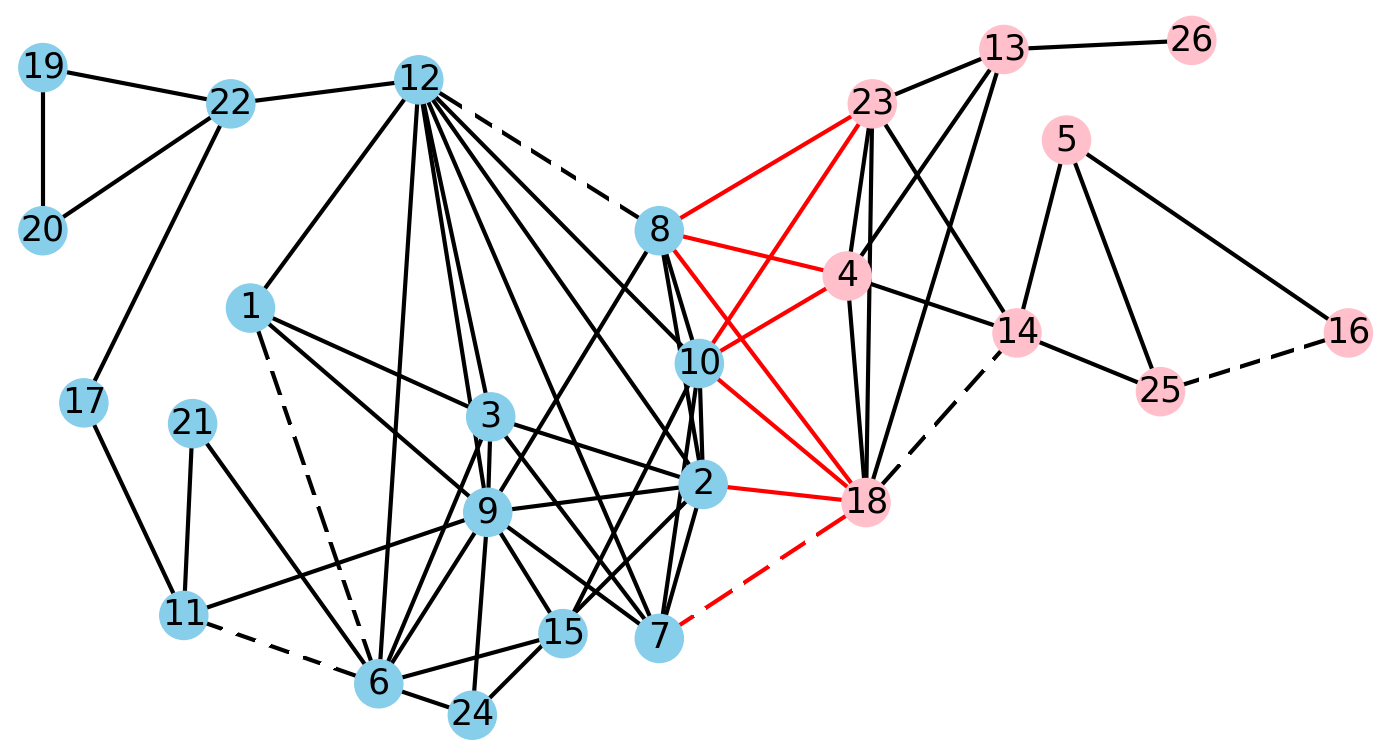}
         \caption{}
     \end{subfigure}
     \\ \vspace{4mm}
     \begin{subfigure}[b]{0.48\textwidth}
         \centering
         \includegraphics[width=\textwidth]{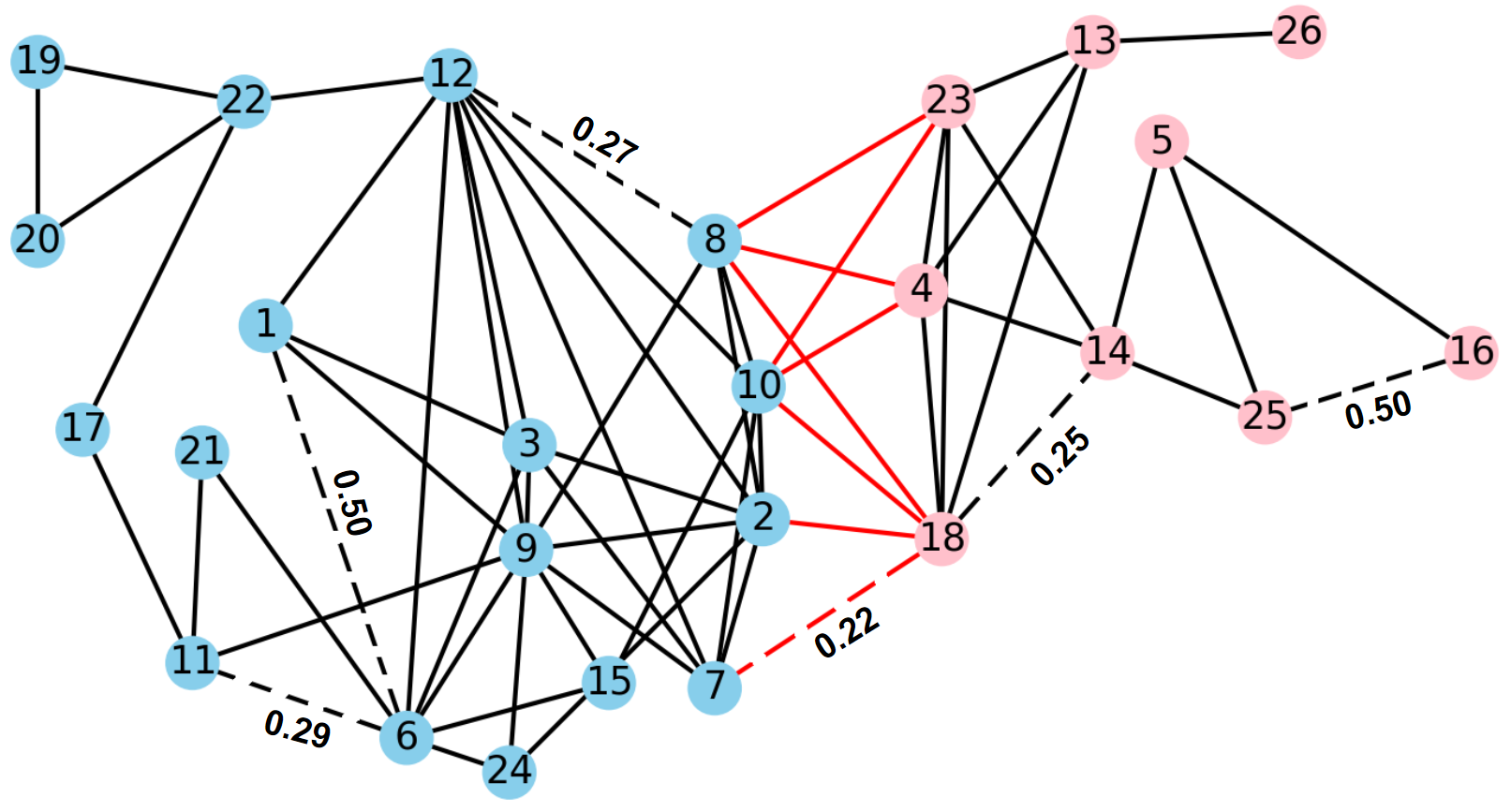}
         \caption{}
         \label{}
     \end{subfigure}
     \hfill
     \begin{subfigure}[b]{0.48\textwidth}
         \centering
         \includegraphics[width=\textwidth]{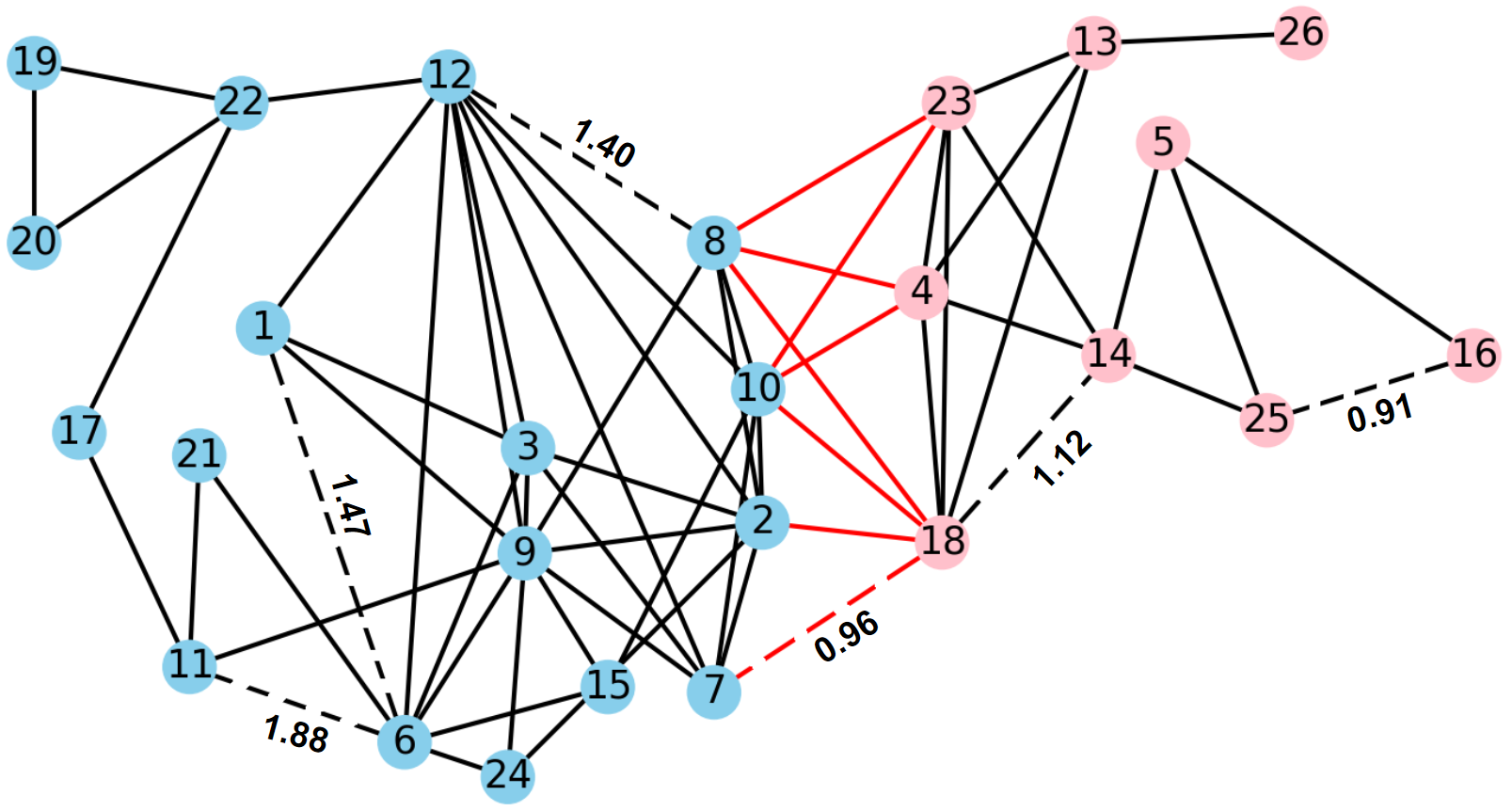}
         \caption{}
         \label{}
     \end{subfigure}
    \caption{(a). Dutch School Social network having two groups where blue nodes represent girls and pink nodes represent boys, (b). Dashed lines represent around 10\% u.a.r. removed links from the network, (c). and (d). Values corresponding to dashed lines show the Jaccard coefficient and Adamic Adar scores for predicting the missing links, respectively.} 
    \label{lp_exa}
\end{figure}

In Fig.~\ref{lp_exa}, we show a small example of unfairness in link prediction using the Dutch School social network \cite{knecht2010friendship} that has 26 nodes (17 girls and 9 boys), and 63 edges. The homophily value of the network is 0.7 \cite{newman2003mixing}. In Fig. ~\ref{lp_exa} (a), the network is shown, and the nodes are divided into two groups based on gender; blue nodes represent girls and pink nodes represent boys. Next, we remove around 10\% of intra-community and inter-community edges uniformly at random, and the missing links are shown using dashed lines in Fig. \ref{lp_exa} (b). Now, we compute the similarity scores for predicting the missing links using two heuristics methods, (i) Jaccard Coefficient \cite{liben2007link}, and (ii) Adamic Adar Index \cite{adamic2003friends}; and similarity scores are shown corresponding to the missing links in Fig.~\ref{lp_exa} (c) and (d), respectively. We can observe that the value of similarity scores for inter-community links is lower than the intra-community links. Besides this, similarity scores are lower for small and sparse communities. For example, in Fig.~\ref{lp_exa} (d), Adamic Adar coefficient values for the links from the pink community are smaller than the blue community. Saxena et al. \cite{saxena2021hm} computed similarity scores for link prediction using nine different heuristic methods on many real-world scale-free networks and showed that inter-community links have lower scores than intra-community links. An in-depth understanding of the performance of different link prediction methods for sparse and dense communities is still an open research question. In fair link prediction, the aim is to efficiently predict all kinds of links with high accuracy, irrespective of users' attributes, their communities, or community sizes.

\subsubsection{Fairness Definition.}

Next, we discuss fairness constraints considered in link prediction.

\begin{enumerate} 
    \item Statistical Parity: In link prediction, if the sensitive attribute is gender having two values $\{m, f\}$, then the recommended links can be divided into four groups $G_{mm}, G_{mf}, G_{fm}, G_{ff}$; $G_{mf}$ denotes the links recommended to group $m$ from group $f$. Similarly, if a sensitive attribute $A$ has $z$ values $(\{1, 2, \cdots, i, j, \cdots, z\})$, then there will be $z^2$ groups. Let the link recommendation probability of positive predictions for link class $G_{ij}$ is, $P(G_{ij})=|\{p(u,v) =1 : (u,v) \in G_{ij} \}|/|G_{ij}|$, where $p(u,v) =1$ if the link is likely to exist (positively recommended) between nodes $u$ and $v$ irrespective of ground-truth label. In the case of similarity score-based methods, such as the Jaccard Coefficient and Adamic Adar index, $p(u,v) =1$, if the computed similarity score between node pair $(u, v)$ is higher than the given threshold \cite{al2011survey}. The statistical parity or bias for two groups $G_{ij}$ and $G_{jk}$ is the difference between $P(G_{ij})$ and $P(G_{jk})$. For analyzing bias among multiple groups, the variance between the link recommendation probabilities of each group can be calculated \cite{rahman2019fairwalk}. So, the bias with respect to statistical parity, i.e. denoted by $bias^{SP}(G)$, is computed as,
    
    \begin{center}
    $bias^{SP}(G) = Var(\{P(G_{ij})\} : (i,j) \in A)$
    \end{center} 

    \item Accuracy Parity: The accuracy parity focuses on correct predictions (for both existing and non-existing links) instead of positive predictions given a pair of nodes. The Accuracy Parity for two groups $G_{ij}$ and $G_{jk}$ is the difference between $P_c(G_{ij})$ and $P_c(G_{jk})$, where $P_c(G_{ij})$ is the probability of correctly predicted existing and non-existing links in group $G_{ij}$. Finally, the bias with respect to accuracy parity is computed as,
    
    \begin{center}
    $bias^{AP}(G) = Var(\{P_c(G_{ij})\} : (i,j) \in A)$
    \end{center} 
    
    \item Equality of Recommendation at Group Level \cite{rahman2019fairwalk}: In link recommendation, the equality of recommendation at the group level (denoted by $ERG$) with respect to a sensitive attribute $A$ can be computed using the total number of recommendation for a pair of groups. If $N(G_{ij})$ denotes the number of recommendation for a group $G_{ij}$, and $N(G_{ij})= |\{p(u,v) =1 : (u,v) \in G_{ij} \}|$, then the bias with respect to $ERG$ is computed as,
    
    \begin{center}
     $bias^{ERG}(G,A)= Var(\{N(G_{ij}) \} : \{i,j\} \in A)$ 
    \end{center}
    
    \item Equality of Recommendation at User Level \cite{rahman2019fairwalk}: To compute the Equality of Recommendation at user level for a user $u$ (denoted by $ERu$), we compute the fraction of recommended users to $u$ having attribute value $j$ and denote it as $j$-$share(u)$. Now, the bias with respect to $ERu$ can be computed for each sensitive attribute value $j$, which is the difference between a fair fraction where the recommended links from each group have the same fraction and the average $j$-$share$ over all users. It is computed as,
    
    \begin{center}
     $bias^{ERu}(G,j)= \frac{1}{|A|} - \frac{\sum_{u \in V} j \mhyphen share(u)}{|V|}$
    \end{center}
    
\end{enumerate}

\subsubsection{State-of-the-Art.} 

In this section, we discuss state-of-the-art for fairness-aware link prediction methods. \\
\textbf{Heuristics Methods.} In heuristic or similarity-based link prediction methods, such as Jaccard Coefficient (JC), Adamic-Adar (AA), and Resource Allocation (RA) Index \cite{al2011survey}, the similarity score of a given pair of nodes is computed based on their structural similarity, that is further used for link prediction. Saxena et al. \cite{saxena2021hm} studied the fairness in link prediction with respect to two classes, intra-community and inter-community links. They observed that inter-community links have lower structural similarities than intra-community links. Based on this observation, the authors proposed the HM-EIICT (\textbf{H}euristic \textbf{M}ethod-\textbf{E}xtended using \textbf{I}ntra and \textbf{I}nter \textbf{C}ommunity \textbf{T}hresholds) framework, where a different threshold value is considered for intra-community and inter-community node pairs for link prediction. The proposed method was verified on several real-world networks, and even the simple heuristic methods extended using the given approach, such as JC-EIICT and RA-EIICT, provide better results as compared to all nine considered baselines. Specifically, the proposed method highly improves the accuracy of inter-community link prediction. 

\textbf{Network Embedding-based Methods.} Besides heuristic methods, network embedding-based methods have also been proposed for fair link prediction. Rahman et al. \cite{rahman2019fairwalk} observed a high group-level bias in link recommendation using the Node2Vec embedding \cite{grover2016node2vec}, i.e., a well-known network embedding method. They proposed a fairness-aware network embedding method, called Fairwalk, in which a biased random walk is proposed that explores diverse neighborhoods. In the proposed fairwalk, the neighbors are partitioned based on the sensitive attribute, and each group gets the same probability of being chosen by the random walker irrespective of its size. The analysis showed that the presentation of each group is fair in the explored fairwalk traces, and Fairwalk reduced biases under the statistical parity and equality of representation fairness constraints. 
\cite{khajehnejad2021crosswalk} also proposed a reweighting method that can be applied to any random walk based network embedding method, such as deepwalk, node2vec, to generate fair network embedding. The proposed method, called CrossWalk, updates the transition probabilities such that the random walker is more likely to cross group boundaries by up-weighing edges closer to the periphery of the groups or inter-group edges. The crosswalk method, first, computes the proximity of each node with respect to other groups, i.e., the expected number of times the nodes from other groups will be visited in random walks starting from the given node, and then the proximity is used to reweight the transition probabilities of the edges. Therefore, it will embed the periphery nodes from different groups closer if they are more similar while preserving the network structure information. The method is evaluated for influence maximization, node classification, and link prediction using total accuracy (non-fairness based metric) and disparity (fairness based metric), and the results on real-world as well as synthetic networks show that the method achieves a high disparity on the cost of a very small decrease in the accuracy. One limitation of this work is that the method has been verified on small datasets having 2-3 groups based on the attributes, such as age or political inclination, and the performance of the model for large-size networks having multiple communities is not discussed.

In a recent work, Saxena et al. \cite{saxena2021nodesim} proposed a network embedding based method, called NodeSim, for fair link prediction, where the links are categorized into two classes (i) intra-community links and (ii) inter-community links. The authors proposed NodeSim random walk that explores intra as well as inter-community neighborhood of a node based on nodes' structural similarities. In the NodeSim random walk, the probability of moving to the next node depends on both the community label and its similarity with the current node. The authors train a logistic regression model for link prediction using vector representation of nodes and their community information. The proposed method outperforms baseline methods for both intra and inter-community link prediction. This is the first work that uses community information while training the link prediction model for achieving fairness. 

In network embedding methods, adversarial approach has also been used to generate a fair embedding that is invariant to protected attributes \cite{bose2019compositional}. The adversarial learning framework has two important components (i) generator and (ii) discriminator. The generator generates network embedding while the discriminator aims to predict the sensitive attribute value based on the generated output. When the discriminator is failed to predict the sensitive attribute, then it is considered that the generated embedding is decoupled from the sensitive attributes and used for further downstream tasks. Masrour et al. \cite{masrour2020bursting} proposed an adversarial learning based network embedding method that provides fairness while recommending links between people belonging to the same and different genders. The method was verified on small networks having a limited number of groups; therefore, the scalability of the proposed method is questionable for large-size networks having multiple protected attributes with multiple values.

All the above-discussed methods provide fairness using in-processing or post-processing techniques to provide a fair output. However, pre-processing steps can also be taken to remove structural bias from a network dataset. Laclau et al. \cite{laclau2021all} proposed an embedding-agnostic repairing method that repairs the adjacency matrix by adding edges to obfuscate the dependency on the sensitive attribute, and the repaired adjacency matrix can be further used for the network embedding. The group-fairness aware repair will add edges that obfuscate the original network structure both within and across the sensitive groups; however, the individual-fairness aware repair will keep the within-group structure almost intact. The authors tested the proposed method for binary as well as multi-class settings and showed that it could control the trade-off between individual and group fairness. As we discussed, the pre-processing methods to repair adjacency matrices are yet to be extended for dynamic streaming networks. Otherwise, the computational complexity will be very high if we compute the repaired adjacency matrix for every change in a dynamic network.

\subsubsection{Future Directions.}
The fairness-aware link prediction methods are helpful in increasing the diversity in a network. However, there are some limitations that might be hurdles in designing and testing fair solutions and should be considered for improving the research in this direction. Lichtnwalter and Chawla \cite{lichtnwalter2012link} discussed various limitations of fair link prediction methods. One of the main problems is the imbalance of both the classes (i) existing links, and (ii) non-existing links. The empirical analysis showed that the current evaluation methods might not provide practical conclusions if evaluated on imbalanced datasets, and we can get more robust results by considering the imbalance of classes in training and testing datasets. Another important point to note is that the currently available data of OSNs is based on the evolution of networks, where the users formed connections based on their personal choice plus based on the existing friend recommendation system used by the platform. Therefore, there might be inherited bias in the data due to the algorithmic unfairness of link recommendation methods, and the network might have less diverse links that might impact the training and testing of the proposed methods. However, this has not been looked at in the past and might affect the training and testing of the model. A promising solution in this direction will be well appreciated by the scientific community that might be the formation of a fair dataset or better testing methods.

We also observed that there are very minimal works on fairness-aware link prediction methods for other types of networks, such as directed networks \cite{bianconi2008local}, weighted networks \cite{saxena2022evolving}, temporal networks \cite{holme2012temporal}, multilayered networks \cite{kivela2014multilayer}, and so on. The methods proposed for undirected networks might not work well for other types of networks. For example, in directed networks, the direction of an edge is an important driven factor in their evolution; on Twitter, if user $a$ follows user $b$, then user $b$ may or may not follow user $a$. Therefore, it is essential to consider the structural bias of different types of networks and then propose specific methods for specific types of networks. Apart from these, the fairness of the proposed methods has not been verified with respect to different types of communities, such as weak and sparse communities, varying size communities, and multiple protected attributes. What if the methods are unfair for some specific communities or specific types of users, and if yes, how the accuracy for them can be further improved? A theoretical and empirical analysis of the bias of different link prediction methods on diverse types of groups will be interesting to study.

\subsection{Fair Centrality Ranking}\label{seccr}

In a social network, each user has some unique characteristics that define its importance in accordance with the given application context. There exist several centrality measures, such as degree centrality, closeness centrality, betweenness centrality, and pagerank, that assign a centrality value to each user based on its characteristics, its position in the network, and the network structure \cite{saxena2020centrality}. The assigned centrality value is further used to compute the centrality ranking of the nodes. The inequalities and biases among individuals in a society with respect to different parameters, such as race, gender, or ethnicity \cite{schwartz1980welfare}, have not been considered while computing the centrality value or centrality ranking of users in a given social network.

\begin{figure}[h]
     \centering
     \begin{subfigure}[b]{0.48\textwidth}
         \centering
         \includegraphics[width=\textwidth]{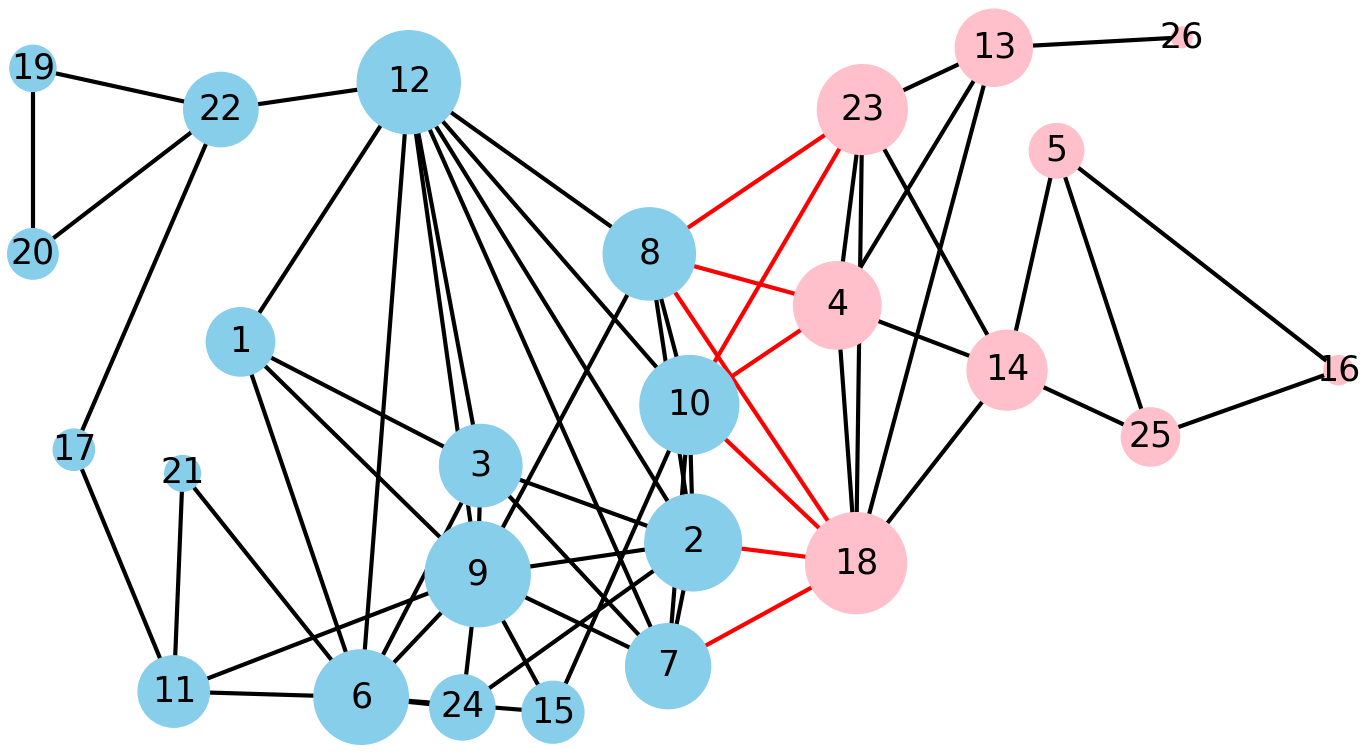}
         \caption{Degree Ranking}
         \label{}
     \end{subfigure}
     \hfill
     \begin{subfigure}[b]{0.48\textwidth}
         \centering
         \includegraphics[width=\textwidth]{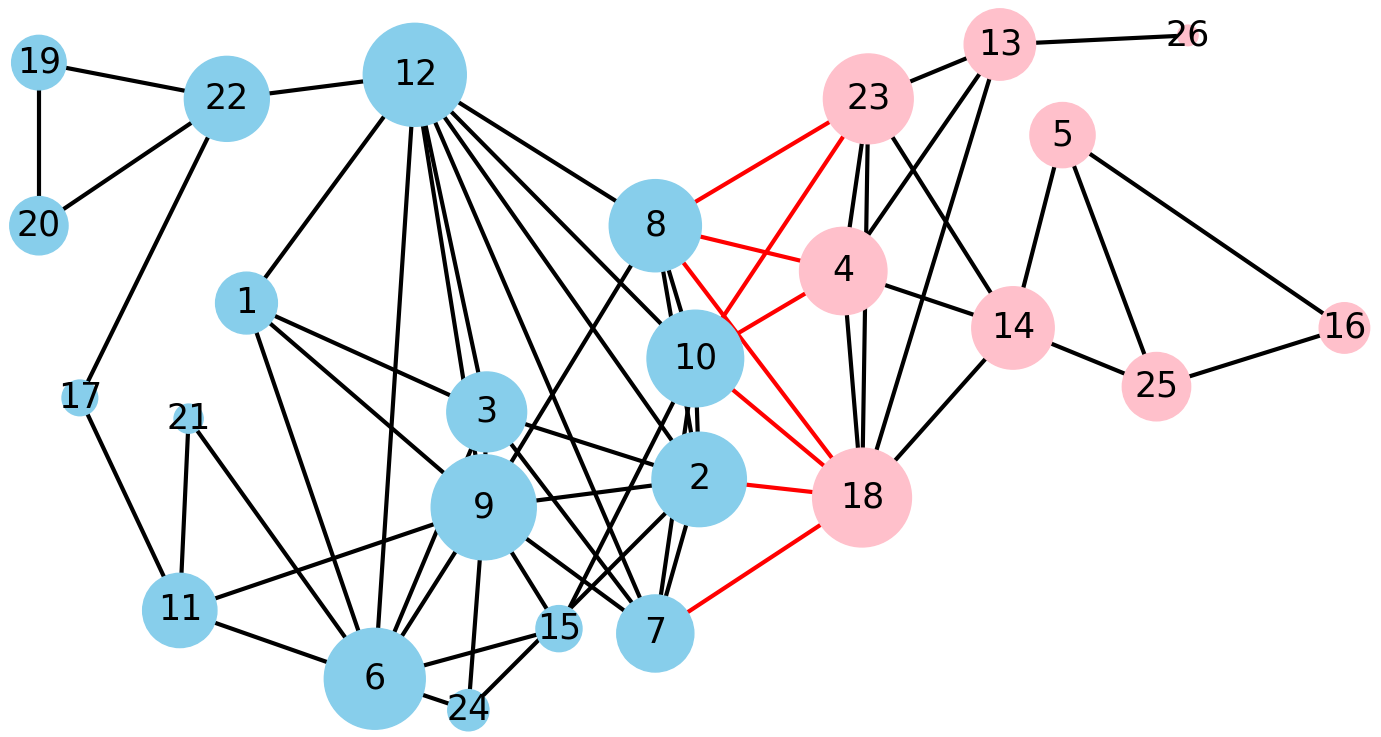}
         \caption{PageRank}
     \end{subfigure}
     \\ \vspace{4mm}
     \begin{subfigure}[b]{0.48\textwidth}
         \centering
         \includegraphics[width=\textwidth]{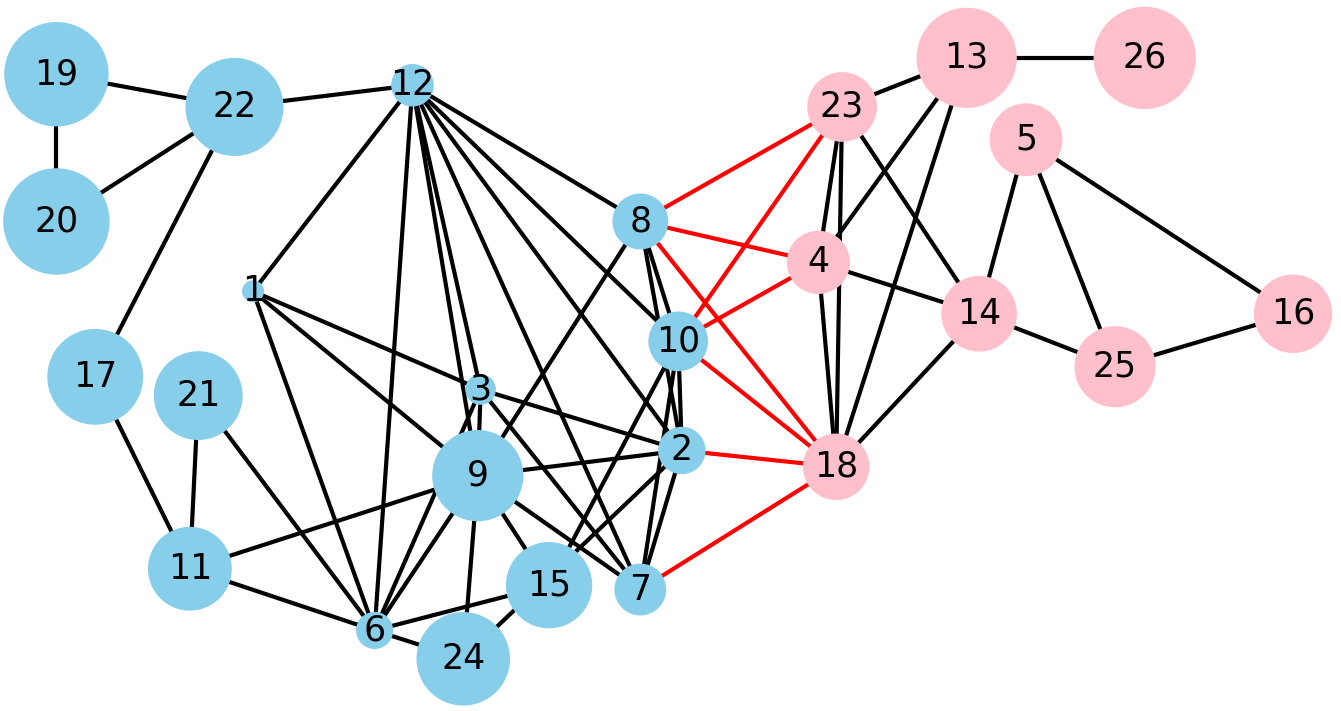}
         \caption{Locally Fair PageRank} 
         \label{}
     \end{subfigure}
     \hfill
     \begin{subfigure}[b]{0.48\textwidth}
         \centering
         \includegraphics[width=\textwidth]{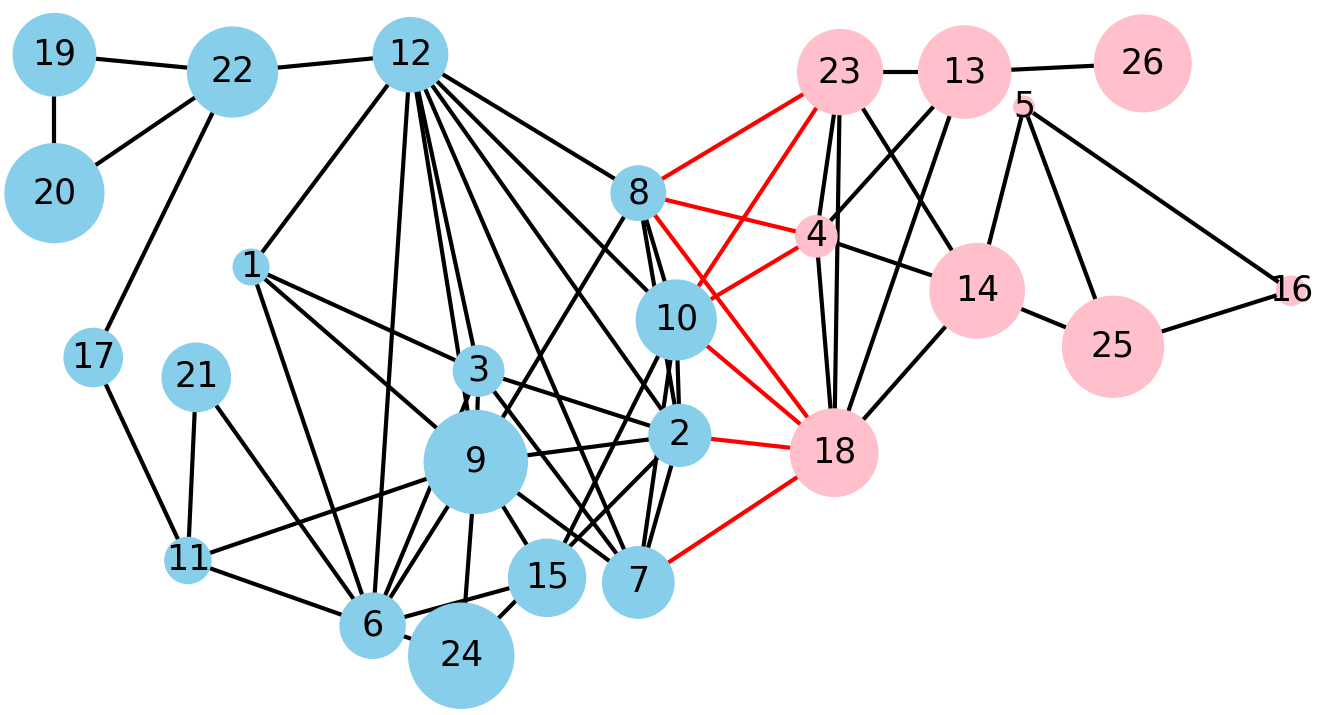}
         \caption{Fairness Sensitive PageRank}
         \label{}
     \end{subfigure}
    \caption{Nodes are ranked based on four different centrality rankings, and nodes' size corresponds to their centrality rank, where (a) Degree Ranking and (b) PageRank are fairness-oblivious centrality ranking, and (c) Locally Fair Pagerank and (d) Fairness sensitive PageRank are fairness-aware PageRank from \cite{tsioutsiouliklis2021fairness}.} 
    \label{cr_exa}
\end{figure}

In Fig. \ref{cr_exa}, we plot the dutch school social network using two fairness-oblivious and two fair centrality rankings, where the node's size corresponds to its centrality rank. Fig. (a) and (b) show the degree and pagerank centrality of the nodes, respectively, and one can observe that large-size communities have more highly ranked nodes that negatively affect the social capital of nodes belonging to smaller communities. In Fig. \ref{cr_exa} (c) and (d), centrality ranking is computed using locally fair pagerank and fairness sensitive pagerank, respectively, which were proposed by Tsioutsiouliklis et al. \cite{tsioutsiouliklis2021fairness}. These fair methods allocate pagerank fairly to both the communities, and one can observe that the distribution of pagerank is more balanced using fair methods. The locally fair pagerank method assigns higher pagerank values to the nodes that are local to the group, and the fairness sensitive pagerank considers both local and global connectivity of the nodes. 

In ML, fairness has been explored in ranking to generate an ordered list of items, and several fairness definitions have been proposed to assess and quantify fairness in the ranking. These methods focus on group or individual fairness-aware ranking \cite{geyik2019fairness, celis2018ranking}, pair-wise orderings \cite{beutel2019fairness}, and exposure bias \cite{singh2018fairness}. However, the case is a bit different in social networks, as each user's importance depends on other users as all users are interconnected to form a complex network. Therefore, it is required to define fairness constraints for nodes' ranking in complex social networks so that the ranking is invariant to structural bias. 
Next, we discuss how group fairness proposed for ranking in ML \cite{geyik2019fairness} can be extended to apply to OSNs where users are grouped into communities. 

\subsubsection{Fairness Definitions.}

Let's assume that a given network $G(V, E)$ has $C=\{C_1, C_2, \cdots , C_l\}$ communities, and $\tau^k$ shows top-$k$ ranked users based on a given centrality measure. 

\begin{enumerate}
    \item Demographic Parity: A group fairness ranking, with respect to communities $C$, should satisfy the following constraint to achieve the demographic parity \cite{geyik2019fairness}.
\begin{equation*}
    |\{u \in \tau^k and \; u \in C_i\}| \leq \left \lceil \frac{|C_i|}{|V|} \cdot k \right \rceil, \forall k \; \& \; \forall C_i \in C
\end{equation*}
and
\begin{equation*}
    |\{u \in \tau^k and \; u \in C_i\}| \geq \left \lfloor \frac{|C_i|}{|V|} \cdot k \right \rfloor, \forall k \; \& \; \forall C_i \in C
\end{equation*}
This fairness constraint assures that the representation of candidates from different communities is balanced in the top-$k$ list. 

\item $\phi \mhyphen fairness$: Tsioutsiouliklis et al. \cite{tsioutsiouliklis2021fairness} defined $\phi \mhyphen fairness$ for pagerank, where a generated ranking is fair if the fraction of the total mass allocated to the members of the protected group is $\phi$. It is defined as, given a graph $G= (V, E)$ having a protected group $R \subset V$, a generated pagerank $PR$ is $\phi \mhyphen fair$ if $PR(R) = \phi$, where $PR(R)$ denotes the total pagerank mass allocated to the nodes of group $R$. $\phi \mhyphen fairness$ is similar to the statistical parity when $\phi$ is equal to the fraction of protected nodes in the network.

Initially, \cite{tsioutsiouliklis2021fairness} defined $\phi \mhyphen fairness$ for the network having two groups, though it can be further extended for networks having multiple groups, where a $\phi$ vector can be defined to denote the mass allocation of pagerank to all groups in the network. 

\end{enumerate}

\subsubsection{State-of-the-Art and Future Directions.} 

Fairness constraints for centrality ranking are not yet well defined. Given that there exist several centrality measures suitable for different applications, the respective fairness constraints for centrality rankings are yet to be discussed. How a biased ranked list of nodes might affect minor communities should be looked at in-depth. 
Karimi et al. \cite{karimi2018homophily} showed that the degree ranking of users from the minority community depends on the relative community sizes and the homophilic index. In a homophilic network, the nodes from minor communities suffer and are not able to achieve a higher degree rank as compared to the nodes from major communities. Therefore, it is important to define methods that can generate a fair centrality ranking of nodes irrespective of their community sizes. 

In one recent work, Tsioutsiouliklis et al. \cite{tsioutsiouliklis2021fairness} studied the bias in the PageRank and observed that the pagerank is not equally distributed among groups based on different sensitive attributes, such as demographics or gender. They proposed a parity-based definition of fairness ($\phi$-fairness) that focuses on maintaining the proportion of pagerank allocated to the candidates of each group. Fairness can be achieved using two methods, (i) Fairness-Sensitive Pagerank and (ii) Locally Fair Pagerank. Fairness-sensitive pagerank method modifies the jump factor of the pagerank method so that the random walker will reach the nodes of the underrepresented group with a higher probability and will achieve the fair pagerank. Locally Fair Pagerank aims to achieve fairness by imposing the fairness constraint on each node so that each node acts fairly in the process of computing pagerank. In the locally fair pagerank method, each node will assign its pagerank to both groups fairly equally; thus, the process is fair at each step and not only on the convergence. Overall the gist is that the proposed methods manipulate jump vector or transition probabilities for the nodes so that nodes from both the groups are fairly represented in the ranking, and finally, each node gets a fair rank. The authors further proposed a link recommendation method that aims to recommend links that will maximize the fairness gain in pagerank \cite{tsioutsiouliklis2022link}. The proposed method will reduce network bias and lead toward a diverse and fair network evolution. 

One simple solution to achieve group fairness might be to use representative ranking that has been applied in various recommendation methods \cite{geyik2019fairness}. The representative ranking maintains the representation ratio of each protected group in the final ranking. In recommendation, the quality of an item/member is quantified using its characteristics; however, in social networks, the quality needs to be quantified using local as well as global influence of the users and their characteristics. The local and global influence of a user depends on its local and global connectivity. These influence values can be further used to generate a final representative ranking. However, centrality measures to generate a meaningful local ranking of the nodes in their communities that can be further used to generate a fair global ranking are yet to be studied, especially with respect to different applications. \cite{saxena2021fair} shows the bias in fairness-aware representative ranking if a group is sub-active on a platform, and this should also be considered while generating a fair ranking. 

Given the dynamic nature of large-scale real-world networks, researchers have proposed several methods to update centrality value in dynamic networks \cite{kas2013incremental, yen2013efficient}, approximate centrality value \cite{wehmuth2013daccer, bader2007approximating, saxena2018k}, or estimate centrality ranking using local neighborhood information \cite{saxena2019heuristic, saxena2017fast, saxena2017observe, saxena2017global} for fairness-oblivious centrality measures (that has been defined without using any fairness constraints) \cite{saxena2020centrality}. One should further consider extending these methods to estimate, update, or approximate fair centrality ranking. As we already discussed that the pre-processing step to repair the adjacency matrix of a network seems like an infeasible option for large-scale dynamic networks, one can further explore efficient methods to update fair-centrality ranking in dynamic networks when the fair ranking was computed by incorporating fairness constraints during pre-processing of the data.

\subsection{Fair Influence Maximization}\label{secim}

Influence maximization (IM) is a well-known problem in network science, in which a group of users, called ``influencers'' (also ``seed nodes''), is chosen to spread the information who can influence a maximum number of people in the network \cite{li2018influence}. IM methods have been applied in many applications, including marketing, news spread, awareness spread, and online trend-setting \cite{huang2019community, yadav2018bridging, azaouzi2021new}. Initially proposed methods for social influence maximization only focused on maximizing the total outreach of information in a given social network and did not consider the fairness in outreach \cite{li2018influence}. Therefore, the achieved outreach might be biased towards large size communities. 

\begin{figure}[h]
     \centering
     \begin{subfigure}[b]{0.48\textwidth}
         \centering
         \includegraphics[width=\textwidth]{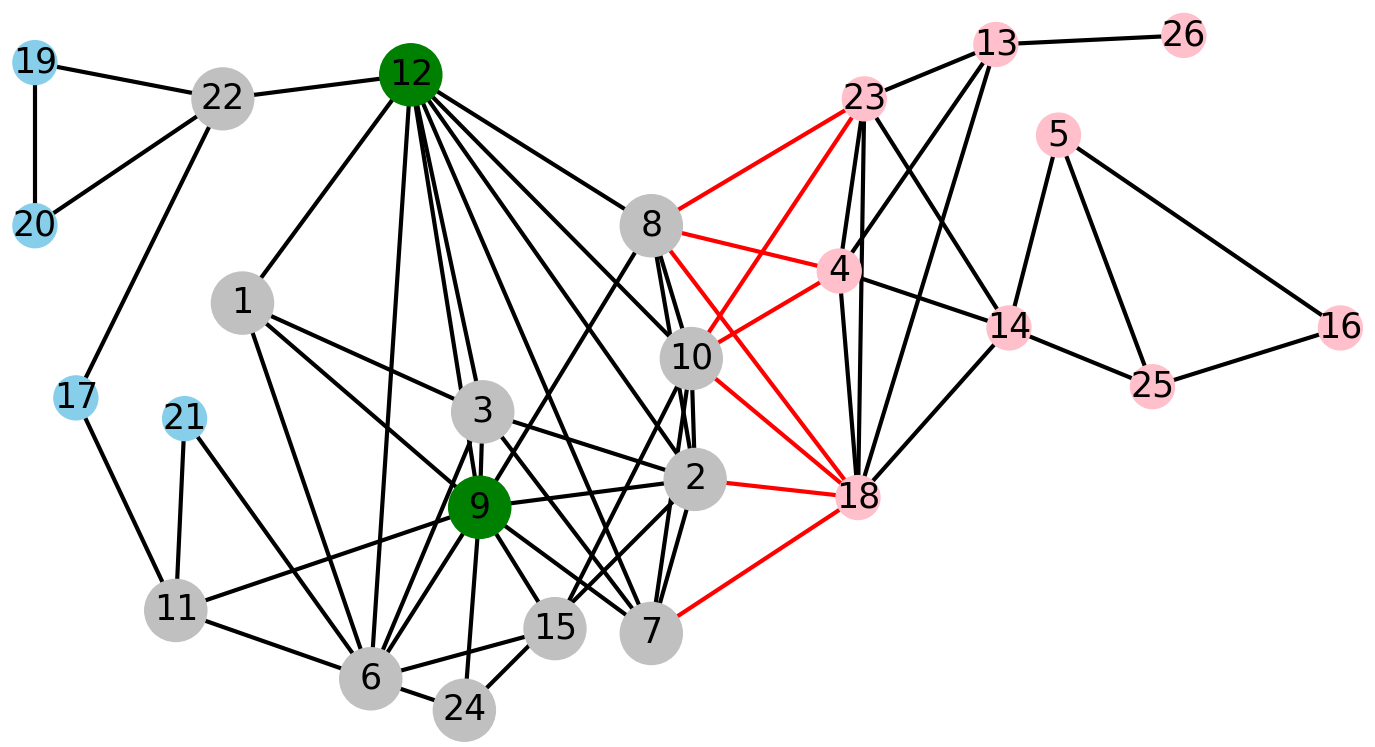}
         \caption{Degree Centrality}
         \label{}
     \end{subfigure}
     \hfill
     \begin{subfigure}[b]{0.48\textwidth}
         \centering
         \includegraphics[width=\textwidth]{figures/im_pagerank.png}
         \caption{PageRank}
     \end{subfigure}
     \\ \vspace{4mm}
     \begin{subfigure}[b]{0.48\textwidth}
         \centering
         \includegraphics[width=\textwidth]{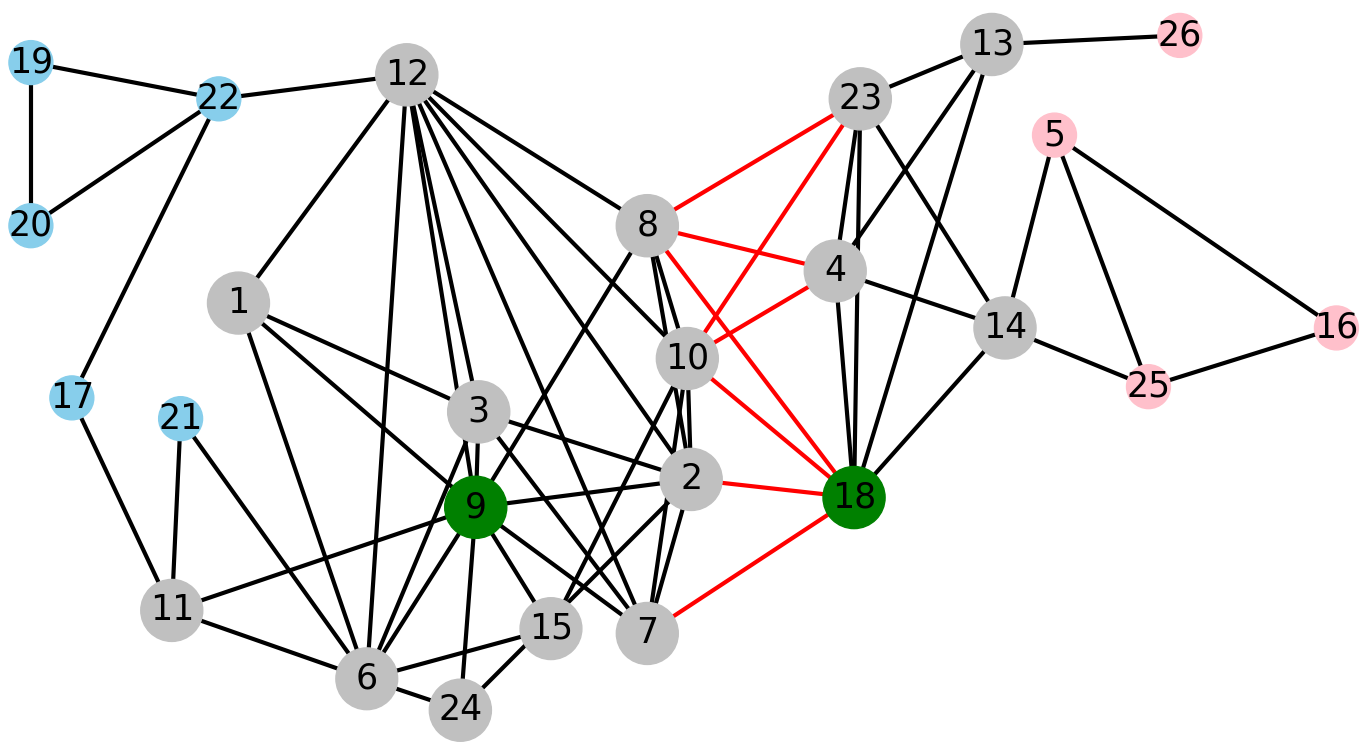}
         \caption{CELF Method}
         \label{}
     \end{subfigure}
     \hfill
     \begin{subfigure}[b]{0.48\textwidth}
         \centering
         \includegraphics[width=\textwidth]{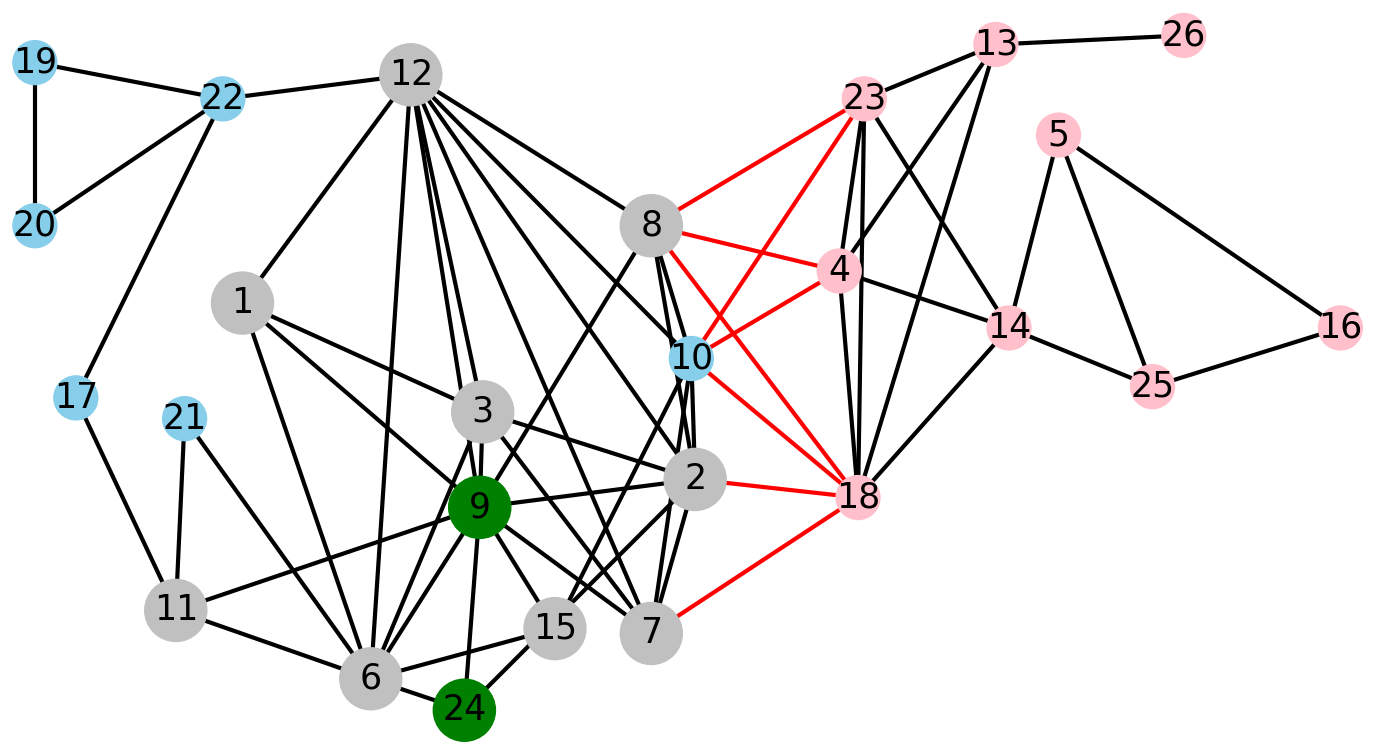}
         \caption{Fairness Sensitive PageRank}
         \label{}
     \end{subfigure}
    \caption{Influence propagation spread using Independent Cascade model where top-2 seed nodes are chosen using four different methods, including degree, pagerank, CELF, and fairness sensitive pagerank. Seed nodes are shown in green color and influenced nodes are shown in grey color. A node's size is proportional to the probability of it getting infected.}
    \label{im_exa}
\end{figure}

For example, in Fig. \ref{im_exa}, we show an example of influence maximization on the dutch school social network. In this example, influence propagation is modeled using the Independent Cascade model, and the probability of information propagation for intra-community edges is $0.1$ and for inter-community edges is $0.05$. We choose top-$2$ nodes for influence propagation using four methods (i) degree centrality, (ii) pagerank, (iii) CELF (Cost-Effective Lazy Forward selection) method \cite{leskovec2007cost}, and (iv) fairness sensitive pagerank \cite{tsioutsiouliklis2021fairness}, and the influence spread is shown in Fig. \ref{im_exa} (a), (b), (c), and (d), respectively. In the figure, top-$2$ chosen seed nodes are shown in green color and the influenced nodes are shown in grey color. One can observe that in most of the cases, influence has not reached the small community, and it shows the bias in outreach. Therefore, it is important to consider the information access equality while selecting the top-$k$ seed nodes.

Wang et al. \cite{wang2021information} defined information access equality as, ``for a given process and seeds, the majority and minority nodes should receive information at similar rates across various stages of the spreading processes''. They analyzed simple vs. complex information spreading processes on the networks evolved with different evolution mechanisms, including majority/minority dichotomy, homophily, preferential attachment, and diversity. The authors observed that the equality in information access depends on both the network structure as well as on the spreading process. They further observed that homophily and preferential attachment based network growth highly affect the information access quality. However, introducing diversity in the network can help in gaining information access equality. 
Teng et al. \cite{teng2020influencing} also studied the bias and glass ceiling effects for the minor community (mainly the community having females) in influence propagation on Instagram and Facebook networks where each user belongs to males or females group. They also observed that the fairness in influence propagation is affected by both factors (i) structure homophily (people of the same kind are more likely to be connected), and (ii) influence diffusion homophily (people are more likely to be influenced by other people of the same kind or having same attributes) \cite{anwar2021balanced}. 

Recently, researchers have focused on designing fairness-aware methods for influence maximization as they can help achieve a diverse outreach \cite{stoica2019fairness, stoica2020seeding, anwar2021balanced, tsang2019group}. The proposed methods show that the feature-aware methods can achieve more diverse outcomes in outreach and seed selection than being feature-blind \cite{stoica2019fairness}.

\subsubsection{Fairness Definitions.}

Next, we discuss different group fairness constraints concerning the influence maximization problem.

\begin{enumerate}
    \item Equality \cite{farnad2020unifying}: It considers that a proposed IM method is fair in choosing the influencers if the number of influencers from each community is proportional to its size. If $\{C_1, C_2, ..., C_l\}$ are communities, and $S$ is the initial chosen set of influencers, then it is defined as,
    
    \begin{equation*}
       \frac{|\{u \in S | u \in C_i\}|}{|C_i|} = \frac{|\{u \in S | u \in C_j\}|}{|C_j|} , \forall i,j
    \end{equation*}
    Equality is also referred to as ``fairness for early-adopters''.
    
    \item Equity \cite{farnad2020unifying}: The equity constraint suggests that the proposed method is fair in outreach if the expected number of influenced nodes for all communities are proportional to their size. 
    
    \begin{equation*}
        \frac{E|I(S, C_i)|}{|C_i|} = \frac{E|I(S, C_j)|}{|C_j|} , \forall i,j
    \end{equation*}
    where $I(S, C_i)$ is the influence outreach achieved by seed nodes $S$ in community $C_i$. This type of fair methods ensure that the information or product awareness reaches the population in a calibrated manner and each community has an equal fraction of aware people. This is important in the news or health-awareness spread, where unequal distribution can manipulate one type of population or can lead to amplifying the echo-chamber effect. This type of fairness is also referred to as ``fairness in outreach''.
    
    The following fairness measures aim to minimize the outreach gap between different groups with respect to their size.
    
    \item Maximin \cite{farnad2020unifying}: The maximin fairness aims to maximize the influence of a minimally influenced group as compared to others with respect to their size in the network. Given a set of communities, the maximin fairness will maximize $\min_{C_i \in C}E|I(S, C_i)|/|C_i|$.
    
    The maximin fairness focuses on the least influenced community, and therefore, the overall outreach is impacted and is lower than the achieved by other fairness-aware methods.
    
    \item Disparity \cite{ali2019fairness}: The disparity of a method is computed by the maximum disparity in the normalized outreach across all communities. It is defined as,
    
    \begin{equation*}
        \max_{i,j \in \{1, 2, ... , |C|\}}\left | \frac{E|I(S,C_i)|}{|C_i|} - \frac{E|I(S,C_j)|}{|C_j|} \right |
    \end{equation*}
    
    \item Diversity \cite{farnad2020unifying}: The diversity measure allocates resources according to the network structure of each community. It follows two steps. In the first step, a fraction of initial influencers are chosen proportional to the size of the community, and the influence is propagated in the induced sub-graph of that community. In the second step, the outreach of influence in each community is used as a lower bound on the influence received by the nodes of that community. In simple words, it ensures that each community receives outreach at least equal to the internal outreach of that community. The steps are summarized below.
    \begin{enumerate}
        \item Let $G_{C_i}$ denotes the induced subgraph for a community $C_i$, and $k_{C_i}=\left \lceil |S| \cdot |C_i|/|V| \right \rceil$ and \\ $Outreach_{C_i}=\max_{S \subset C_i :|S|=k_{C_i}} |I(S,G_{C_i})|$.

        \item For each community $C_i$, the diversity constraint ensures that it holds $E|I(S, C_i)| \geq Outreach_{C_i}$.
    \end{enumerate}

\end{enumerate}

\subsubsection{State-of-the-Art.} 
Most of the fairness-aware IM methods have been proposed in the past three years. 

\textbf{Group Fairness.} Stoica and Chaintreau \cite{stoica2019fairness} considered two types of fairness in IM, (i) fairness in early adopters (called seed nodes) and (ii) fairness in outreach. Early adopters play an important role in the diffusion process, and having early access to the information or products may help establish their role as influencers and gain people's trust. These days, influencers earn money, which can help them gain social capital or financial benefits \cite{kazienko2006social}. For example, influencers on Instagram are paid by companies to promote products. However, a calibrated outreach is more useful in spreading awareness information, where we aim that all communities should be equally aware of the given information. Both kinds of fairness are important and which one should be chosen depends on the application. In this work, the authors compared two techniques for choosing early adopters (i) greedy approach, i.e., a strategic method to choose the next node that maximizes the outreach, and (ii) highest degree node, i.e., a heuristic method. They assessed the performance of both methods using the independent cascade (IC) model \cite{kempe2005influential} for influence propagation and showed that the strategic method achieves fair outreach as compared to the heuristic method on the Instagram dataset. The authors also theoretically proved that degree heuristics in some conditions having differentiated thresholds for choosing top-$k$ seed nodes in each community provides higher outreach with more diversity. In simple words, the authors suggest that the equal number of chosen nodes from each community, irrespective of their degree, provides a better diverse outreach. 

Stoica et al. \cite{stoica2020seeding} further compared three different methods for selecting early adopters (i) agnostic seeding, i.e., feature blind and choose all nodes above the given threshold irrespective of their community membership, (ii) parity seeding, which maintains the ratio of early adopters with respect to their community size, or we can say, it uses differentiated degree thresholds for both groups, and (iii) diversity seeding that uses differentiated thresholds based on the influence outreach. The authors theoretically showed that if the budget is less than a specific threshold (depending on the network), then the diversity seeding achieves a larger expected outreach than the agnostic seeding and gets close to the parity outreach. The authors verified the theoretical results on the DBLP dataset, having 81\% males and 19\% females with a high homophilic structure. The authors used degree heuristic for all three kinds of seeding. For diversity seeding, the authors used a relaxation parameter that selects the seed nodes in between agnostic and parity seeding, and the best results were considered. The diversity seeding showed a slight improvement over other methods to get an efficient and diverse outreach using the independent cascade model, especially for the minor community. However, an increase in the outreach of minor community is at the cost of a decrease in the outreach of major community. Teng et al. \cite{teng2021influencing} further proposed disparity seeding based methods for a different IM problem setting, where top-k seed nodes are selected such that a given target gender ratio $\zeta$ should be maintained in the influenced nodes with an error margin $e$.  

Tsang et al. \cite{tsang2019group} studied fair resource allocation for IM from a group perspective using Maximin and Diversity constraints. The authors showed that fairness-aware IM is non-submodular as opposed to the classic IM problem (refer to Section 3 in \cite{tsang2019group} for the detailed proof). The authors proposed a method using multi-objective submodular optimization that can satisfy either fairness constraints and provides an asymptotic approximation guarantee of $1 -1/e$. The authors theoretically proved that the price of group fairness can be very high under a range of network structures for both fairness constraints. One notable finding is that the price of fairness gets worse if nodes join multiple communities (overlapping community structures). The proposed solution with diversity constraint performs 55-65\% better as compared to the greedy approach and also gives competitive performance for maximin constraint. However, when the algorithm was applied with the maximin constraint, it provided the best maximin value, as expected. In the experiments, the price of fairness ranges from 1.05-1.15, which is comparatively very small as provided by the theoretical bounds. It suggests that applying fairness-aware methods in practice is less costly than observed in the theoretical worst-case scenario.

Becker et al. \cite{becker2020fairness} proposed to use probabilistic strategies instead of deterministic strategies to achieve group fairness using maximin. The results are promising and show that randomization can be further used for achieving efficient and fair solutions for other fairness constraints in IM. 
Farnadi et al. \cite{farnad2020unifying} formulated fairness in IM using Mixed Integer Programming (MIP), where the fairness requirements were enforced using the linear constraints or updating the objective function. The proposed framework is a generalized framework that can achieve four types of group fairness (equity, equality, maximin, and diversity) with competitive results. The authors also mentioned that each kind of fairness has certain properties, and a proposed method might not be able to satisfy all these properties using a single fairness metric. 

Khajehnejad et al. \cite{khajehnejad2020adversarial} proposed the first work that used adversarial learning based network embedding to identify seed nodes for achieving fair IM. The adversarial graph embedding is generated by co-training an auto-encoder for network embedding and a discriminator to discern sensitive attributes. Therefore, in the generated embedding, the nodes belonging to different sensitive attribute groups are similarly distributed. Once the embedding is generated, a clustering algorithm is applied, and then the seed nodes are chosen fairly from the identified clusters. A simple way is to choose the centroids of k-clusters as k-seed nodes, though it might not be fair for different groups. For achieving fairness, the identified k-clusters are again divided into subgroups based on the sensitive attribute; for example, each cluster will be divided into two groups (males and females) if the sensitive attribute is gender having two values. Then, the number of seed nodes selected from each subgroup is proportional to its size, according to the given budget $k$. The proposed method reduces disparity, given that it is competitive in total outreach obtained. This is the first work that uses network embedding to choose seed nodes for fair IM. Similar network embeddings can also be explored for other SNA downstream tasks, such as fair node classification or fair link prediction.

\textbf{Balance Trade-off.} A series of works have focused on designing IM methods that aim to balance the trade-off between maximum outreach and fairness; next, we will discuss them. Teng et al. \cite{teng2020influencing} proposed disparity seeding, which focuses on both factors, maximizing the outreach and influencing the required fraction of the target community. In disparity seeding, the influential users are ranked using PageRank, Target HI-index, and Embedding index, and the seed nodes are selected using simulation-based learning. The proposed method outperformed \cite{stoica2020seeding} on the considered datasets; however, the running time can be a concern on large-scale datasets as the seed nodes are identified using a simulation-based approach. 

Anwar et al. \cite{anwar2021balanced} also proposed a method that greedily chooses seed nodes by maximizing an objective function depending on both the total outreach and outreach balance. The algorithm outperforms \cite{stoica2020seeding} in achieving a better total outreach for the given seed nodes size. Rahmattalabi et al. \cite{rahmattalabi2020fair} proposed a family of welfare functions with an inequity aversion parameter that can be varied to get the required trade-off between fairness and maximum outreach. They also observed that high welfare could be obtained without a significant reduction in the total outreach, as observed in \cite{tsang2019group}. 

\textbf{Time Critical Fair IM.} All above-discussed works did not consider the deadline or time constraint while maximizing the influence. However, in real-life applications, receiving the information before the deadline is crucial \cite{chen2012time}. For example, spreading the information about any marketing discount that is valid for a short period of time is time-critical. In Time Critical IM (TCIM), the aim is that within the given time deadline, the fraction of influenced nodes across different groups should follow the fairness constraint. Ali et al. \cite{ali2019fairness} studied group-fairness in TCIM, where the fraction of influenced nodes should be equal across all groups. The authors mainly worked on two variants of the TCIM problem, (i) FAIRTCIM-BUDGET, where the budget (i.e., the number of seed nodes) is fixed, and the aim is to choose seed nodes that maximize the time-critical influence, and (ii) FAIRTCIM-COVER, where the aim is to find a minimal size set of seed nodes such that the given fraction of the population is influenced within the given time deadline. As the problem is NP-hard, the authors proposed a greedy approximation solution with provable guarantees. The empirical results showed that fairness comes with the price of reduced total outreach though it is bounded as guaranteed by the proposed solution. 

\textbf{Individual Fairness.} In Individual fairness, the aim is to balance the probability of receiving the information for each individual. Simply put, the probability of receiving the information for each individual should be the same irrespective of their connectivity and community. 
Fish et al. \cite{fish2019gaps} investigated individual fairness using the notion of information access gap, where they aim to maximize the minimum probability of receiving the information by an individual to reduce this gap. The authors showed that maximizing the minimum probability of information access is NP-hard. The authors used the maximin social welfare function as an objective function and proved that it reduces the access gap. They further proposed various greedy-based algorithms that provide good results. In \textbf{greedy} solution, the probabilities are computed for each node, and the node that maximizes the objective function is added to the solution. However, the proposed solution is slower as the probabilities will be computed for each node. Therefore, the authors further proposed a faster solution, called \textbf{Myopic}, in which, in each round, the node with the currently smallest probability will be added to the seed nodes without evaluating the objective function. Another faster variation is \textbf{Naive Myopic}, in which the probabilities are estimated initially, and then top-$k$ nodes with the smallest probabilities are chosen as the seed nodes. An alternative fast solution, called \textbf{Gonzalez}, avoids estimating the probabilities and chooses seed nodes that are distant from each other. This method will pick the next seed node that is furthest from the current seed nodes. The empirical results on real-world networks showed that both the Myopic and Naive Myopic methods provide better total outreach with improved information access for each node in the network. These greedy methods have high time complexity, and a faster heuristic method will always be well received. 

Jalali et al. \cite{jalali2020information} observed that the information unfairness increases with increasing the number of seed nodes $k$ and influence probability $p$ in real-world networks. The reason is that as $k$ will increase, information cascades will reach nodes of the small size community though the number of cascades passing through the local neighborhood in large size community will increase exponentially. Therefore, the difference in distributions of information access between both groups will increase. However, if $p$ is small, even if $k$ is high, the cascades will reach a very small number of nodes and will die out. 
The authors proposed a link recommendation method, called MaxFair, that predicts the given number of links which will minimize the information unfairness in the network. The edges receive a positive score for increasing information flow between group pairs with below-average flow and a negative score for increasing flow between above-average group pairs. Then a final score is computed for all pairs of unconnected nodes, and the pair having the highest score is selected in each iteration. The proposed method shows a decrease in information unfairness and can be adopted for link recommendation in OSNs for fair information flow. The diverse link recommendation methods will reduce structural bias and will also impact other network phenomena, such as opinion formation of users and forming echo chambers in OSNs.  

In Table \ref{tableim}, we summarize works on fairness-aware IM. In the table, columns are as follows: (i) Ref - reference of the paper, (ii) name of the proposed methods, (iii) complexity of the problem statement and if there is any other remark to mention, (iv) considered fairness constraints, (v) top-k - if the proposed method aims to choose top-k seed nodes based on the given budget, (vi) Target Nodes - if the method aims to save the given target nodes, (viii) $\alpha$ - decontamination ratio that includes constraints, such as maintain the ratio of influenced nodes in different communities or maintain the influence ratio for the minimally influenced community, (ix) diffusion model - the used influence propagation model, and (x) baselines considered in the work. All of the works in Fair-IM have considered Independent Cascade (IC) model.

\afterpage{
\begin{landscape}
\begin{table}[]
\centering
\caption{Fairness-aware Influence Maximization methods.}
\label{tableim}
\begin{tabular}{|p{.6cm}|p{2.1cm}|p{5cm}|p{2.5cm}|C{.5cm}|C{.8cm}|C{.4cm}|C{.4cm}|C{1.2cm}|p{2.2cm}|}
\hline
\textbf{Ref} & \textbf{Proposed Methods} & \textbf{Complexity \& Remarks} & \textbf{Fairness Constraints} & \textbf{Top-k} & \textbf{Target Nodes} & \textbf{$\alpha$} & \textbf{$t$} & \textbf{Diffusion Model} & \textbf{Baselines} \\ \hline

\cite{stoica2019fairness}    &  Differentiated Seeding & -  &  Equity, Equality, Maximin, Diversity & \cmark  & -  &  -  &  - & IC  & Greedy  \\ \hline

\cite{fish2019gaps}  &   Greedy, Myopic, Naive Myopic, Gonzalez   &     NP-hard  &   Maximin    &    \cmark          &  - & -  & - &   IC  &  	TIM+ \cite{tang2014influence}, Random          \\ \hline
	
\cite{farnad2020unifying}    & MIP  &  Used Mixed Integer Programming encodings of fairness measures in influence maximization & Equity, Equality, Maximin, Diversity  & \cmark  & -  & -  & -  &  IC & CELF \cite{leskovec2007cost}, Simpath \cite{goyal2011simpath}, TIM \cite{tang2014influence} , IMM \cite{tang2015influence}, FairIM \cite{tsang2019group}  \\ \hline 
	    
\cite{tsang2019group}   & Tsang Method  & fairness-aware IM is non-submodular; approximation ratio $\to (1-1/e)$ as $k \to \infty $  & Maximin, Diversity  &  \cmark &  - & -  & - & IC &  Greedy \cite{kempe2003maximizing} \\ \hline 
			
\cite{rahmattalabi2020fair}    & DC and Maximin Fair IM  & Prove monotonicity and submodularity of the resulting optimization problem; Greedy algorithm for fair IM through welfare maximization  & Maximin, Diversity  & \cmark  & -  & -  & -  & IC  & fairness-oblivious IM  \\ \hline
    
\cite{stoica2020seeding}    & Parity \& Diversity Seeding  & -  &  Equality & \cmark  &  - & -  &  - & IC  & Agnostic seeding  \\ \hline 
    
\cite{khajehnejad2020adversarial}    & Fair-Emb &  - & Equality  & \cmark  & -  & -  & -  & IC  & Greedy \cite{kempe2003maximizing}, Tsang et al. \cite{tsang2019group}  \\ \hline
    
\cite{ali2019fairness}    & FAIRTCIM-COVER  &  $k \leq ln(1+|V|)(\sum_{i=1}^k |S_i^*|) $, where $S_i^*$ is an optimal solution & Disparity  & -  & -  & \cmark  & \cmark  & IC &  - \\ \hline

\cite{ali2019fairness}    & FAIRTCIM-BUDGET  &  Lower bound of total Influence $\geq (1-1/e) \cdot \frac{H(z)}{z}$ where $z$ is the optimal solution and $H$ is a monotone concave function. & Disparity  & \cmark  & -  & -  & \cmark  & IC  &  - \\ \hline
    
\cite{becker2020fairness}    & Set-based and Node-based solution  & approximate within a factor of $(1 - 1/e)$ plus an additive small error $\epsilon$  & Maximin  & \cmark   &  - &  - &  - &  IC & Myopic \cite{fish2019gaps}, Greedy, Tsang et al. \cite{tsang2019group}  \\ \hline   
    
\cite{anwar2021balanced}    & Balanced IM  &  $(1-1/e)$ approximation guarantee. problem is monotone and submodular &  Categorical Balance in outreach & \cmark  &  - &  - & -  &  IC &  Diversity Seeding \cite{stoica2020seeding} \\ \hline 
    
\cite{teng2021influencing}    &  Disparity Seeding &  NP-hard & Disparity (ratio of influenced users is $\zeta$ within an error margin $e$) & \cmark  & -  & \cmark  & -  & IC  & Diversity seeding \cite{stoica2020seeding}, IM-balanced \cite{gershtein2018balanced}  \\ \hline
    
\end{tabular}
\end{table}
\end{landscape}
}

\subsubsection{Future Directions.}

Fairness-aware IM methods for various constraints applicable in real-life applications still have not been proposed. For example, In IM with Priority (IMP), the aim is to find a set of $k$ seed nodes that obtains the maximum outreach, given that the influence should spread to a given target set of nodes, called priority set, with at least the given threshold $T$ \cite{pham2020influence}. The fairness-aware solution for IMP is still an open question. The solutions will have to consider various cases, such as whether the priority set is equally distributed over all communities or not, and if not, what is the best achievable solution. Other interesting constraints are time constraints for time-critical information, outreach threshold that should be achieved, and rigid users that might not be propagating the information. The generic solutions that might achieve fairness given any combination of constraints will also be very appreciated.

Individual fairness in IM is not much explored and still requires researchers' attention. Fish et al. \cite{fish2019gaps} proposed a method to reduce the information access gap to improve individual fairness. However, the applicability of the method for time-constrained information is not explored. In the case of social interventions, or awareness programs, the information accessibility must be fair for each individual. Therefore, It is essential to propose methods that can maintain individual fairness for time-critical information propagation.  
Another important point to note is that all the proposed fairness-aware methods have used the Independent Cascade model for information propagation. Therefore, the achieved fairness with respect to other influence propagation models, such as the Linear threshold model or SIR model, is still not explored \cite{azaouzi2021new, sumith2018influence}. Besides these, there are several other extensions of ICM to model influence propagation in real-world, such as shortest path model \cite{kimura2006tractable}, penta-level spreading model \cite{saxena2015understanding, gupta2016modeling}, trust-based latency-aware independent cascade \cite{mohamadi2015trust}, conformity-aware cascade model \cite{li2015conformity}, continuous-time markov chain-independent cascade model \cite{zhu2014maximizing}, and dynamic independent cascade model \cite{tong2016adaptive}. An in-depth comparative analysis of different IM methods for different influence propagation models from a fairness perspective will help get a deeper understanding of the problem and further design promising solutions. For example, in the case of the Linear Threshold model \cite{kempe2003maximizing}, the existing IM methods \cite{chen2010scalable,goyal2011simpath, bozorgi2016incim} focus on identifying a critical group of nodes that maximize the influence propagation locally, and therefore, the information might be constrained within the major community and may not reach to the minor community. Therefore, the bias in such critical mass models, including the Linear Threshold model \cite{kempe2003maximizing}, Tripping model \cite{wu2016mining}, and Multiple Adoption Linear Threshold model \cite{pham2019multiple}, should be further studied. 

Apart from these, in recent years, researchers have focused on identifying topic-based influential users \cite{panchendrarajan2023topic, varshney2014modeling}. At the same time, some studies have shown gender-based biases in online communication on social media \cite{nilizadeh2016twitter, manzano2019women, messias2017white, macedo2024gender}. However, these two topics have been studied separately and should be analyzed together to highlight the impact of communication biases and following patterns on the (perceived) influential ranking of different types of users in a range of topics from home decor to STEM. Another practical topic has been to compute influence probabilities in real-world networks, and most of these works have yet to consider nodes' characteristics based on sensitive attributes \cite{varshney2017predicting}, and one can look further to understand it better.

\subsection{Fair Influence Blocking Maximization}\label{secibm}

Social Media is beneficial in spreading valuable information and awareness among people, though at the same time, it has also been misused to propagate rumors, fake news, propaganda, or misinformation. The adverse impact of fake information propagation has been seen in several major events, such as the USA and Indonesian presidential elections \cite{bovet2019influence, grinberg2019fake, santosa2018digital}. It motivated researchers to study influence blocking maximization (IBM), also known as influence minimization, on social media. In IBM, the aim is to identify a minimal set of users whose blocking or immunization will minimize the spread of misinformation in the network \cite{he2012influence, wu2017scalable, erd2021generalized}. The immunized or blocked nodes do not spread the information further.

\subsubsection{Fairness Definitions.}

Fairness in IBM is yet to be defined than being solved. Here, we define some fairness constraints that can be considered in IBM. Let's assume that $I(R, G)$ denotes the rumor outreach (set of influenced nodes) from rumor starters $R$ on the network $G$ and $I(S, R, G)$ denotes the set of influenced nodes by rumor outreach in the network $G$, given $R$ is the set of rumor starters and $S$ is the set of immunized nodes. A node $u$ is called \textit{saved node} by immunized nodes $S$ if $u \in I(R, G)$ and $u \notin I(S, R, G)$. 

\begin{enumerate}
    \item Equality: It ensures that the chosen immunized nodes in each community should be proportional to the number of nodes that might be affected by misinformation in the absence of the immunized nodes. This definition is based on the fact that the communities that might be more affected should have more immunized nodes to save them. It is defined as, 

\begin{equation*}
        \frac{|\{u \in S | u \in C_i\}|}{E|\{u \in I(R, G) | u \in C_i\}|} = \frac{|\{u \in S | u \in C_j\}|}{E|\{u \in I(R, G) | u \in C_j\}|}, \forall i,j
\end{equation*}

\item Equity: The equity constraint suggests that the proposed solution is fair in saving nodes if the expected number of saved nodes for all communities are proportional to their sizes.  
    \begin{equation*}
        \frac{E|\{u \in I(R, G) \; \& \; u \notin I(S, R, G) \; | \; u \in C_i\}|}{|C_i|} = \frac{E|\{u \in I(R, G) \; \& \; u \notin I(S, R, G) \; | \; u \in C_j\}|}{|C_j|}, \forall i,j
    \end{equation*}
The equity constraint can also be defined as the expected number of saved nodes for all communities should be proportional to the nodes influenced by the rumor in respective communities. This will ensure that each community gets the resources compared to the number of people who would have been affected by the rumor spread.

Maximin \cite{saxena2023fairness}: The maximin fairness aims to maximize the fraction of saved nodes for a community that has minimum proportion of saved nodes. It will maximize the maximin value that is computed as, 
    \begin{equation*}
        \min_{C_i \in C}\frac{E|\{u \in I(R, G) \; \& \; u \notin I(S, R, G) \; | \; u \in C_i\}|}{E|\{u \in I(R, G) | \; u \in C_i\}|}
    \end{equation*}
    
\end{enumerate}

Other group and individual fairness definitions discussed for IM can also be similarly extended for IBM concerning the saved nodes.

\subsubsection{State-of-the-art and Future Directions.}

First, we would like to highlight that the solutions for fairness-aware influence maximization can not directly be applied to achieve fair IBM, as the solution in IBM is also dependent on the set of rumor starters. Therefore, given the same network and the same budget for immunizing nodes, the solution might differ for different sets of rumor starters; that is not the case in IM. A practical solution to minimize the misinformation spread is by propagating its counter-true information, i.e., also known as ``truth-campaigning'' \cite{saxena2022fakeprop}. The psychology-based studies have shown that users are more probed to find the correct information and believe in this once they have been exposed to both fake and true information \cite{saxena2022fakeprop}. Most of the existing truth-campaigning methods have focused on maximizing the total number of saved nodes \cite{saxena2020mitigating}. The saved nodes refer to the nodes who believe in true information if the true information is propagated and would have been believed in misinformation in the absence of true information. Fair truth-campaigning methods should ensure that a community can not be manipulated, given its small size or connectivity with the network. In \cite{saxena2023fairness}, Saxena et al. proposed a fairness-aware influence blocking method, called FWRRS (\textbf{F}airness-aware \textbf{W}eighted \textbf{R}eversible \textbf{R}eachable \textbf{S}ystem), that aims to maximize maximin fairness. The proposed method uses weighted reversible reachable trees to compute the blocking power of each node with respect to different communities and then chooses top-k truth campaigning nodes using six selection steps that maximize the fairness-aware optimization function. According to the experimental results, the proposed FWRRS method outperforms both fairness-oblivious and fairness-aware baselines in terms of both fairness and saved nodes. These results also highlighted that fairness does not always have a cost in terms of efficiency, and in many cases, it serves as a catalyst for improving overall effectiveness in the future. The main limitations of this work are that the whole network structure and the community information should be known in advance to achieve group fairness.

Given an abundance of literature on IBM \cite{saxena2022fakeprop}, it is surprising to see that no other work has concerned fairness and structural bias, though there are many interesting questions to be explored. A comparative study of state-of-the-art theoretical and simulation-based methods from a fairness perspective will provide in-depth insights into the impact of structural biases on these methods. These insights can help further propose fair methods for IBM, truth-campaigning based misinformation mitigation, and competitive information propagation maximization. Fair solutions for competitive information propagation will also be interesting for several applications, such as marketing or political parties' agenda sharing.

\subsection{Fair Community Detection}\label{seccd}

In social networks, the nodes are organized into communities. As per the definition proposed by Barab{\'a}si in his book, ``In network science, we call a community a group of nodes that have a higher likelihood of connecting to each other than to nodes from other communities'' \cite{barabasi2014network}. The understanding of community structure has played an essential role in understanding the network evolution as nodes join communities, and these communities are further connected with each other to form a large-scale complex network. As we already discussed, communities play a crucial role in defining fairness for other discussed problems, and therefore, their fairness is dependent on fair communities. For example, if the ground-truth community labels are not known, and communities are not well identified using community detection methods (or small communities are ignored), then feature-aware fair link prediction methods might not be fair for the nodes of minor communities. Ghasemian et al. \cite{ghasemian2019evaluating} compared 16 community detection methods and showed that the number of identified communities varies a lot across different methods. They further studied the impact of community detection methods on link prediction and link description tasks, and observed that no method is always the best for such downstream tasks across all networks. The authors did not study the impact of different community detection methods on different types of nodes or communities having varying sizes and densities. 
Fairness is not yet well defined and studied for community detection (CD) methods, given that it has a vast literature \cite{fortunato2016community}.

\subsubsection{Fairness Definitions.}

Here, we discuss some fairness aspects that should be considered while identifying communities. One important aspect is that small as well as large size communities should be identified equally well. Similarly, sparse as well as dense communities should be identified well if they co-exist in a network. Let's assume that the ground truth communities in a network $G$ are $C=\{C_1, C_2, \cdots ,C_i, \cdots \}$, and the communities identified using a community detection method are $C'= \{C'_1, C'_2, \cdots ,C'_i, \cdots\}$. 

A fairness constraint for community detection based on \textit{\textbf{demographic parity}} can be defined as,

\begin{equation*}
    P(C'(u,v) = 1 | u, v \in C_i) = P(C'(u,v) = 1 | u, v \in C_j), \; \forall i,j
\end{equation*}
where $C'(u,v)$ is $1$ if both nodes $u$ and $v$ belong to the same community in $C'$, otherwise $0$. This constraint ensures that the nodes belonging to the same community in the ground truth should be identified in the same community by the given community detection method, irrespective of their community sizes. However, there is one issue with this definition: if the community detection method returns the entire network as one community, even then, the fairness constraint is satisfied. The reason is that it does not consider if the nodes belong to two different communities. Therefore, the quality of the identified communities should also be considered while maximizing fairness.

Another important fairness constraint based on \textit{\textbf{equity}} can be defined as the ratio of misclassified nodes for all communities should be maintained. First, the ground truth communities should be mapped to the identified communities. An identified community $C'_j$ is mapped to $C_i$ if $|C'_j \cap C_i| \geq |C'_j \cap C_k| \; \forall \; k$. All ties while mapping can be resolved using a heuristic or uniformly at random. If a ground truth community is not mapped with any identified community, then map it to an empty set. Now, the fairness constraint can be defined as,

\begin{equation*}
\frac{|C'_i \cap C_i|}{|C_i|} = \frac{|C'_j \cap C_j|}{|C_j|}, \; \forall i,j
\end{equation*}

For simplicity, in the formula, we assumed that $C_i$ is mapped to $C'_i$ community. The K-L divergence and EMD (refer to Appendix \ref{metrics}) can be used to compute the fairness of a proposed method by comparing achieved and desired distribution. Other methods, such as Jaccard similarity, can also be used for community mapping. 

Khajehnejad et al. \cite{khajehnejad2021crosswalk} used \textit{\textbf{disparity}} to compare the fairness of node classification methods.

\subsubsection{State-of-the-art and Future Directions.}

Mehrabi et al. \cite{mehrabi2019debiasing} highlighted the bias of community detection methods for small-size communities. They showed that well-known methods, such as Louvain \cite{blondel2008fast} and CESNA \cite{yang2013community}, fail to assign lowly connected nodes to proper communities or assign them to various small size communities and exclude them from being included in different analyses. They proposed a method, called CLAN, for attributed networks that assigns lowly connected nodes to large size communities than assigning them to several small meaningless communities. The method works in two steps. In the first step, an unsupervised community detection method that only uses network structure, such as the Louvain method \cite{blondel2008fast}, can be applied. The next step is supervised in which the nodes assigned to insignificant communities are re-assigned to major communities using nodes' attributes that were not used in the first step. This work does not define fairness constraints as it mainly focuses on the bias aspect. Another important point to note is that this is not the only way to reduce bias towards lowly connected nodes. One should aim to identify small communities correctly, and then they should be treated equally well as large communities in other downstream tasks. 

Since fairness has almost never been considered in the community detection context, a simple in-depth analysis of community detection methods from a fairness perspective will be very interesting. In literature, plenty of community detection methods of different types, including modularity optimization based methods, label propagation based methods, betweenness centrality based methods, representation learning based methods, spectral properties based methods, and genetic algorithm based methods, have been defined \cite{plantie2013survey, fortunato2016community, arya2022node}. An exhaustive comparative analysis of different types of methods is required to understand which kind of methods are fairer for which kind of networks or which method is fairer. This analysis will help understand which kind of methods are more respectful to small-size communities and identify them well. A better understanding will provide insights for designing fair community detection methods. Fair community detection methods will be useful in better interpreting results for other research problems as they will ensure that all kinds of communities are well-identified irrespective of their size, density, or connectivity. Apart from crisp communities, where a node only belongs to one community, a fairness analysis of overlapping communities \cite{amelio2014overlapping} will also be of great interest as this will highlight if the lowly connected nodes are well assigned to all communities they belong to, and if not then what is the reason and how it can be further improved. Fair community detection in other types of networks, such as directed networks, weighted networks, multilayer networks, and hypergraphs, should also be looked at given their real-life applications \cite{malliaros2013clustering, lu2014algorithms, huang2021survey, chien2019minimax}. 
Another important point is that several evaluation metrics exist to compare the quality of communities identified using different community detection methods; however, none has considered fairness \cite{chakraborty2017metrics}. How do different evaluation metrics perform compared to each other from fairness perspectives? Evaluation metrics for fair community detection are still an open research direction. 

Fairness has been explored in machine learning based clustering methods and is mainly defined from two different perspectives. From the first perspective, clustering of nodes is called fair if each protected group is equally present in each cluster \cite{chierichetti2018fair, kleindessner2019guarantees}. The principle behind this definition is that each identified cluster should be a good representation of the data and should reflect its diversity. Another well-known fairness constraint is that the cluster center should be a good representative of the members of the cluster by being ``close'' to the points assigned to it \cite{abbasi2021fair,ghadiri2021socially}. In this fairness constraint, the centers should represent different (protected) groups very well. This definition is much closer to the definition of communities in network science and can be applied to identify fair communities using a low-dimensional vector representation of the nodes learned using a network embedding technique.

\subsection{Other Research Topics}\label{secothers}

There are several other topics in SNA for which the network structure-based methods have been designed. Next, we briefly discuss some of these topics and highlight why fairness is an essential factor to be considered. 

\subsubsection{Fairness in Opinion Formation.}

In social networks, a user's opinion is dependent on the opinion of other neighboring users and the user's personal bias. Therefore, opinion formation models take the network structure into consideration while modeling the opinion dynamics \cite{wu2004social}. Nguyen et al. \cite{nguyen2020dynamics} theoretically showed that the minor community nodes converge to the opinion of the majority in a dense network; however, it has not been verified on real-world data. The authors also did not consider the impact of community structures, users' bias, and several other factors that might impact the opinion of users. \cite{wu2004social} also observed that highly connected nodes have a stronger impact on opinion formation dynamics. It is important to understand the opinion formation for minor communities given that we know that minor communities are not able to achieve a higher rank in the network due to the homophily \cite{karimi2018homophily}. As per the best of our knowledge, the fairness of opinion formation models for minor communities is not yet studied, and one main limitation might be the lack of real-world datasets.

\subsubsection{Fairness in Anomaly Detection.}

In anomaly detection, we aim to identify unusual instances in different applications, including malicious users detection in OSNs, fraud detection, and suspicious bank transaction detection \cite{anand2017anomaly, kumar2013detection}. Most of the proposed anomaly detection methods are dependent on network structure as some specific structural pattern can convey abnormal behavior \cite{akoglu2009anomaly, chen2012community}. Unfairness in such systems might affect some particular communities, for example, targeting users from a specific community while identifying suspicious users. Davidson and Ravi \cite{davidson2020framework} compared five classic anomaly detection methods and showed that their outputs are unfair; however, these works might mislead someone to conclude that their results are fair, especially when the number of outliers and the number of protected status variables are small. 
The such analysis raises the question of whether anomalous nodes and links detection methods \cite{ur2018detecting, wani2018mutual} are fair or not for different protected groups in complex networks. If not, then anomaly detection methods for complex network data should address these issues and focus on the fairness of all protected groups. 

\subsubsection{Fairness in Network Anonymization.}

In the past 10-15 years, a massive amount of social networking data has been released publicly and analyzed to better understand complex networks and their different applications. However, ensuring the privacy of the released data has been a primary concern. Most of the graph anonymization techniques can be categorized as (i) graph modification methods and (ii) clustering-based methods \cite{zhou2008brief, casas2017survey}. Briefly, we would like to highlight whether the graph anonymization affects the analysis for different protected groups using anonymized data is not yet studied. Besides, linkability is used to obtain useful information by mapping the data collected from different sources. This is a privacy threat, and the extent of its impact on different types of user groups in anonymized data should be analyzed to propose better fair methods.

\subsubsection{Graph Coverage Problem.}

Fairness has also been explored in graph theory problems. Halabi et al. \cite{halabi2020fairness} proposed approximation algorithms for submodular maximization under fairness constraints for both monotone and non-monotone functions. The authors verified the proposed solutions on various real-life problems, including the maximum coverage problem that is of our interest. 
The results showed a huge improvement in the bias (around 15\% difference) given that the objective value obtained by the fair solution is similar to the unfair baseline methods. The fair solution of maximum coverage can be used further in influence maximization to obtain a fair seed set for starting the spread. However, the application of such solutions for influence maximization is still an open question. The theoretical as well as empirical results in this direction will be interesting for the scientific community.

Recently network structure has also been used to design solutions for role analytics, fake news detection, echo chamber detection, hate spreading user detection, and mental health prediction \cite{saxena2022fake, ribeiro2018characterizing, villa2021echo, kawachi2001social, forestier2012roles}, and the proposed methods might also inherit the bias of network structure that should be addressed.

In Table \ref{fairsnatable}, we summarize state-of-the-art fairSNA methods that includes LP (link prediction), CR (centrality ranking), IM (influence maximization), IBM (influence blocking maximization), and CD (community detection). It shows that most fair methods are feature-aware and use in-processing, and individual fairness is not well studied in SNA. This table highlights the gap in research on fairness in SNA. In this work, we have discussed ample open research directions, and hopefully, in the future, it will help in bridging this research gap.

In this section, we also discussed metrics used to evaluate fairness for different problems; they are further summarized in Appendix \ref{metrics} for the sake of completeness. 


\begin{table}[]
\caption{FairSNA State-of-the-art Summary.} 
\label{fairsnatable}
\begin{tabular}{p{1.1cm}C{1.2cm}C{2cm}C{2.1cm}C{.9cm}C{1.3cm}C{2.1cm}C{1.3cm}}
\hline
\textbf{Research Topic} & \textbf{Individual fairness}               & \textbf{Group Fairness}             & \textbf{Feature-Aware}    & \textbf{Feature-Blind} & \textbf{Pre-Processing} & \textbf{In-Processing} & \textbf{Post-Processing}                                  \\ \hline
\textbf{LP}            & \cite{rahman2019fairwalk, laclau2021all} & \cite{rahman2019fairwalk, saxena2021hm, khajehnejad2021crosswalk, saxena2021nodesim, masrour2020bursting, laclau2021all}   & \cite{rahman2019fairwalk, saxena2021hm, khajehnejad2021crosswalk, saxena2021nodesim, masrour2020bursting, laclau2021all} &  \textendash  &  \cite{laclau2021all}   & \cite{rahman2019fairwalk, khajehnejad2021crosswalk, saxena2021nodesim, masrour2020bursting} & \cite{saxena2021hm, masrour2020bursting} \\ \hline
\textbf{CR}            & \textendash                                          & \cite{tsioutsiouliklis2021fairness}                                            & \cite{tsioutsiouliklis2021fairness}  & \textendash                      & \textendash                       & \cite{tsioutsiouliklis2021fairness}  & \textendash                                                     \\ \hline
\textbf{IM}            & \cite{fish2019gaps}                               & \cite{farnad2020unifying, ali2019fairness, stoica2019fairness, stoica2020seeding, teng2020influencing, anwar2021balanced, tsang2019group, rahmattalabi2020fair, khajehnejad2020adversarial, becker2020fairness, khajehnejad2021crosswalk} &  \cite{fish2019gaps,farnad2020unifying, stoica2019fairness, stoica2020seeding, teng2020influencing, anwar2021balanced, tsang2019group, rahmattalabi2020fair, khajehnejad2020adversarial, becker2020fairness, ali2019fairness, khajehnejad2021crosswalk}  & \textendash                     & \textendash                       & \cite{fish2019gaps,farnad2020unifying, ali2019fairness, stoica2019fairness, stoica2020seeding, teng2020influencing, anwar2021balanced, tsang2019group, rahmattalabi2020fair, khajehnejad2020adversarial, becker2020fairness, khajehnejad2021crosswalk}                                                      & \textendash                                                         \\ \hline
\textbf{IBM}           & \textendash                                          & \cite{saxena2023fairness}          & \cite{saxena2023fairness}   & \textendash                      & \textendash                       & \cite{saxena2023fairness}    & \textendash   \\ \hline
\textbf{CD}                                              & \textendash  & \cite{khajehnejad2021crosswalk}                                          & \cite{mehrabi2019debiasing, khajehnejad2021crosswalk} & \textendash                      & \textendash                      & \cite{khajehnejad2021crosswalk}                                                                                                   & \cite{mehrabi2019debiasing}                                      \\ \hline
\end{tabular}
\end{table}

\section{Datasets}\label{datasets}

One main limitation in designing fairness-aware methods in SNA is the limited availability of datasets, as the labeling of the dataset is a difficult task. In Table \ref{datasettable}, we summarize real-world datasets used in different works, i.e., mentioned in the last column. 
Some methods were also verified on other complex networks than social networks; mentioned in the last three rows of Table \ref{datasettable}. 
In most research works, researchers have used synthetic network generating models to create benchmark datasets for an in-depth understanding of the impact of different parameters, such as homophily, size of the minor community, or network density. The synthetic models to generate homophilic networks having minor major communities are (i) Homophily BA Model \cite{karimi2018homophily, lee2019homophily}, (ii) Diversified Homophily BA \cite{wang2021information}, (iii) Directed Homophily model \cite{anwar2021balanced}, and (iv) Organic Growth Model \cite{stoica2018algorithmic}; refer Appendix \ref{appendixsynmodel} for detailed synthetic models.

\begin{table}[!]
\caption{Real-World Network Datasets.}
\begin{tabular}{p{3.35cm}C{1cm}C{1cm}C{2.3cm}C{4.2cm}C{1.2cm}}
\hline
\textbf{Dataset}                            & \textbf{\#Nodes}   & \textbf{\#Edges}   & \textbf{Protected Attributes}                                      & \textbf{Remarks}                                                 & \textbf{Works}                                                             \\
\hline
Instagram 1 \cite{stoica2019fairness}                      & 539,023                     & 640,211                     & Gender                                                   & 45.57\% males \& 54.43\% females                                                  & \cite{stoica2019fairness}                      \\ \hline
Instagram-London   \cite{rahman2019fairwalk}                & 53,902                      & 165,184                     & Gender, Race                                             & Races: African, Caucasian, Asian                                                 & \cite{rahman2019fairwalk}                      \\ \hline
Instagram-Los Angeles \cite{rahman2019fairwalk}             & 82,607                      & 482,305                     & Gender, Race                                             & African, Caucasian or Asian                                                 & \cite{rahman2019fairwalk}                      \\ \hline 
DBLP \cite{stoica2019fairness}                              & 53,307                      & 288,864                     & Gender                                                   & 81\% males and 19\% females                                                        & \cite{stoica2019fairness, jalali2020information, tsioutsiouliklis2021fairness, wang2021information}                      \\ \hline
DBLP-Data Mining and Database \cite{tsioutsiouliklis2022link}                              & 16,501                      & 66,613                     & Gender                                                   & 25.7\% females; Considered publications from 2011 to 2020  & \cite{tsioutsiouliklis2022link}                      \\ \hline 
DBLP-continent \cite{stoica2019fairness}                              & 3,980                      & 6,965                     & Continent                                                   & labels have 5 continent values                                                       & \cite{laclau2021all, spinelli2021biased}                      \\ \hline
Enron \cite{shetty2004enron}                             & 144                         & 1,344                        & Gender                                                   & 76\% men and 24\% women                                                            & \cite{jalali2020information} 
\\ \hline
Norwegian Interlocking Directorate Network \cite{seierstad2011few} & 1,421                        & 3,855                        & Gender                                                   & 63\% men and 37\% women                                                            & \cite{jalali2020information} 
\\ \hline
Github Follower network \cite{karimi2016inferring}           & 89,630                      & 167,359                      & Gender                                                   & 5.7\% females                                                                      & \cite{wang2021information} 
\\ \hline
Brazilian network \cite{rocha2011simulated}                 & 16,730                      & 39,044                      & Gender                                                   & 10,106 sex workers \& 6,624 sex buyers; 40\% minority nodes                    & \cite{lee2019homophily}                      \\ \hline
POK Swedish Dating Network \cite{holme2004structure}        & 29,341                      & -      & Gender                                                   & minority group has 44\% nodes                                                      & \cite{lee2019homophily}                      \\ \hline 
Dutch School Friendship Network \cite{snijders2010introduction}       & 26                       & 221       & Gender                                                   & 17 Girls \& 9 Boys                                                       & \cite{masrour2020bursting}                         \\ \hline
Google+ Network \cite{mcauley2012learning}       & 4,938                       & 547,923       & Gender                                                   & anonymized gender data                                                       & \cite{masrour2020bursting}                         \\ \hline
Facebook Network \cite{mcauley2012learning}       & 4,039                       & 88,234       & Gender                                                   & anonymized gender data                                                       & \cite{laclau2021all, spinelli2021biased, masrour2020bursting}                         \\ \hline
Rice University Facebook \cite{mislove2010you}          & 1,205                       & 42,443                      & Gender, Age                                              & Nodes' Attributes: student id, age (18 to 22), and major                           & \cite{khajehnejad2020adversarial, ali2019fairness}                          \\ \hline
Pokec social network \cite{takac2012data}              & 1,632,803                   & 30,622,564                  & Gender, Age                                              & Nodes' attributes: gender, age, hobbies, interest, education                       & \cite{halabi2020fairness, dai2021say, deniz2021fairness}                           \\ \hline
Twitter network \cite{rossi2015network}              & 18,470                   & 61,157                  & Political Inclination                                              & 61\% nodes in Political (left)                       & \cite{tsioutsiouliklis2021fairness}                           \\ \hline
Twitter network-2 \cite{babaei2016efficiency}              & 3,560                   & 6,677                  & Political Inclination                                              & 2598 nodes in neutrals, 782 nodes in liberals, and 180 nodes in conservatives group                       & \cite{khajehnejad2021crosswalk}                           \\ \hline
HIV Prevention for homeless youth \cite{wilder2020clinical}           & 60-70   & -       & birth sex, gender identity, race, and sexual orientation & 4 datasets each having 60-70 nodes                                                 & \cite{tsang2019group}                          \\ \hline
Homeless Youth Social Networks \cite{barman2016sociometric}                   & 124-296 & 111-326 & Race                                                     & six social networks of homeless youths from US                                     & \cite{rahmattalabi2020fair}  \\      \hline    
APS citation network \cite{lee2019homophily}                    & 1853                        & 3,627                       & Research Fields                                          & 37\% Classical Statistical Mechanics, and 63\% Quantum Statistical Mechanics & \cite{wang2021information}                        \\ \hline 
US Politics Blogs network \cite{adamic2005political}                    & 1,222                        & 19,089                       & Political (left)                                          & 52\% nodes in Political (left)  & \cite{tsioutsiouliklis2021fairness, laclau2021all}                        \\ \hline 
US Politics Books Co-purchasing network \cite{networkdata}                    & 92                        & 748                       & Political (left)                                          & 47\% nodes in Political (left)  & \cite{tsioutsiouliklis2021fairness}                        \\ \hline
\end{tabular}
\label{datasettable}
\end{table}

\section{Conclusion}\label{conclusion} 

In the past two decades, social network analysis has been highly used to understand different complex phenomena, ranging from individual to group-based human behavior. Structural properties of nodes and networks provide insights into understanding their roles in the system. Therefore, several solutions in this domain, including for influence maximization, recommendation systems, and centrality ranking, have used structural properties of the nodes and nodes' attributes. However, due to homophily and different sizes of protected groups, these networks have structural biases that might impact the analysis for some specific groups based on their size, density, or protected attribute. It is therefore important to account for these biases and consider fairness while designing complex network structure-based solutions for such problems as they might unfairly impact some specific groups of the society.

In this paper, we discussed the taxonomy of fairSNA, followed by various SNA-related research topics, including why fairness is an important aspect to be considered, how fairness can be defined, state-of-the-art, and our vision as future directions. We highlighted in Section \ref{section2} that most of the fair methods had been proposed in the past three years, and still, the research is in its infancy state. Fairness has not yet been explored for some problems, and we defined fairness constraints that can be considered for such problems. Most of the proposed methods in FairSNA are feature-aware and use in-processing techniques. However, feature-blind fair methods will be much more useful in SNA due to privacy concerns as well as the availability of limited features. Another important point to note is that individual fairness is mostly unexplored in SNA, and the price of fairness for achieving individual fairness is unknown.  
We observed that different network embedding techniques have been used to achieve fairness in downstream tasks, such as link prediction and influence maximization, and provide promising results. This direction can also be further explored for achieving fairness for other downstream tasks. 

Last but not least, we would like to highlight that within the wider fairness research community, the interdisciplinary nature of tackled problems has been widely acknowledged \cite{mulligan2019thing, de2024canon}. There are attempts to bridge the understanding gaps in terminology, formulation of research questions and used research methods in computing sciences, social sciences and humanities at large. 
Research community is asking foundational questions - \textit {How to determine the most suitable fairness metric for a particular use case?} \cite{hellman2020measuring, hertweck2021moral, weerts2023can, gajane2022survey}, \textit{Can we morally justify the ways how fairness metrics are optimized?} \cite{weerts2022does}, \textit{What types of bias and unfairness does the law address when it prohibits discrimination?}, \textit{What role can fairness metrics play in establishing legal compliance?} \cite{weerts2023algorithmic}, and \textit{Can fairness be automated?} \cite{weerts2023can}. 
Through this paper, we would like to get the attention of researchers, including network scientists, data scientists, and machine learning experts, towards this gap in research. We have discussed several open research directions that one can look at. In future, this paper should motivate researchers to consider fairness while proposing solutions in social network analysis and computational social science. 

\bibliographystyle{ACM-Reference-Format}
\bibliography{mybib}


\begin{thebibliography}{206}


\ifx \showCODEN    \undefined \def \showCODEN     #1{\unskip}     \fi
\ifx \showDOI      \undefined \def \showDOI       #1{#1}\fi
\ifx \showISBNx    \undefined \def \showISBNx     #1{\unskip}     \fi
\ifx \showISBNxiii \undefined \def \showISBNxiii  #1{\unskip}     \fi
\ifx \showISSN     \undefined \def \showISSN      #1{\unskip}     \fi
\ifx \showLCCN     \undefined \def \showLCCN      #1{\unskip}     \fi
\ifx \shownote     \undefined \def \shownote      #1{#1}          \fi
\ifx \showarticletitle \undefined \def \showarticletitle #1{#1}   \fi
\ifx \showURL      \undefined \def \showURL       {\relax}        \fi
\providecommand\bibfield[2]{#2}
\providecommand\bibinfo[2]{#2}
\providecommand\natexlab[1]{#1}
\providecommand\showeprint[2][]{arXiv:#2}

\bibitem[Abbasi et~al\mbox{.}(2021)]%
        {abbasi2021fair}
\bibfield{author}{\bibinfo{person}{Mohsen Abbasi}, \bibinfo{person}{Aditya
  Bhaskara}, {and} \bibinfo{person}{Suresh Venkatasubramanian}.}
  \bibinfo{year}{2021}\natexlab{}.
\newblock \showarticletitle{Fair clustering via equitable group
  representations}. In \bibinfo{booktitle}{\emph{Proceedings of the 2021 ACM
  Conference on Fairness, Accountability, and Transparency}}.
  \bibinfo{pages}{504--514}.
\newblock


\bibitem[Adamic and Adar(2003)]%
        {adamic2003friends}
\bibfield{author}{\bibinfo{person}{Lada~A Adamic} {and} \bibinfo{person}{Eytan
  Adar}.} \bibinfo{year}{2003}\natexlab{}.
\newblock \showarticletitle{Friends and neighbors on the web}.
\newblock \bibinfo{journal}{\emph{Social networks}} \bibinfo{volume}{25},
  \bibinfo{number}{3} (\bibinfo{year}{2003}), \bibinfo{pages}{211--230}.
\newblock


\bibitem[Adamic and Glance(2005)]%
        {adamic2005political}
\bibfield{author}{\bibinfo{person}{Lada~A Adamic} {and}
  \bibinfo{person}{Natalie Glance}.} \bibinfo{year}{2005}\natexlab{}.
\newblock \showarticletitle{The political blogosphere and the 2004 US election:
  divided they blog}. In \bibinfo{booktitle}{\emph{Proceedings of the 3rd
  international workshop on Link discovery}}. \bibinfo{pages}{36--43}.
\newblock


\bibitem[Akoglu et~al\mbox{.}(2009)]%
        {akoglu2009anomaly}
\bibfield{author}{\bibinfo{person}{Leman Akoglu}, \bibinfo{person}{Mary
  McGlohon}, {and} \bibinfo{person}{Christos Faloutsos}.}
  \bibinfo{year}{2009}\natexlab{}.
\newblock \showarticletitle{Anomaly detection in large graphs}. In
  \bibinfo{booktitle}{\emph{In CMU-CS-09-173 Technical Report}}. Citeseer.
\newblock


\bibitem[Al~Hasan and Zaki(2011)]%
        {al2011survey}
\bibfield{author}{\bibinfo{person}{Mohammad Al~Hasan} {and}
  \bibinfo{person}{Mohammed~J Zaki}.} \bibinfo{year}{2011}\natexlab{}.
\newblock \showarticletitle{A survey of link prediction in social networks}.
\newblock In \bibinfo{booktitle}{\emph{Social network data analytics}}.
  \bibinfo{publisher}{Springer}, \bibinfo{pages}{243--275}.
\newblock


\bibitem[Albanese(2007)]%
        {albanese2007criminal}
\bibfield{author}{\bibinfo{person}{Jay Albanese}.}
  \bibinfo{year}{2007}\natexlab{}.
\newblock \showarticletitle{A criminal network approach to understanding \&
  measuring trafficking in human beings}.
\newblock In \bibinfo{booktitle}{\emph{Measuring human trafficking}}.
  \bibinfo{publisher}{Springer}, \bibinfo{pages}{55--71}.
\newblock


\bibitem[Ali et~al\mbox{.}(2023)]%
        {ali2019fairness}
\bibfield{author}{\bibinfo{person}{Junaid Ali}, \bibinfo{person}{Mahmoudreza
  Babaei}, \bibinfo{person}{Abhijnan Chakraborty}, \bibinfo{person}{Baharan
  Mirzasoleiman}, \bibinfo{person}{Krishna~P Gummadi}, {and}
  \bibinfo{person}{Adish Singla}.} \bibinfo{year}{2023}\natexlab{}.
\newblock \showarticletitle{On the Fairness of Time-Critical Influence
  Maximization in Social Networks}.
\newblock \bibinfo{journal}{\emph{IEEE Transactions on Knowledge and Data
  Engineering}} \bibinfo{volume}{35}, \bibinfo{number}{03} (\bibinfo{date}{mar}
  \bibinfo{year}{2023}), \bibinfo{pages}{2875--2886}.
\newblock
\showISSN{1558-2191}
\urldef\tempurl%
\url{https://doi.org/10.1109/TKDE.2021.3120561}
\showDOI{\tempurl}


\bibitem[Amelio and Pizzuti(2014)]%
        {amelio2014overlapping}
\bibfield{author}{\bibinfo{person}{Alessia Amelio} {and} \bibinfo{person}{Clara
  Pizzuti}.} \bibinfo{year}{2014}\natexlab{}.
\newblock \showarticletitle{Overlapping community discovery methods: a survey}.
\newblock In \bibinfo{booktitle}{\emph{Social networks: Analysis and case
  studies}}. \bibinfo{publisher}{Springer}, \bibinfo{pages}{105--125}.
\newblock


\bibitem[Anand et~al\mbox{.}(2017)]%
        {anand2017anomaly}
\bibfield{author}{\bibinfo{person}{Ketan Anand}, \bibinfo{person}{Jay Kumar},
  {and} \bibinfo{person}{Kunal Anand}.} \bibinfo{year}{2017}\natexlab{}.
\newblock \showarticletitle{Anomaly detection in online social network: A
  survey}. In \bibinfo{booktitle}{\emph{2017 International Conference on
  Inventive Communication and Computational Technologies (ICICCT)}}. IEEE,
  \bibinfo{pages}{456--459}.
\newblock


\bibitem[Anwar et~al\mbox{.}(2021)]%
        {anwar2021balanced}
\bibfield{author}{\bibinfo{person}{Md~Sanzeed Anwar}, \bibinfo{person}{Martin
  Saveski}, {and} \bibinfo{person}{Deb Roy}.} \bibinfo{year}{2021}\natexlab{}.
\newblock \showarticletitle{Balanced Influence Maximization in the Presence of
  Homophily}. In \bibinfo{booktitle}{\emph{WSDM}}. \bibinfo{pages}{175--183}.
\newblock


\bibitem[Arenas et~al\mbox{.}(2004)]%
        {arenas2004community}
\bibfield{author}{\bibinfo{person}{Alex Arenas}, \bibinfo{person}{Leon Danon},
  \bibinfo{person}{Albert Diaz-Guilera}, \bibinfo{person}{Pablo~M Gleiser},
  {and} \bibinfo{person}{Roger Guimera}.} \bibinfo{year}{2004}\natexlab{}.
\newblock \showarticletitle{Community analysis in social networks}.
\newblock \bibinfo{journal}{\emph{The European Physical Journal B}}
  \bibinfo{volume}{38}, \bibinfo{number}{2} (\bibinfo{year}{2004}),
  \bibinfo{pages}{373--380}.
\newblock


\bibitem[Arya et~al\mbox{.}(2022)]%
        {arya2022node}
\bibfield{author}{\bibinfo{person}{Aikta Arya}, \bibinfo{person}{Pradumn~Kumar
  Pandey}, {and} \bibinfo{person}{Akrati Saxena}.}
  \bibinfo{year}{2022}\natexlab{}.
\newblock \showarticletitle{Node classification using deep learning in social
  networks}.
\newblock In \bibinfo{booktitle}{\emph{Deep Learning for Social Media Data
  Analytics}}. \bibinfo{publisher}{Springer}, \bibinfo{pages}{3--26}.
\newblock


\bibitem[Avin et~al\mbox{.}(2015)]%
        {avin2015homophily}
\bibfield{author}{\bibinfo{person}{Chen Avin}, \bibinfo{person}{Barbara
  Keller}, \bibinfo{person}{Zvi Lotker}, \bibinfo{person}{Claire Mathieu},
  \bibinfo{person}{David Peleg}, {and} \bibinfo{person}{Yvonne-Anne Pignolet}.}
  \bibinfo{year}{2015}\natexlab{}.
\newblock \showarticletitle{Homophily and the glass ceiling effect in social
  networks}. In \bibinfo{booktitle}{\emph{Proceedings of the 2015 conference on
  innovations in theoretical computer science}}. \bibinfo{pages}{41--50}.
\newblock


\bibitem[Avin et~al\mbox{.}(2017)]%
        {avin2017modeling}
\bibfield{author}{\bibinfo{person}{Chen Avin}, \bibinfo{person}{Zvi Lotker},
  \bibinfo{person}{Yinon Nahum}, {and} \bibinfo{person}{David Peleg}.}
  \bibinfo{year}{2017}\natexlab{}.
\newblock \showarticletitle{Modeling and Analysis of Glass Ceiling and Power
  Inequality in Bi-populated Societies}. In
  \bibinfo{booktitle}{\emph{International Conference and School on Network
  Science}}. Springer, \bibinfo{pages}{61--73}.
\newblock


\bibitem[Azaouzi et~al\mbox{.}(2021)]%
        {azaouzi2021new}
\bibfield{author}{\bibinfo{person}{Mehdi Azaouzi}, \bibinfo{person}{Wassim
  Mnasri}, {and} \bibinfo{person}{Lotfi~Ben Romdhane}.}
  \bibinfo{year}{2021}\natexlab{}.
\newblock \showarticletitle{New trends in influence maximization models}.
\newblock \bibinfo{journal}{\emph{Computer Science Review}}
  \bibinfo{volume}{40} (\bibinfo{year}{2021}), \bibinfo{pages}{100393}.
\newblock


\bibitem[Babaei et~al\mbox{.}(2016)]%
        {babaei2016efficiency}
\bibfield{author}{\bibinfo{person}{Mahmoudreza Babaei},
  \bibinfo{person}{Przemyslaw Grabowicz}, \bibinfo{person}{Isabel Valera},
  \bibinfo{person}{Krishna~P Gummadi}, {and} \bibinfo{person}{Manuel
  Gomez-Rodriguez}.} \bibinfo{year}{2016}\natexlab{}.
\newblock \showarticletitle{On the efficiency of the information networks in
  social media}. In \bibinfo{booktitle}{\emph{Proceedings of the Ninth ACM
  International Conference on Web Search and Data Mining}}.
  \bibinfo{pages}{83--92}.
\newblock


\bibitem[Bader et~al\mbox{.}(2007)]%
        {bader2007approximating}
\bibfield{author}{\bibinfo{person}{David~A Bader}, \bibinfo{person}{Shiva
  Kintali}, \bibinfo{person}{Kamesh Madduri}, {and} \bibinfo{person}{Milena
  Mihail}.} \bibinfo{year}{2007}\natexlab{}.
\newblock \showarticletitle{Approximating betweenness centrality}. In
  \bibinfo{booktitle}{\emph{International Workshop on Algorithms and Models for
  the Web-Graph}}. Springer, \bibinfo{pages}{124--137}.
\newblock


\bibitem[Barab{\'a}si(2014)]%
        {barabasi2014network}
\bibfield{author}{\bibinfo{person}{Albert-L{\'a}szl{\'o} Barab{\'a}si}.}
  \bibinfo{year}{2014}\natexlab{}.
\newblock \showarticletitle{Network science book}.
\newblock \bibinfo{journal}{\emph{Network Science}}  \bibinfo{volume}{625}
  (\bibinfo{year}{2014}).
\newblock


\bibitem[Barman-Adhikari et~al\mbox{.}(2016)]%
        {barman2016sociometric}
\bibfield{author}{\bibinfo{person}{Anamika Barman-Adhikari},
  \bibinfo{person}{Stephanie Begun}, \bibinfo{person}{Eric Rice},
  \bibinfo{person}{Amanda Yoshioka-Maxwell}, {and} \bibinfo{person}{Andrea
  Perez-Portillo}.} \bibinfo{year}{2016}\natexlab{}.
\newblock \showarticletitle{Sociometric network structure and its association
  with methamphetamine use norms among homeless youth}.
\newblock \bibinfo{journal}{\emph{Social science research}}
  \bibinfo{volume}{58} (\bibinfo{year}{2016}), \bibinfo{pages}{292--308}.
\newblock


\bibitem[Barocas et~al\mbox{.}(2017)]%
        {barocas2017fairness}
\bibfield{author}{\bibinfo{person}{Solon Barocas}, \bibinfo{person}{Moritz
  Hardt}, {and} \bibinfo{person}{Arvind Narayanan}.}
  \bibinfo{year}{2017}\natexlab{}.
\newblock \showarticletitle{Fairness in machine learning}.
\newblock \bibinfo{journal}{\emph{Nips tutorial}}  \bibinfo{volume}{1}
  (\bibinfo{year}{2017}), \bibinfo{pages}{2}.
\newblock


\bibitem[Becker et~al\mbox{.}(2022)]%
        {becker2020fairness}
\bibfield{author}{\bibinfo{person}{Ruben Becker}, \bibinfo{person}{Gianlorenzo
  D’angelo}, \bibinfo{person}{Sajjad Ghobadi}, {and} \bibinfo{person}{Hugo
  Gilbert}.} \bibinfo{year}{2022}\natexlab{}.
\newblock \showarticletitle{Fairness in influence maximization through
  randomization}.
\newblock \bibinfo{journal}{\emph{Journal of Artificial Intelligence Research}}
   \bibinfo{volume}{73} (\bibinfo{year}{2022}), \bibinfo{pages}{1251--1283}.
\newblock


\bibitem[Beutel et~al\mbox{.}(2019)]%
        {beutel2019fairness}
\bibfield{author}{\bibinfo{person}{Alex Beutel}, \bibinfo{person}{Jilin Chen},
  \bibinfo{person}{Tulsee Doshi}, \bibinfo{person}{Hai Qian},
  \bibinfo{person}{Li Wei}, \bibinfo{person}{Yi Wu}, \bibinfo{person}{Lukasz
  Heldt}, \bibinfo{person}{Zhe Zhao}, \bibinfo{person}{Lichan Hong},
  \bibinfo{person}{Ed~H Chi}, {et~al\mbox{.}}} \bibinfo{year}{2019}\natexlab{}.
\newblock \showarticletitle{Fairness in recommendation ranking through pairwise
  comparisons}. In \bibinfo{booktitle}{\emph{Proceedings of the 25th ACM SIGKDD
  International Conference on Knowledge Discovery \& Data Mining}}.
  \bibinfo{pages}{2212--2220}.
\newblock


\bibitem[Bianconi et~al\mbox{.}(2008)]%
        {bianconi2008local}
\bibfield{author}{\bibinfo{person}{Ginestra Bianconi}, \bibinfo{person}{Natali
  Gulbahce}, {and} \bibinfo{person}{Adilson~E Motter}.}
  \bibinfo{year}{2008}\natexlab{}.
\newblock \showarticletitle{Local structure of directed networks}.
\newblock \bibinfo{journal}{\emph{Physical review letters}}
  \bibinfo{volume}{100}, \bibinfo{number}{11} (\bibinfo{year}{2008}),
  \bibinfo{pages}{118701}.
\newblock


\bibitem[Blondel et~al\mbox{.}(2008)]%
        {blondel2008fast}
\bibfield{author}{\bibinfo{person}{Vincent~D Blondel},
  \bibinfo{person}{Jean-Loup Guillaume}, \bibinfo{person}{Renaud Lambiotte},
  {and} \bibinfo{person}{Etienne Lefebvre}.} \bibinfo{year}{2008}\natexlab{}.
\newblock \showarticletitle{Fast unfolding of communities in large networks}.
\newblock \bibinfo{journal}{\emph{Journal of statistical mechanics: theory and
  experiment}} \bibinfo{volume}{2008}, \bibinfo{number}{10}
  (\bibinfo{year}{2008}), \bibinfo{pages}{P10008}.
\newblock


\bibitem[Bollob{\'a}s et~al\mbox{.}(2003)]%
        {bollobas2003directed}
\bibfield{author}{\bibinfo{person}{B{\'e}la Bollob{\'a}s},
  \bibinfo{person}{Christian Borgs}, \bibinfo{person}{Jennifer~T Chayes}, {and}
  \bibinfo{person}{Oliver Riordan}.} \bibinfo{year}{2003}\natexlab{}.
\newblock \showarticletitle{Directed scale-free graphs.}. In
  \bibinfo{booktitle}{\emph{SODA}}, Vol.~\bibinfo{volume}{3}.
  \bibinfo{pages}{132--139}.
\newblock


\bibitem[Bolotaeva and Cata(2010)]%
        {bolotaeva2010marketing}
\bibfield{author}{\bibinfo{person}{Victoria Bolotaeva} {and}
  \bibinfo{person}{Teuta Cata}.} \bibinfo{year}{2010}\natexlab{}.
\newblock \showarticletitle{Marketing opportunities with social networks}.
\newblock \bibinfo{journal}{\emph{Journal of Internet Social Networking and
  Virtual Communities}}  \bibinfo{volume}{2010} (\bibinfo{year}{2010}),
  \bibinfo{pages}{1--8}.
\newblock


\bibitem[Bordons et~al\mbox{.}(2015)]%
        {bordons2015relationship}
\bibfield{author}{\bibinfo{person}{Mar{\'\i}a Bordons}, \bibinfo{person}{Javier
  Aparicio}, \bibinfo{person}{Borja Gonz{\'a}lez-Albo}, {and}
  \bibinfo{person}{Adri{\'a}n~A D{\'\i}az-Faes}.}
  \bibinfo{year}{2015}\natexlab{}.
\newblock \showarticletitle{The relationship between the research performance
  of scientists and their position in co-authorship networks in three fields}.
\newblock \bibinfo{journal}{\emph{Journal of informetrics}}
  \bibinfo{volume}{9}, \bibinfo{number}{1} (\bibinfo{year}{2015}),
  \bibinfo{pages}{135--144}.
\newblock


\bibitem[Bose and Hamilton(2019)]%
        {bose2019compositional}
\bibfield{author}{\bibinfo{person}{Avishek Bose} {and} \bibinfo{person}{William
  Hamilton}.} \bibinfo{year}{2019}\natexlab{}.
\newblock \showarticletitle{Compositional fairness constraints for graph
  embeddings}. In \bibinfo{booktitle}{\emph{International Conference on Machine
  Learning}}. PMLR, \bibinfo{pages}{715--724}.
\newblock


\bibitem[Bottero(2007)]%
        {bottero2007social}
\bibfield{author}{\bibinfo{person}{Wendy Bottero}.}
  \bibinfo{year}{2007}\natexlab{}.
\newblock \showarticletitle{Social inequality and interaction}.
\newblock \bibinfo{journal}{\emph{Sociology Compass}} \bibinfo{volume}{1},
  \bibinfo{number}{2} (\bibinfo{year}{2007}), \bibinfo{pages}{814--831}.
\newblock


\bibitem[Bovet and Makse(2019)]%
        {bovet2019influence}
\bibfield{author}{\bibinfo{person}{Alexandre Bovet} {and}
  \bibinfo{person}{Hern{\'a}n~A Makse}.} \bibinfo{year}{2019}\natexlab{}.
\newblock \showarticletitle{Influence of fake news in Twitter during the 2016
  US presidential election}.
\newblock \bibinfo{journal}{\emph{Nature communications}} \bibinfo{volume}{10},
  \bibinfo{number}{1} (\bibinfo{year}{2019}), \bibinfo{pages}{1--14}.
\newblock


\bibitem[Bozorgi et~al\mbox{.}(2016)]%
        {bozorgi2016incim}
\bibfield{author}{\bibinfo{person}{Arastoo Bozorgi}, \bibinfo{person}{Hassan
  Haghighi}, \bibinfo{person}{Mohammad~Sadegh Zahedi}, {and}
  \bibinfo{person}{Mojtaba Rezvani}.} \bibinfo{year}{2016}\natexlab{}.
\newblock \showarticletitle{INCIM: A community-based algorithm for influence
  maximization problem under the linear threshold model}.
\newblock \bibinfo{journal}{\emph{Information Processing \& Management}}
  \bibinfo{volume}{52}, \bibinfo{number}{6} (\bibinfo{year}{2016}),
  \bibinfo{pages}{1188--1199}.
\newblock


\bibitem[Carrington(2011)]%
        {carrington2011crime}
\bibfield{author}{\bibinfo{person}{Peter~J Carrington}.}
  \bibinfo{year}{2011}\natexlab{}.
\newblock \showarticletitle{Crime and social network analysis}.
\newblock \bibinfo{journal}{\emph{The SAGE handbook of social network
  analysis}} (\bibinfo{year}{2011}), \bibinfo{pages}{236--255}.
\newblock


\bibitem[Casas-Roma et~al\mbox{.}(2017)]%
        {casas2017survey}
\bibfield{author}{\bibinfo{person}{Jordi Casas-Roma}, \bibinfo{person}{Jordi
  Herrera-Joancomart{\'\i}}, {and} \bibinfo{person}{Vicen{\c{c}} Torra}.}
  \bibinfo{year}{2017}\natexlab{}.
\newblock \showarticletitle{A survey of graph-modification techniques for
  privacy-preserving on networks}.
\newblock \bibinfo{journal}{\emph{Artificial Intelligence Review}}
  \bibinfo{volume}{47}, \bibinfo{number}{3} (\bibinfo{year}{2017}),
  \bibinfo{pages}{341--366}.
\newblock


\bibitem[Caton and Haas(2020)]%
        {caton2020fairness}
\bibfield{author}{\bibinfo{person}{Simon Caton} {and}
  \bibinfo{person}{Christian Haas}.} \bibinfo{year}{2020}\natexlab{}.
\newblock \showarticletitle{Fairness in machine learning: A survey}.
\newblock \bibinfo{journal}{\emph{arXiv preprint arXiv:2010.04053}}
  (\bibinfo{year}{2020}).
\newblock


\bibitem[Celis et~al\mbox{.}(2018)]%
        {celis2018ranking}
\bibfield{author}{\bibinfo{person}{L~Elisa Celis}, \bibinfo{person}{Damian
  Straszak}, {and} \bibinfo{person}{Nisheeth~K Vishnoi}.}
  \bibinfo{year}{2018}\natexlab{}.
\newblock \showarticletitle{Ranking with Fairness Constraints}. In
  \bibinfo{booktitle}{\emph{45th International Colloquium on Automata,
  Languages, and Programming (ICALP 2018)}}. Schloss Dagstuhl-Leibniz-Zentrum
  fuer Informatik.
\newblock


\bibitem[Chakraborty et~al\mbox{.}(2017)]%
        {chakraborty2017metrics}
\bibfield{author}{\bibinfo{person}{Tanmoy Chakraborty}, \bibinfo{person}{Ayushi
  Dalmia}, \bibinfo{person}{Animesh Mukherjee}, {and} \bibinfo{person}{Niloy
  Ganguly}.} \bibinfo{year}{2017}\natexlab{}.
\newblock \showarticletitle{Metrics for community analysis: A survey}.
\newblock \bibinfo{journal}{\emph{ACM Computing Surveys (CSUR)}}
  \bibinfo{volume}{50}, \bibinfo{number}{4} (\bibinfo{year}{2017}),
  \bibinfo{pages}{1--37}.
\newblock


\bibitem[Chen et~al\mbox{.}(2012b)]%
        {chen2012time}
\bibfield{author}{\bibinfo{person}{Wei Chen}, \bibinfo{person}{Wei Lu}, {and}
  \bibinfo{person}{Ning Zhang}.} \bibinfo{year}{2012}\natexlab{b}.
\newblock \showarticletitle{Time-critical influence maximization in social
  networks with time-delayed diffusion process}. In
  \bibinfo{booktitle}{\emph{Twenty-Sixth AAAI Conference on Artificial
  Intelligence}}.
\newblock


\bibitem[Chen et~al\mbox{.}(2010)]%
        {chen2010scalable}
\bibfield{author}{\bibinfo{person}{Wei Chen}, \bibinfo{person}{Chi Wang}, {and}
  \bibinfo{person}{Yajun Wang}.} \bibinfo{year}{2010}\natexlab{}.
\newblock \showarticletitle{Scalable influence maximization for prevalent viral
  marketing in large-scale social networks}. In
  \bibinfo{booktitle}{\emph{Proceedings of the 16th ACM SIGKDD international
  conference on Knowledge discovery and data mining}}.
  \bibinfo{pages}{1029--1038}.
\newblock


\bibitem[Chen et~al\mbox{.}(2012a)]%
        {chen2012community}
\bibfield{author}{\bibinfo{person}{Zhengzhang Chen}, \bibinfo{person}{William
  Hendrix}, {and} \bibinfo{person}{Nagiza~F Samatova}.}
  \bibinfo{year}{2012}\natexlab{a}.
\newblock \showarticletitle{Community-based anomaly detection in evolutionary
  networks}.
\newblock \bibinfo{journal}{\emph{Journal of Intelligent Information Systems}}
  \bibinfo{volume}{39}, \bibinfo{number}{1} (\bibinfo{year}{2012}),
  \bibinfo{pages}{59--85}.
\newblock


\bibitem[Chien et~al\mbox{.}(2019)]%
        {chien2019minimax}
\bibfield{author}{\bibinfo{person}{I~Eli Chien}, \bibinfo{person}{Chung-Yi
  Lin}, {and} \bibinfo{person}{I-Hsiang Wang}.}
  \bibinfo{year}{2019}\natexlab{}.
\newblock \showarticletitle{On the minimax misclassification ratio of
  hypergraph community detection}.
\newblock \bibinfo{journal}{\emph{IEEE Transactions on Information Theory}}
  \bibinfo{volume}{65}, \bibinfo{number}{12} (\bibinfo{year}{2019}),
  \bibinfo{pages}{8095--8118}.
\newblock


\bibitem[Chierichetti et~al\mbox{.}(2018)]%
        {chierichetti2018fair}
\bibfield{author}{\bibinfo{person}{Flavio Chierichetti}, \bibinfo{person}{Ravi
  Kumar}, \bibinfo{person}{Silvio Lattanzi}, {and} \bibinfo{person}{Sergei
  Vassilvitskii}.} \bibinfo{year}{2018}\natexlab{}.
\newblock \showarticletitle{Fair clustering through fairlets}.
\newblock \bibinfo{journal}{\emph{arXiv preprint arXiv:1802.05733}}
  (\bibinfo{year}{2018}).
\newblock


\bibitem[Cho et~al\mbox{.}(2012)]%
        {cho2012identification}
\bibfield{author}{\bibinfo{person}{Youngsang Cho}, \bibinfo{person}{Junseok
  Hwang}, {and} \bibinfo{person}{Daeho Lee}.} \bibinfo{year}{2012}\natexlab{}.
\newblock \showarticletitle{Identification of effective opinion leaders in the
  diffusion of technological innovation: A social network approach}.
\newblock \bibinfo{journal}{\emph{Technological Forecasting and Social Change}}
  \bibinfo{volume}{79}, \bibinfo{number}{1} (\bibinfo{year}{2012}),
  \bibinfo{pages}{97--106}.
\newblock


\bibitem[Choudhary et~al\mbox{.}(2022)]%
        {choudhary2022survey}
\bibfield{author}{\bibinfo{person}{Manvi Choudhary}, \bibinfo{person}{Charlotte
  Laclau}, {and} \bibinfo{person}{Christine Largeron}.}
  \bibinfo{year}{2022}\natexlab{}.
\newblock \showarticletitle{A Survey on Fairness for Machine Learning on
  Graphs}.
\newblock \bibinfo{journal}{\emph{arXiv preprint arXiv:2205.05396}}
  (\bibinfo{year}{2022}).
\newblock


\bibitem[Cisse and Koyejo(2019)]%
        {cisse2019fairness}
\bibfield{author}{\bibinfo{person}{Moustapha Cisse} {and}
  \bibinfo{person}{Sanmi Koyejo}.} \bibinfo{year}{2019}\natexlab{}.
\newblock \showarticletitle{Fairness and representation learning}.
\newblock \bibinfo{journal}{\emph{NeurIPS Invited Talk}}
  \bibinfo{volume}{2019} (\bibinfo{year}{2019}).
\newblock


\bibitem[Corbett-Davies and Goel(2018)]%
        {corbett2018measure}
\bibfield{author}{\bibinfo{person}{Sam Corbett-Davies} {and}
  \bibinfo{person}{Sharad Goel}.} \bibinfo{year}{2018}\natexlab{}.
\newblock \showarticletitle{The measure and mismeasure of fairness: A critical
  review of fair machine learning}.
\newblock \bibinfo{journal}{\emph{arXiv preprint arXiv:1808.00023}}
  (\bibinfo{year}{2018}).
\newblock


\bibitem[Dai and Wang(2021)]%
        {dai2021say}
\bibfield{author}{\bibinfo{person}{Enyan Dai} {and} \bibinfo{person}{Suhang
  Wang}.} \bibinfo{year}{2021}\natexlab{}.
\newblock \showarticletitle{Say no to the discrimination: Learning fair graph
  neural networks with limited sensitive attribute information}. In
  \bibinfo{booktitle}{\emph{Proceedings of the 14th ACM International
  Conference on Web Search and Data Mining}}. \bibinfo{pages}{680--688}.
\newblock


\bibitem[data(2022)]%
        {networkdata}
\bibfield{author}{\bibinfo{person}{Network data}.}
  \bibinfo{year}{2022}\natexlab{}.
\newblock \showarticletitle{http://www-personal.umich.edu/~mejn/netdata/}.
\newblock \bibinfo{journal}{\emph{Accessed on 9-Sep-2022}}
  (\bibinfo{year}{2022}).
\newblock


\bibitem[Davidson and Ravi(2020)]%
        {davidson2020framework}
\bibfield{author}{\bibinfo{person}{Ian Davidson} {and}
  \bibinfo{person}{Selvan~Suntiha Ravi}.} \bibinfo{year}{2020}\natexlab{}.
\newblock \showarticletitle{A framework for determining the fairness of outlier
  detection}.
\newblock In \bibinfo{booktitle}{\emph{ECAI 2020}}. \bibinfo{publisher}{IOS
  Press}, \bibinfo{pages}{2465--2472}.
\newblock


\bibitem[de~Groot et~al\mbox{.}(2024)]%
        {de2024canon}
\bibfield{author}{\bibinfo{person}{Aviva de Groot}, \bibinfo{person}{George~HL
  Fletcher}, \bibinfo{person}{Gijs van Manen}, \bibinfo{person}{Akrati Saxena},
  \bibinfo{person}{Alexander Serebrenik}, {and} \bibinfo{person}{LEM Taylor}.}
  \bibinfo{year}{2024}\natexlab{}.
\newblock \showarticletitle{A canon is a blunt force instrument: data science,
  canons, and generative frictions}.
\newblock In \bibinfo{booktitle}{\emph{Dialogues in Data Power Shifting
  Response-abilities in a Datafied World}}. \bibinfo{publisher}{Bristol
  University Press}.
\newblock


\bibitem[Deniz~K{\"o}se and Shen(2021)]%
        {deniz2021fairness}
\bibfield{author}{\bibinfo{person}{{\"O}yk{\"u} Deniz~K{\"o}se} {and}
  \bibinfo{person}{Yanning Shen}.} \bibinfo{year}{2021}\natexlab{}.
\newblock \showarticletitle{Fairness-Aware Node Representation Learning}.
\newblock \bibinfo{journal}{\emph{arXiv e-prints}} (\bibinfo{year}{2021}),
  \bibinfo{pages}{arXiv--2106}.
\newblock


\bibitem[DiMaggio and Garip(2012)]%
        {dimaggio2012network}
\bibfield{author}{\bibinfo{person}{Paul DiMaggio} {and} \bibinfo{person}{Filiz
  Garip}.} \bibinfo{year}{2012}\natexlab{}.
\newblock \showarticletitle{Network effects and social inequality}.
\newblock \bibinfo{journal}{\emph{Annual review of sociology}}
  \bibinfo{volume}{38} (\bibinfo{year}{2012}), \bibinfo{pages}{93--118}.
\newblock


\bibitem[Dong et~al\mbox{.}(2022)]%
        {dong2022fairness}
\bibfield{author}{\bibinfo{person}{Yushun Dong}, \bibinfo{person}{Jing Ma},
  \bibinfo{person}{Chen Chen}, {and} \bibinfo{person}{Jundong Li}.}
  \bibinfo{year}{2022}\natexlab{}.
\newblock \showarticletitle{Fairness in Graph Mining: A Survey}.
\newblock \bibinfo{journal}{\emph{arXiv preprint arXiv:2204.09888}}
  (\bibinfo{year}{2022}).
\newblock


\bibitem[Doyle(2007)]%
        {doyle2007role}
\bibfield{author}{\bibinfo{person}{Shaun Doyle}.}
  \bibinfo{year}{2007}\natexlab{}.
\newblock \showarticletitle{The role of social networks in marketing}.
\newblock \bibinfo{journal}{\emph{Journal of Database Marketing \& Customer
  Strategy Management}} \bibinfo{volume}{15}, \bibinfo{number}{1}
  (\bibinfo{year}{2007}), \bibinfo{pages}{60--64}.
\newblock


\bibitem[Du et~al\mbox{.}(2020)]%
        {du2020fairness}
\bibfield{author}{\bibinfo{person}{Mengnan Du}, \bibinfo{person}{Fan Yang},
  \bibinfo{person}{Na Zou}, {and} \bibinfo{person}{Xia Hu}.}
  \bibinfo{year}{2020}\natexlab{}.
\newblock \showarticletitle{Fairness in deep learning: A computational
  perspective}.
\newblock \bibinfo{journal}{\emph{IEEE Intelligent Systems}}
  \bibinfo{volume}{36}, \bibinfo{number}{4} (\bibinfo{year}{2020}),
  \bibinfo{pages}{25--34}.
\newblock


\bibitem[Dwork et~al\mbox{.}(2012)]%
        {dwork2012fairness}
\bibfield{author}{\bibinfo{person}{Cynthia Dwork}, \bibinfo{person}{Moritz
  Hardt}, \bibinfo{person}{Toniann Pitassi}, \bibinfo{person}{Omer Reingold},
  {and} \bibinfo{person}{Richard Zemel}.} \bibinfo{year}{2012}\natexlab{}.
\newblock \showarticletitle{Fairness through awareness}. In
  \bibinfo{booktitle}{\emph{Proceedings of the 3rd innovations in theoretical
  computer science conference}}. \bibinfo{pages}{214--226}.
\newblock


\bibitem[Erd et~al\mbox{.}(2021)]%
        {erd2021generalized}
\bibfield{author}{\bibinfo{person}{Fernando~C Erd},
  \bibinfo{person}{Andr{\'e}~L Vignatti}, {and} \bibinfo{person}{Murilo~VG da
  Silva}.} \bibinfo{year}{2021}\natexlab{}.
\newblock \showarticletitle{The generalized influence blocking maximization
  problem}.
\newblock \bibinfo{journal}{\emph{Social Network Analysis and Mining}}
  \bibinfo{volume}{11}, \bibinfo{number}{1} (\bibinfo{year}{2021}),
  \bibinfo{pages}{1--17}.
\newblock


\bibitem[Eshghi et~al\mbox{.}(2019)]%
        {eshghi2019efficient}
\bibfield{author}{\bibinfo{person}{Soheil Eshghi}, \bibinfo{person}{Setareh
  Maghsudi}, \bibinfo{person}{Valerio Restocchi}, \bibinfo{person}{Sebastian
  Stein}, {and} \bibinfo{person}{Leandros Tassiulas}.}
  \bibinfo{year}{2019}\natexlab{}.
\newblock \showarticletitle{Efficient influence maximization under network
  uncertainty}. In \bibinfo{booktitle}{\emph{IEEE INFOCOM 2019-IEEE Conference
  on Computer Communications Workshops (INFOCOM WKSHPS)}}. IEEE,
  \bibinfo{pages}{365--371}.
\newblock


\bibitem[Farnadi et~al\mbox{.}(2020)]%
        {farnad2020unifying}
\bibfield{author}{\bibinfo{person}{Golnoosh Farnadi}, \bibinfo{person}{Behrouz
  Babaki}, {and} \bibinfo{person}{Michel Gendreau}.}
  \bibinfo{year}{2020}\natexlab{}.
\newblock \showarticletitle{A Unifying Framework for Fairness-Aware Influence
  Maximization}. In \bibinfo{booktitle}{\emph{Companion Proceedings of the Web
  Conference 2020}}. \bibinfo{pages}{714--722}.
\newblock


\bibitem[Finneran and Kelly(2003)]%
        {finneran2003social}
\bibfield{author}{\bibinfo{person}{Lisa Finneran} {and} \bibinfo{person}{Morgan
  Kelly}.} \bibinfo{year}{2003}\natexlab{}.
\newblock \showarticletitle{Social networks and inequality}.
\newblock \bibinfo{journal}{\emph{Journal of urban economics}}
  \bibinfo{volume}{53}, \bibinfo{number}{2} (\bibinfo{year}{2003}),
  \bibinfo{pages}{282--299}.
\newblock


\bibitem[Fish et~al\mbox{.}(2019)]%
        {fish2019gaps}
\bibfield{author}{\bibinfo{person}{Benjamin Fish}, \bibinfo{person}{Ashkan
  Bashardoust}, \bibinfo{person}{Danah Boyd}, \bibinfo{person}{Sorelle
  Friedler}, \bibinfo{person}{Carlos Scheidegger}, {and}
  \bibinfo{person}{Suresh Venkatasubramanian}.}
  \bibinfo{year}{2019}\natexlab{}.
\newblock \showarticletitle{Gaps in Information Access in Social Networks?}. In
  \bibinfo{booktitle}{\emph{The World Wide Web Conference}}.
  \bibinfo{pages}{480--490}.
\newblock


\bibitem[Forestier et~al\mbox{.}(2012)]%
        {forestier2012roles}
\bibfield{author}{\bibinfo{person}{Mathilde Forestier}, \bibinfo{person}{Anna
  Stavrianou}, \bibinfo{person}{Julien Velcin}, {and} \bibinfo{person}{Djamel~A
  Zighed}.} \bibinfo{year}{2012}\natexlab{}.
\newblock \showarticletitle{Roles in social networks: Methodologies and
  research issues}.
\newblock \bibinfo{journal}{\emph{Web Intelligence and Agent Systems: An
  international Journal}} \bibinfo{volume}{10}, \bibinfo{number}{1}
  (\bibinfo{year}{2012}), \bibinfo{pages}{117--133}.
\newblock


\bibitem[Fortunato and Hric(2016)]%
        {fortunato2016community}
\bibfield{author}{\bibinfo{person}{Santo Fortunato} {and}
  \bibinfo{person}{Darko Hric}.} \bibinfo{year}{2016}\natexlab{}.
\newblock \showarticletitle{Community detection in networks: A user guide}.
\newblock \bibinfo{journal}{\emph{Physics reports}}  \bibinfo{volume}{659}
  (\bibinfo{year}{2016}), \bibinfo{pages}{1--44}.
\newblock


\bibitem[Freeman(2004)]%
        {freeman2004development}
\bibfield{author}{\bibinfo{person}{Linton Freeman}.}
  \bibinfo{year}{2004}\natexlab{}.
\newblock \showarticletitle{The development of social network analysis}.
\newblock \bibinfo{journal}{\emph{A Study in the Sociology of Science}}
  \bibinfo{volume}{1}, \bibinfo{number}{687} (\bibinfo{year}{2004}),
  \bibinfo{pages}{159--167}.
\newblock


\bibitem[Gajane and Pechenizkiy(2017)]%
        {gajane2017formalizing}
\bibfield{author}{\bibinfo{person}{Pratik Gajane} {and} \bibinfo{person}{Mykola
  Pechenizkiy}.} \bibinfo{year}{2017}\natexlab{}.
\newblock \showarticletitle{On formalizing fairness in prediction with machine
  learning}.
\newblock \bibinfo{journal}{\emph{arXiv preprint arXiv:1710.03184}}
  (\bibinfo{year}{2017}).
\newblock


\bibitem[Gajane et~al\mbox{.}(2022)]%
        {gajane2022survey}
\bibfield{author}{\bibinfo{person}{Pratik Gajane}, \bibinfo{person}{Akrati
  Saxena}, \bibinfo{person}{Maryam Tavakol}, \bibinfo{person}{George Fletcher},
  {and} \bibinfo{person}{Mykola Pechenizkiy}.} \bibinfo{year}{2022}\natexlab{}.
\newblock \showarticletitle{Survey on fair reinforcement learning: Theory and
  practice}.
\newblock \bibinfo{journal}{\emph{arXiv preprint arXiv:2205.10032}}
  (\bibinfo{year}{2022}).
\newblock


\bibitem[Gallivan and Ahuja(2015)]%
        {gallivan2015co}
\bibfield{author}{\bibinfo{person}{Michael Gallivan} {and}
  \bibinfo{person}{Manju Ahuja}.} \bibinfo{year}{2015}\natexlab{}.
\newblock \showarticletitle{Co-authorship, homophily, and scholarly influence
  in information systems research}.
\newblock \bibinfo{journal}{\emph{Journal of the Association for Information
  Systems}} \bibinfo{volume}{16}, \bibinfo{number}{12} (\bibinfo{year}{2015}),
  \bibinfo{pages}{2}.
\newblock


\bibitem[Gershtein et~al\mbox{.}(2018)]%
        {gershtein2018balanced}
\bibfield{author}{\bibinfo{person}{Shay Gershtein}, \bibinfo{person}{Tova
  Milo}, \bibinfo{person}{Brit Youngmann}, {and} \bibinfo{person}{Gal Zeevi}.}
  \bibinfo{year}{2018}\natexlab{}.
\newblock \showarticletitle{IM balanced: influence maximization under balance
  constraints}. In \bibinfo{booktitle}{\emph{Proceedings of the 27th ACM
  International Conference on Information and Knowledge Management}}.
  \bibinfo{pages}{1919--1922}.
\newblock


\bibitem[Geyik et~al\mbox{.}(2019)]%
        {geyik2019fairness}
\bibfield{author}{\bibinfo{person}{Sahin~Cem Geyik}, \bibinfo{person}{Stuart
  Ambler}, {and} \bibinfo{person}{Krishnaram Kenthapadi}.}
  \bibinfo{year}{2019}\natexlab{}.
\newblock \showarticletitle{Fairness-aware ranking in search \& recommendation
  systems with application to linkedin talent search}. In
  \bibinfo{booktitle}{\emph{Proceedings of the 25th acm sigkdd international
  conference on knowledge discovery \& data mining}}.
  \bibinfo{pages}{2221--2231}.
\newblock


\bibitem[Ghadiri et~al\mbox{.}(2021)]%
        {ghadiri2021socially}
\bibfield{author}{\bibinfo{person}{Mehrdad Ghadiri}, \bibinfo{person}{Samira
  Samadi}, {and} \bibinfo{person}{Santosh Vempala}.}
  \bibinfo{year}{2021}\natexlab{}.
\newblock \showarticletitle{Socially fair k-means clustering}. In
  \bibinfo{booktitle}{\emph{Proceedings of the 2021 ACM Conference on Fairness,
  Accountability, and Transparency}}. \bibinfo{pages}{438--448}.
\newblock


\bibitem[Ghasemian et~al\mbox{.}(2019)]%
        {ghasemian2019evaluating}
\bibfield{author}{\bibinfo{person}{Amir Ghasemian}, \bibinfo{person}{Homa
  Hosseinmardi}, {and} \bibinfo{person}{Aaron Clauset}.}
  \bibinfo{year}{2019}\natexlab{}.
\newblock \showarticletitle{Evaluating overfit and underfit in models of
  network community structure}.
\newblock \bibinfo{journal}{\emph{IEEE Transactions on Knowledge and Data
  Engineering}} \bibinfo{volume}{32}, \bibinfo{number}{9}
  (\bibinfo{year}{2019}), \bibinfo{pages}{1722--1735}.
\newblock


\bibitem[Goyal et~al\mbox{.}(2011)]%
        {goyal2011simpath}
\bibfield{author}{\bibinfo{person}{Amit Goyal}, \bibinfo{person}{Wei Lu}, {and}
  \bibinfo{person}{Laks~VS Lakshmanan}.} \bibinfo{year}{2011}\natexlab{}.
\newblock \showarticletitle{Simpath: An efficient algorithm for influence
  maximization under the linear threshold model}. In
  \bibinfo{booktitle}{\emph{2011 IEEE 11th international conference on data
  mining}}. IEEE, \bibinfo{pages}{211--220}.
\newblock


\bibitem[Grinberg et~al\mbox{.}(2019)]%
        {grinberg2019fake}
\bibfield{author}{\bibinfo{person}{Nir Grinberg}, \bibinfo{person}{Kenneth
  Joseph}, \bibinfo{person}{Lisa Friedland}, \bibinfo{person}{Briony
  Swire-Thompson}, {and} \bibinfo{person}{David Lazer}.}
  \bibinfo{year}{2019}\natexlab{}.
\newblock \showarticletitle{Fake news on Twitter during the 2016 US
  presidential election}.
\newblock \bibinfo{journal}{\emph{Science}} \bibinfo{volume}{363},
  \bibinfo{number}{6425} (\bibinfo{year}{2019}), \bibinfo{pages}{374--378}.
\newblock


\bibitem[Grover and Leskovec(2016)]%
        {grover2016node2vec}
\bibfield{author}{\bibinfo{person}{Aditya Grover} {and} \bibinfo{person}{Jure
  Leskovec}.} \bibinfo{year}{2016}\natexlab{}.
\newblock \showarticletitle{node2vec: Scalable feature learning for networks}.
  In \bibinfo{booktitle}{\emph{Proceedings of the 22nd ACM SIGKDD international
  conference on Knowledge discovery and data mining}}.
  \bibinfo{pages}{855--864}.
\newblock


\bibitem[Gupta et~al\mbox{.}(2016)]%
        {gupta2016modeling}
\bibfield{author}{\bibinfo{person}{Yayati Gupta}, \bibinfo{person}{Akrati
  Saxena}, \bibinfo{person}{Debarati Das}, {and} \bibinfo{person}{SRS
  Iyengar}.} \bibinfo{year}{2016}\natexlab{}.
\newblock \showarticletitle{Modeling memetics using edge diversity}.
\newblock In \bibinfo{booktitle}{\emph{Complex Networks VII}}.
  \bibinfo{publisher}{Springer}, \bibinfo{pages}{187--198}.
\newblock


\bibitem[Halabi et~al\mbox{.}(2020)]%
        {halabi2020fairness}
\bibfield{author}{\bibinfo{person}{Marwa~El Halabi}, \bibinfo{person}{Slobodan
  Mitrovi{\'c}}, \bibinfo{person}{Ashkan Norouzi-Fard}, \bibinfo{person}{Jakab
  Tardos}, {and} \bibinfo{person}{Jakub Tarnawski}.}
  \bibinfo{year}{2020}\natexlab{}.
\newblock \showarticletitle{Fairness in Streaming Submodular Maximization:
  Algorithms and Hardness}.
\newblock \bibinfo{journal}{\emph{arXiv preprint arXiv:2010.07431}}
  (\bibinfo{year}{2020}).
\newblock


\bibitem[He et~al\mbox{.}(2012)]%
        {he2012influence}
\bibfield{author}{\bibinfo{person}{Xinran He}, \bibinfo{person}{Guojie Song},
  \bibinfo{person}{Wei Chen}, {and} \bibinfo{person}{Qingye Jiang}.}
  \bibinfo{year}{2012}\natexlab{}.
\newblock \showarticletitle{Influence blocking maximization in social networks
  under the competitive linear threshold model}. In
  \bibinfo{booktitle}{\emph{Proceedings of the 2012 siam international
  conference on data mining}}. SIAM, \bibinfo{pages}{463--474}.
\newblock


\bibitem[Hellman(2020)]%
        {hellman2020measuring}
\bibfield{author}{\bibinfo{person}{Deborah Hellman}.}
  \bibinfo{year}{2020}\natexlab{}.
\newblock \showarticletitle{Measuring algorithmic fairness}.
\newblock \bibinfo{journal}{\emph{Virginia Law Review}} \bibinfo{volume}{106},
  \bibinfo{number}{4} (\bibinfo{year}{2020}), \bibinfo{pages}{811--866}.
\newblock


\bibitem[Hertweck et~al\mbox{.}(2021)]%
        {hertweck2021moral}
\bibfield{author}{\bibinfo{person}{Corinna Hertweck},
  \bibinfo{person}{Christoph Heitz}, {and} \bibinfo{person}{Michele Loi}.}
  \bibinfo{year}{2021}\natexlab{}.
\newblock \showarticletitle{On the moral justification of statistical parity}.
  In \bibinfo{booktitle}{\emph{Proceedings of the 2021 ACM Conference on
  Fairness, Accountability, and Transparency}}. \bibinfo{pages}{747--757}.
\newblock


\bibitem[Holme et~al\mbox{.}(2004)]%
        {holme2004structure}
\bibfield{author}{\bibinfo{person}{Petter Holme}, \bibinfo{person}{Christofer~R
  Edling}, {and} \bibinfo{person}{Fredrik Liljeros}.}
  \bibinfo{year}{2004}\natexlab{}.
\newblock \showarticletitle{Structure and time evolution of an Internet dating
  community}.
\newblock \bibinfo{journal}{\emph{Social Networks}} \bibinfo{volume}{26},
  \bibinfo{number}{2} (\bibinfo{year}{2004}), \bibinfo{pages}{155--174}.
\newblock


\bibitem[Holme and Saram{\"a}ki(2012)]%
        {holme2012temporal}
\bibfield{author}{\bibinfo{person}{Petter Holme} {and} \bibinfo{person}{Jari
  Saram{\"a}ki}.} \bibinfo{year}{2012}\natexlab{}.
\newblock \showarticletitle{Temporal networks}.
\newblock \bibinfo{journal}{\emph{Physics reports}} \bibinfo{volume}{519},
  \bibinfo{number}{3} (\bibinfo{year}{2012}), \bibinfo{pages}{97--125}.
\newblock


\bibitem[Huang et~al\mbox{.}(2019)]%
        {huang2019community}
\bibfield{author}{\bibinfo{person}{Huimin Huang}, \bibinfo{person}{Hong Shen},
  \bibinfo{person}{Zaiqiao Meng}, \bibinfo{person}{Huajian Chang}, {and}
  \bibinfo{person}{Huaiwen He}.} \bibinfo{year}{2019}\natexlab{}.
\newblock \showarticletitle{Community-based influence maximization for viral
  marketing}.
\newblock \bibinfo{journal}{\emph{Applied Intelligence}} \bibinfo{volume}{49},
  \bibinfo{number}{6} (\bibinfo{year}{2019}), \bibinfo{pages}{2137--2150}.
\newblock


\bibitem[Huang et~al\mbox{.}(2021)]%
        {huang2021survey}
\bibfield{author}{\bibinfo{person}{Xinyu Huang}, \bibinfo{person}{Dongming
  Chen}, \bibinfo{person}{Tao Ren}, {and} \bibinfo{person}{Dongqi Wang}.}
  \bibinfo{year}{2021}\natexlab{}.
\newblock \showarticletitle{A survey of community detection methods in
  multilayer networks}.
\newblock \bibinfo{journal}{\emph{Data Mining and Knowledge Discovery}}
  \bibinfo{volume}{35}, \bibinfo{number}{1} (\bibinfo{year}{2021}),
  \bibinfo{pages}{1--45}.
\newblock


\bibitem[Hymowitz and Schellhardt(1986)]%
        {hymowitz1986women}
\bibfield{author}{\bibinfo{person}{Carol Hymowitz} {and}
  \bibinfo{person}{Timothy~D Schellhardt}.} \bibinfo{year}{1986}\natexlab{}.
\newblock \showarticletitle{Why women can’t seem to break the invisible
  barrier that blocks them from the top jobs}.
\newblock \bibinfo{journal}{\emph{The Wall Street Journal}}
  \bibinfo{volume}{1} (\bibinfo{year}{1986}).
\newblock


\bibitem[Jalali et~al\mbox{.}(2020)]%
        {jalali2020information}
\bibfield{author}{\bibinfo{person}{Zeinab~S Jalali}, \bibinfo{person}{Weixiang
  Wang}, \bibinfo{person}{Myunghwan Kim}, \bibinfo{person}{Hema Raghavan},
  {and} \bibinfo{person}{Sucheta Soundarajan}.}
  \bibinfo{year}{2020}\natexlab{}.
\newblock \showarticletitle{On the Information Unfairness of Social Networks}.
  In \bibinfo{booktitle}{\emph{Proceedings of the 2020 SIAM International
  Conference on Data Mining}}. SIAM, \bibinfo{pages}{613--521}.
\newblock


\bibitem[Jeon and Kim(2017)]%
        {jeon2017community}
\bibfield{author}{\bibinfo{person}{Hyoungjun Jeon} {and}
  \bibinfo{person}{Taewhan Kim}.} \bibinfo{year}{2017}\natexlab{}.
\newblock \showarticletitle{Community-adaptive link prediction}. In
  \bibinfo{booktitle}{\emph{Proceedings of the 2017 International Conference on
  Data Mining, Communications and Information Technology}}.
  \bibinfo{pages}{1--5}.
\newblock


\bibitem[Kang and Tong(2021)]%
        {kang2021fair}
\bibfield{author}{\bibinfo{person}{Jian Kang} {and} \bibinfo{person}{Hanghang
  Tong}.} \bibinfo{year}{2021}\natexlab{}.
\newblock \showarticletitle{Fair graph mining}. In
  \bibinfo{booktitle}{\emph{Proceedings of the 30th ACM International
  Conference on Information \& Knowledge Management}}.
  \bibinfo{pages}{4849--4852}.
\newblock


\bibitem[Karimi et~al\mbox{.}(2018)]%
        {karimi2018homophily}
\bibfield{author}{\bibinfo{person}{Fariba Karimi}, \bibinfo{person}{Mathieu
  G{\'e}nois}, \bibinfo{person}{Claudia Wagner}, \bibinfo{person}{Philipp
  Singer}, {and} \bibinfo{person}{Markus Strohmaier}.}
  \bibinfo{year}{2018}\natexlab{}.
\newblock \showarticletitle{Homophily influences ranking of minorities in
  social networks}.
\newblock \bibinfo{journal}{\emph{Scientific reports}} \bibinfo{volume}{8},
  \bibinfo{number}{1} (\bibinfo{year}{2018}), \bibinfo{pages}{1--12}.
\newblock


\bibitem[Karimi et~al\mbox{.}(2016)]%
        {karimi2016inferring}
\bibfield{author}{\bibinfo{person}{Fariba Karimi}, \bibinfo{person}{Claudia
  Wagner}, \bibinfo{person}{Florian Lemmerich}, \bibinfo{person}{Mohsen
  Jadidi}, {and} \bibinfo{person}{Markus Strohmaier}.}
  \bibinfo{year}{2016}\natexlab{}.
\newblock \showarticletitle{Inferring gender from names on the web: A
  comparative evaluation of gender detection methods}. In
  \bibinfo{booktitle}{\emph{Proceedings of the 25th International conference
  companion on World Wide Web}}. \bibinfo{pages}{53--54}.
\newblock


\bibitem[Kas et~al\mbox{.}(2013)]%
        {kas2013incremental}
\bibfield{author}{\bibinfo{person}{Miray Kas}, \bibinfo{person}{Matthew Wachs},
  \bibinfo{person}{Kathleen~M Carley}, {and} \bibinfo{person}{L~Richard
  Carley}.} \bibinfo{year}{2013}\natexlab{}.
\newblock \showarticletitle{Incremental algorithm for updating betweenness
  centrality in dynamically growing networks}. In
  \bibinfo{booktitle}{\emph{Proceedings of the 2013 IEEE/ACM international
  conference on advances in social networks analysis and mining}}.
  \bibinfo{pages}{33--40}.
\newblock


\bibitem[Kawachi and Berkman(2001)]%
        {kawachi2001social}
\bibfield{author}{\bibinfo{person}{Ichiro Kawachi} {and}
  \bibinfo{person}{Lisa~F Berkman}.} \bibinfo{year}{2001}\natexlab{}.
\newblock \showarticletitle{Social ties and mental health}.
\newblock \bibinfo{journal}{\emph{Journal of Urban health}}
  \bibinfo{volume}{78}, \bibinfo{number}{3} (\bibinfo{year}{2001}),
  \bibinfo{pages}{458--467}.
\newblock


\bibitem[Kazienko and Musia{\l}(2006)]%
        {kazienko2006social}
\bibfield{author}{\bibinfo{person}{Przemys{\l}aw Kazienko} {and}
  \bibinfo{person}{Katarzyna Musia{\l}}.} \bibinfo{year}{2006}\natexlab{}.
\newblock \showarticletitle{Social capital in online social networks}. In
  \bibinfo{booktitle}{\emph{International Conference on Knowledge-Based and
  Intelligent Information and Engineering Systems}}. Springer,
  \bibinfo{pages}{417--424}.
\newblock


\bibitem[Kempe et~al\mbox{.}(2003)]%
        {kempe2003maximizing}
\bibfield{author}{\bibinfo{person}{David Kempe}, \bibinfo{person}{Jon
  Kleinberg}, {and} \bibinfo{person}{{\'E}va Tardos}.}
  \bibinfo{year}{2003}\natexlab{}.
\newblock \showarticletitle{Maximizing the spread of influence through a social
  network}. In \bibinfo{booktitle}{\emph{Proceedings of the ninth ACM SIGKDD
  international conference on Knowledge discovery and data mining}}.
  \bibinfo{pages}{137--146}.
\newblock


\bibitem[Kempe et~al\mbox{.}(2005)]%
        {kempe2005influential}
\bibfield{author}{\bibinfo{person}{David Kempe}, \bibinfo{person}{Jon
  Kleinberg}, {and} \bibinfo{person}{{\'E}va Tardos}.}
  \bibinfo{year}{2005}\natexlab{}.
\newblock \showarticletitle{Influential nodes in a diffusion model for social
  networks}. In \bibinfo{booktitle}{\emph{International Colloquium on Automata,
  Languages, and Programming}}. Springer, \bibinfo{pages}{1127--1138}.
\newblock


\bibitem[Khajehnejad et~al\mbox{.}(2021)]%
        {khajehnejad2021crosswalk}
\bibfield{author}{\bibinfo{person}{Ahmad Khajehnejad}, \bibinfo{person}{Moein
  Khajehnejad}, \bibinfo{person}{Mahmoudreza Babaei},
  \bibinfo{person}{Krishna~P Gummadi}, \bibinfo{person}{Adrian Weller}, {and}
  \bibinfo{person}{Baharan Mirzasoleiman}.} \bibinfo{year}{2021}\natexlab{}.
\newblock \showarticletitle{CrossWalk: Fairness-enhanced Node Representation
  Learning}.
\newblock \bibinfo{journal}{\emph{arXiv preprint arXiv:2105.02725}}
  (\bibinfo{year}{2021}).
\newblock


\bibitem[Khajehnejad et~al\mbox{.}(2020)]%
        {khajehnejad2020adversarial}
\bibfield{author}{\bibinfo{person}{Moein Khajehnejad},
  \bibinfo{person}{Ahmad~Asgharian Rezaei}, \bibinfo{person}{Mahmoudreza
  Babaei}, \bibinfo{person}{Jessica Hoffmann}, \bibinfo{person}{Mahdi Jalili},
  {and} \bibinfo{person}{Adrian Weller}.} \bibinfo{year}{2020}\natexlab{}.
\newblock \showarticletitle{Adversarial graph embeddings for fair influence
  maximization over social networks}.
\newblock \bibinfo{journal}{\emph{arXiv preprint arXiv:2005.04074}}
  (\bibinfo{year}{2020}).
\newblock


\bibitem[Kimura and Saito(2006)]%
        {kimura2006tractable}
\bibfield{author}{\bibinfo{person}{Masahiro Kimura} {and}
  \bibinfo{person}{Kazumi Saito}.} \bibinfo{year}{2006}\natexlab{}.
\newblock \showarticletitle{Tractable models for information diffusion in
  social networks}. In \bibinfo{booktitle}{\emph{Knowledge Discovery in
  Databases: PKDD 2006: 10th European Conference on Principles and Practice of
  Knowledge Discovery in Databases Berlin, Germany, September 18-22, 2006
  Proceedings 10}}. Springer, \bibinfo{pages}{259--271}.
\newblock


\bibitem[Kivel{\"a} et~al\mbox{.}(2014)]%
        {kivela2014multilayer}
\bibfield{author}{\bibinfo{person}{Mikko Kivel{\"a}}, \bibinfo{person}{Alex
  Arenas}, \bibinfo{person}{Marc Barthelemy}, \bibinfo{person}{James~P
  Gleeson}, \bibinfo{person}{Yamir Moreno}, {and} \bibinfo{person}{Mason~A
  Porter}.} \bibinfo{year}{2014}\natexlab{}.
\newblock \showarticletitle{Multilayer networks}.
\newblock \bibinfo{journal}{\emph{Journal of complex networks}}
  \bibinfo{volume}{2}, \bibinfo{number}{3} (\bibinfo{year}{2014}),
  \bibinfo{pages}{203--271}.
\newblock


\bibitem[Kleindessner et~al\mbox{.}(2019)]%
        {kleindessner2019guarantees}
\bibfield{author}{\bibinfo{person}{Matth{\"a}us Kleindessner},
  \bibinfo{person}{Samira Samadi}, \bibinfo{person}{Pranjal Awasthi}, {and}
  \bibinfo{person}{Jamie Morgenstern}.} \bibinfo{year}{2019}\natexlab{}.
\newblock \showarticletitle{Guarantees for spectral clustering with fairness
  constraints}. In \bibinfo{booktitle}{\emph{International Conference on
  Machine Learning}}. PMLR, \bibinfo{pages}{3458--3467}.
\newblock


\bibitem[Knecht et~al\mbox{.}(2010)]%
        {knecht2010friendship}
\bibfield{author}{\bibinfo{person}{Andrea Knecht}, \bibinfo{person}{Tom~AB
  Snijders}, \bibinfo{person}{Chris Baerveldt}, \bibinfo{person}{Christian~EG
  Steglich}, {and} \bibinfo{person}{Werner Raub}.}
  \bibinfo{year}{2010}\natexlab{}.
\newblock \showarticletitle{Friendship and delinquency: Selection and influence
  processes in early adolescence}.
\newblock \bibinfo{journal}{\emph{Social Development}} \bibinfo{volume}{19},
  \bibinfo{number}{3} (\bibinfo{year}{2010}), \bibinfo{pages}{494--514}.
\newblock


\bibitem[Kullback and Leibler(1951)]%
        {kullback1951information}
\bibfield{author}{\bibinfo{person}{Solomon Kullback} {and}
  \bibinfo{person}{Richard~A Leibler}.} \bibinfo{year}{1951}\natexlab{}.
\newblock \showarticletitle{On information and sufficiency}.
\newblock \bibinfo{journal}{\emph{The annals of mathematical statistics}}
  \bibinfo{volume}{22}, \bibinfo{number}{1} (\bibinfo{year}{1951}),
  \bibinfo{pages}{79--86}.
\newblock


\bibitem[Kumar and Singh(2013)]%
        {kumar2013detection}
\bibfield{author}{\bibinfo{person}{A~Sharath Kumar} {and}
  \bibinfo{person}{Sanjay Singh}.} \bibinfo{year}{2013}\natexlab{}.
\newblock \showarticletitle{Detection of user cluster with suspicious activity
  in online social networking sites}. In \bibinfo{booktitle}{\emph{2013 2nd
  International Conference on Advanced Computing, Networking and Security}}.
  IEEE, \bibinfo{pages}{220--225}.
\newblock


\bibitem[Kundu et~al\mbox{.}(2011)]%
        {kundu2011new}
\bibfield{author}{\bibinfo{person}{Suman Kundu}, \bibinfo{person}{CA Murthy},
  {and} \bibinfo{person}{Sankar~K Pal}.} \bibinfo{year}{2011}\natexlab{}.
\newblock \showarticletitle{A new centrality measure for influence maximization
  in social networks}. In \bibinfo{booktitle}{\emph{International conference on
  pattern recognition and machine intelligence}}. Springer,
  \bibinfo{pages}{242--247}.
\newblock


\bibitem[Laclau et~al\mbox{.}(2021)]%
        {laclau2021all}
\bibfield{author}{\bibinfo{person}{Charlotte Laclau}, \bibinfo{person}{Ievgen
  Redko}, \bibinfo{person}{Manvi Choudhary}, {and} \bibinfo{person}{Christine
  Largeron}.} \bibinfo{year}{2021}\natexlab{}.
\newblock \showarticletitle{All of the Fairness for Edge Prediction with
  Optimal Transport}. In \bibinfo{booktitle}{\emph{International Conference on
  Artificial Intelligence and Statistics}}. PMLR, \bibinfo{pages}{1774--1782}.
\newblock


\bibitem[Lee et~al\mbox{.}(2019)]%
        {lee2019homophily}
\bibfield{author}{\bibinfo{person}{Eun Lee}, \bibinfo{person}{Fariba Karimi},
  \bibinfo{person}{Claudia Wagner}, \bibinfo{person}{Hang-Hyun Jo},
  \bibinfo{person}{Markus Strohmaier}, {and} \bibinfo{person}{Mirta Galesic}.}
  \bibinfo{year}{2019}\natexlab{}.
\newblock \showarticletitle{Homophily and minority-group size explain
  perception biases in social networks}.
\newblock \bibinfo{journal}{\emph{Nature human behaviour}} \bibinfo{volume}{3},
  \bibinfo{number}{10} (\bibinfo{year}{2019}), \bibinfo{pages}{1078--1087}.
\newblock


\bibitem[Leskovec et~al\mbox{.}(2007)]%
        {leskovec2007cost}
\bibfield{author}{\bibinfo{person}{Jure Leskovec}, \bibinfo{person}{Andreas
  Krause}, \bibinfo{person}{Carlos Guestrin}, \bibinfo{person}{Christos
  Faloutsos}, \bibinfo{person}{Jeanne VanBriesen}, {and}
  \bibinfo{person}{Natalie Glance}.} \bibinfo{year}{2007}\natexlab{}.
\newblock \showarticletitle{Cost-effective outbreak detection in networks}. In
  \bibinfo{booktitle}{\emph{Proceedings of the 13th ACM SIGKDD international
  conference on Knowledge discovery and data mining}}.
  \bibinfo{pages}{420--429}.
\newblock


\bibitem[Li et~al\mbox{.}(2015)]%
        {li2015conformity}
\bibfield{author}{\bibinfo{person}{Hui Li}, \bibinfo{person}{Sourav~S
  Bhowmick}, \bibinfo{person}{Aixin Sun}, {and} \bibinfo{person}{Jiangtao
  Cui}.} \bibinfo{year}{2015}\natexlab{}.
\newblock \showarticletitle{Conformity-aware influence maximization in online
  social networks}.
\newblock \bibinfo{journal}{\emph{The VLDB Journal}}  \bibinfo{volume}{24}
  (\bibinfo{year}{2015}), \bibinfo{pages}{117--141}.
\newblock


\bibitem[Li et~al\mbox{.}(2018)]%
        {li2018influence}
\bibfield{author}{\bibinfo{person}{Yuchen Li}, \bibinfo{person}{Ju Fan},
  \bibinfo{person}{Yanhao Wang}, {and} \bibinfo{person}{Kian-Lee Tan}.}
  \bibinfo{year}{2018}\natexlab{}.
\newblock \showarticletitle{Influence maximization on social graphs: A survey}.
\newblock \bibinfo{journal}{\emph{IEEE Transactions on Knowledge and Data
  Engineering}} \bibinfo{volume}{30}, \bibinfo{number}{10}
  (\bibinfo{year}{2018}), \bibinfo{pages}{1852--1872}.
\newblock


\bibitem[Liben-Nowell and Kleinberg(2007)]%
        {liben2007link}
\bibfield{author}{\bibinfo{person}{David Liben-Nowell} {and}
  \bibinfo{person}{Jon Kleinberg}.} \bibinfo{year}{2007}\natexlab{}.
\newblock \showarticletitle{The link-prediction problem for social networks}.
\newblock \bibinfo{journal}{\emph{Journal of the American society for
  information science and technology}} \bibinfo{volume}{58},
  \bibinfo{number}{7} (\bibinfo{year}{2007}), \bibinfo{pages}{1019--1031}.
\newblock


\bibitem[Lichtnwalter and Chawla(2012)]%
        {lichtnwalter2012link}
\bibfield{author}{\bibinfo{person}{Ryan Lichtnwalter} {and}
  \bibinfo{person}{Nitesh~V Chawla}.} \bibinfo{year}{2012}\natexlab{}.
\newblock \showarticletitle{Link prediction: fair and effective evaluation}. In
  \bibinfo{booktitle}{\emph{2012 IEEE/ACM International Conference on Advances
  in Social Networks Analysis and Mining}}. IEEE, \bibinfo{pages}{376--383}.
\newblock


\bibitem[Lu et~al\mbox{.}(2014)]%
        {lu2014algorithms}
\bibfield{author}{\bibinfo{person}{Zongqing Lu}, \bibinfo{person}{Xiao Sun},
  \bibinfo{person}{Yonggang Wen}, \bibinfo{person}{Guohong Cao}, {and}
  \bibinfo{person}{Thomas La~Porta}.} \bibinfo{year}{2014}\natexlab{}.
\newblock \showarticletitle{Algorithms and applications for community detection
  in weighted networks}.
\newblock \bibinfo{journal}{\emph{IEEE Transactions on Parallel and Distributed
  Systems}} \bibinfo{volume}{26}, \bibinfo{number}{11} (\bibinfo{year}{2014}),
  \bibinfo{pages}{2916--2926}.
\newblock


\bibitem[Macedo and Saxena(2024)]%
        {macedo2024gender}
\bibfield{author}{\bibinfo{person}{Mariana Macedo} {and}
  \bibinfo{person}{Akrati Saxena}.} \bibinfo{year}{2024}\natexlab{}.
\newblock \bibinfo{title}{Gender differences in online communication: A case
  study of Soccer}.
\newblock
\newblock
\showeprint[arxiv]{2403.11051}


\bibitem[Malliaros and Vazirgiannis(2013)]%
        {malliaros2013clustering}
\bibfield{author}{\bibinfo{person}{Fragkiskos~D Malliaros} {and}
  \bibinfo{person}{Michalis Vazirgiannis}.} \bibinfo{year}{2013}\natexlab{}.
\newblock \showarticletitle{Clustering and community detection in directed
  networks: A survey}.
\newblock \bibinfo{journal}{\emph{Physics reports}} \bibinfo{volume}{533},
  \bibinfo{number}{4} (\bibinfo{year}{2013}), \bibinfo{pages}{95--142}.
\newblock


\bibitem[Manzano and S{\'a}nchez-Gim{\'e}nez(2019)]%
        {manzano2019women}
\bibfield{author}{\bibinfo{person}{Cristina Manzano} {and}
  \bibinfo{person}{Juan~A S{\'a}nchez-Gim{\'e}nez}.}
  \bibinfo{year}{2019}\natexlab{}.
\newblock \showarticletitle{Women, gender and think tanks: political influence
  network in Twitter 2018}.
\newblock  (\bibinfo{year}{2019}).
\newblock


\bibitem[Masrour et~al\mbox{.}(2020)]%
        {masrour2020bursting}
\bibfield{author}{\bibinfo{person}{Farzan Masrour}, \bibinfo{person}{Tyler
  Wilson}, \bibinfo{person}{Heng Yan}, \bibinfo{person}{Pang-Ning Tan}, {and}
  \bibinfo{person}{Abdol Esfahanian}.} \bibinfo{year}{2020}\natexlab{}.
\newblock \showarticletitle{Bursting the Filter Bubble: Fairness-Aware Network
  Link Prediction}. In \bibinfo{booktitle}{\emph{Proceedings of the AAAI
  Conference on Artificial Intelligence}}, Vol.~\bibinfo{volume}{34}.
  \bibinfo{pages}{841--848}.
\newblock


\bibitem[McAuley and Leskovec(2012)]%
        {mcauley2012learning}
\bibfield{author}{\bibinfo{person}{Julian~J McAuley} {and}
  \bibinfo{person}{Jure Leskovec}.} \bibinfo{year}{2012}\natexlab{}.
\newblock \showarticletitle{Learning to discover social circles in ego
  networks.}. In \bibinfo{booktitle}{\emph{NIPS}}, Vol.~\bibinfo{volume}{2012}.
  Citeseer, \bibinfo{pages}{548--56}.
\newblock


\bibitem[McGuire(2002)]%
        {mcguire2002gender}
\bibfield{author}{\bibinfo{person}{Gail~M McGuire}.}
  \bibinfo{year}{2002}\natexlab{}.
\newblock \showarticletitle{Gender, race, and the shadow structure: A study of
  informal networks and inequality in a work organization}.
\newblock \bibinfo{journal}{\emph{Gender \& society}} \bibinfo{volume}{16},
  \bibinfo{number}{3} (\bibinfo{year}{2002}), \bibinfo{pages}{303--322}.
\newblock


\bibitem[McPherson et~al\mbox{.}(2001)]%
        {mcpherson2001birds}
\bibfield{author}{\bibinfo{person}{Miller McPherson}, \bibinfo{person}{Lynn
  Smith-Lovin}, {and} \bibinfo{person}{James~M Cook}.}
  \bibinfo{year}{2001}\natexlab{}.
\newblock \showarticletitle{Birds of a feather: Homophily in social networks}.
\newblock \bibinfo{journal}{\emph{Annual review of sociology}}
  \bibinfo{volume}{27}, \bibinfo{number}{1} (\bibinfo{year}{2001}),
  \bibinfo{pages}{415--444}.
\newblock


\bibitem[Mehrabi et~al\mbox{.}(2019)]%
        {mehrabi2019debiasing}
\bibfield{author}{\bibinfo{person}{Ninareh Mehrabi}, \bibinfo{person}{Fred
  Morstatter}, \bibinfo{person}{Nanyun Peng}, {and} \bibinfo{person}{Aram
  Galstyan}.} \bibinfo{year}{2019}\natexlab{}.
\newblock \showarticletitle{Debiasing community detection: The importance of
  lowly connected nodes}. In \bibinfo{booktitle}{\emph{2019 IEEE/ACM
  International Conference on Advances in Social Networks Analysis and Mining
  (ASONAM)}}. IEEE, \bibinfo{pages}{509--512}.
\newblock


\bibitem[Mehrabi et~al\mbox{.}(2021)]%
        {mehrabi2021survey}
\bibfield{author}{\bibinfo{person}{Ninareh Mehrabi}, \bibinfo{person}{Fred
  Morstatter}, \bibinfo{person}{Nripsuta Saxena}, \bibinfo{person}{Kristina
  Lerman}, {and} \bibinfo{person}{Aram Galstyan}.}
  \bibinfo{year}{2021}\natexlab{}.
\newblock \showarticletitle{A survey on bias and fairness in machine learning}.
\newblock \bibinfo{journal}{\emph{ACM Computing Surveys (CSUR)}}
  \bibinfo{volume}{54}, \bibinfo{number}{6} (\bibinfo{year}{2021}),
  \bibinfo{pages}{1--35}.
\newblock


\bibitem[Messias et~al\mbox{.}(2017)]%
        {messias2017white}
\bibfield{author}{\bibinfo{person}{Johnnatan Messias},
  \bibinfo{person}{Pantelis Vikatos}, {and} \bibinfo{person}{Fabr{\'\i}cio
  Benevenuto}.} \bibinfo{year}{2017}\natexlab{}.
\newblock \showarticletitle{White, man, and highly followed: Gender and race
  inequalities in Twitter}. In \bibinfo{booktitle}{\emph{Proceedings of the
  International Conference on Web Intelligence}}. \bibinfo{pages}{266--274}.
\newblock


\bibitem[Miller et~al\mbox{.}(2018)]%
        {miller2018discovering}
\bibfield{author}{\bibinfo{person}{Ryan Miller}, \bibinfo{person}{Ralucca
  Gera}, \bibinfo{person}{Akrati Saxena}, {and} \bibinfo{person}{Tanmoy
  Chakraborty}.} \bibinfo{year}{2018}\natexlab{}.
\newblock \showarticletitle{Discovering and leveraging communities in dark
  multi-layered networks for network disruption}. In
  \bibinfo{booktitle}{\emph{2018 IEEE/ACM International Conference on Advances
  in Social Networks Analysis and Mining (ASONAM)}}. IEEE,
  \bibinfo{pages}{1152--1159}.
\newblock


\bibitem[Mislove et~al\mbox{.}(2010)]%
        {mislove2010you}
\bibfield{author}{\bibinfo{person}{Alan Mislove}, \bibinfo{person}{Bimal
  Viswanath}, \bibinfo{person}{Krishna~P Gummadi}, {and} \bibinfo{person}{Peter
  Druschel}.} \bibinfo{year}{2010}\natexlab{}.
\newblock \showarticletitle{You are who you know: inferring user profiles in
  online social networks}. In \bibinfo{booktitle}{\emph{Proceedings of the
  third ACM international conference on Web search and data mining}}.
  \bibinfo{pages}{251--260}.
\newblock


\bibitem[Mitchell et~al\mbox{.}(2021)]%
        {mitchell2021algorithmic}
\bibfield{author}{\bibinfo{person}{Shira Mitchell}, \bibinfo{person}{Eric
  Potash}, \bibinfo{person}{Solon Barocas}, \bibinfo{person}{Alexander
  D'Amour}, {and} \bibinfo{person}{Kristian Lum}.}
  \bibinfo{year}{2021}\natexlab{}.
\newblock \showarticletitle{Algorithmic fairness: Choices, assumptions, and
  definitions}.
\newblock \bibinfo{journal}{\emph{Annual Review of Statistics and Its
  Application}}  \bibinfo{volume}{8} (\bibinfo{year}{2021}),
  \bibinfo{pages}{141--163}.
\newblock


\bibitem[Mohamadi-Baghmolaei et~al\mbox{.}(2015)]%
        {mohamadi2015trust}
\bibfield{author}{\bibinfo{person}{Rezvan Mohamadi-Baghmolaei},
  \bibinfo{person}{Niloofar Mozafari}, {and} \bibinfo{person}{Ali Hamzeh}.}
  \bibinfo{year}{2015}\natexlab{}.
\newblock \showarticletitle{Trust based latency aware influence maximization in
  social networks}.
\newblock \bibinfo{journal}{\emph{Engineering Applications of Artificial
  Intelligence}}  \bibinfo{volume}{41} (\bibinfo{year}{2015}),
  \bibinfo{pages}{195--206}.
\newblock


\bibitem[Mulligan et~al\mbox{.}(2019)]%
        {mulligan2019thing}
\bibfield{author}{\bibinfo{person}{Deirdre~K Mulligan},
  \bibinfo{person}{Joshua~A Kroll}, \bibinfo{person}{Nitin Kohli}, {and}
  \bibinfo{person}{Richmond~Y Wong}.} \bibinfo{year}{2019}\natexlab{}.
\newblock \showarticletitle{This thing called fairness: Disciplinary confusion
  realizing a value in technology}.
\newblock \bibinfo{journal}{\emph{Proceedings of the ACM on Human-Computer
  Interaction}} \bibinfo{volume}{3}, \bibinfo{number}{CSCW}
  (\bibinfo{year}{2019}), \bibinfo{pages}{1--36}.
\newblock


\bibitem[Newman(2003)]%
        {newman2003mixing}
\bibfield{author}{\bibinfo{person}{Mark~EJ Newman}.}
  \bibinfo{year}{2003}\natexlab{}.
\newblock \showarticletitle{Mixing patterns in networks}.
\newblock \bibinfo{journal}{\emph{Physical review E}} \bibinfo{volume}{67},
  \bibinfo{number}{2} (\bibinfo{year}{2003}), \bibinfo{pages}{026126}.
\newblock


\bibitem[Nguyen et~al\mbox{.}(2020)]%
        {nguyen2020dynamics}
\bibfield{author}{\bibinfo{person}{Vu~Xuan Nguyen}, \bibinfo{person}{Gaoxi
  Xiao}, \bibinfo{person}{Xin-Jian Xu}, \bibinfo{person}{Qingchu Wu}, {and}
  \bibinfo{person}{Cheng-Yi Xia}.} \bibinfo{year}{2020}\natexlab{}.
\newblock \showarticletitle{Dynamics of opinion formation under majority rules
  on complex social networks}.
\newblock \bibinfo{journal}{\emph{Scientific reports}} \bibinfo{volume}{10},
  \bibinfo{number}{1} (\bibinfo{year}{2020}), \bibinfo{pages}{1--9}.
\newblock


\bibitem[Nilizadeh et~al\mbox{.}(2016)]%
        {nilizadeh2016twitter}
\bibfield{author}{\bibinfo{person}{Shirin Nilizadeh}, \bibinfo{person}{Anne
  Groggel}, \bibinfo{person}{Peter Lista}, \bibinfo{person}{Srijita Das},
  \bibinfo{person}{Yong-Yeol Ahn}, \bibinfo{person}{Apu Kapadia}, {and}
  \bibinfo{person}{Fabio Rojas}.} \bibinfo{year}{2016}\natexlab{}.
\newblock \showarticletitle{Twitter's glass ceiling: The effect of perceived
  gender on online visibility}. In \bibinfo{booktitle}{\emph{Proceedings of the
  International AAAI Conference on Web and Social Media}},
  Vol.~\bibinfo{volume}{10}. \bibinfo{pages}{289--298}.
\newblock


\bibitem[Nishi et~al\mbox{.}(2015)]%
        {nishi2015inequality}
\bibfield{author}{\bibinfo{person}{Akihiro Nishi}, \bibinfo{person}{Hirokazu
  Shirado}, \bibinfo{person}{David~G Rand}, {and} \bibinfo{person}{Nicholas~A
  Christakis}.} \bibinfo{year}{2015}\natexlab{}.
\newblock \showarticletitle{Inequality and visibility of wealth in experimental
  social networks}.
\newblock \bibinfo{journal}{\emph{Nature}} \bibinfo{volume}{526},
  \bibinfo{number}{7573} (\bibinfo{year}{2015}), \bibinfo{pages}{426--429}.
\newblock


\bibitem[Panchendrarajan and Saxena(2023)]%
        {panchendrarajan2023topic}
\bibfield{author}{\bibinfo{person}{Rrubaa Panchendrarajan} {and}
  \bibinfo{person}{Akrati Saxena}.} \bibinfo{year}{2023}\natexlab{}.
\newblock \showarticletitle{Topic-based influential user detection: a survey}.
\newblock \bibinfo{journal}{\emph{Applied Intelligence}} \bibinfo{volume}{53},
  \bibinfo{number}{5} (\bibinfo{year}{2023}), \bibinfo{pages}{5998--6024}.
\newblock


\bibitem[Pena-L{\'o}pez et~al\mbox{.}(2021)]%
        {pena2021inequality}
\bibfield{author}{\bibinfo{person}{Atilano Pena-L{\'o}pez},
  \bibinfo{person}{Paolo Rungo}, {and} \bibinfo{person}{Jos{\'e}~Manuel
  S{\'a}nchez-Santos}.} \bibinfo{year}{2021}\natexlab{}.
\newblock \showarticletitle{Inequality and individuals’ social networks: the
  other face of social capital}.
\newblock \bibinfo{journal}{\emph{Cambridge Journal of Economics}}
  \bibinfo{volume}{45}, \bibinfo{number}{4} (\bibinfo{year}{2021}),
  \bibinfo{pages}{675--694}.
\newblock


\bibitem[Pessach and Shmueli(2020)]%
        {pessach2020algorithmic}
\bibfield{author}{\bibinfo{person}{Dana Pessach} {and} \bibinfo{person}{Erez
  Shmueli}.} \bibinfo{year}{2020}\natexlab{}.
\newblock \showarticletitle{Algorithmic fairness}.
\newblock \bibinfo{journal}{\emph{arXiv preprint arXiv:2001.09784}}
  (\bibinfo{year}{2020}).
\newblock


\bibitem[Pessach and Shmueli(2022)]%
        {pessach2022review}
\bibfield{author}{\bibinfo{person}{Dana Pessach} {and} \bibinfo{person}{Erez
  Shmueli}.} \bibinfo{year}{2022}\natexlab{}.
\newblock \showarticletitle{A Review on Fairness in Machine Learning}.
\newblock \bibinfo{journal}{\emph{ACM Computing Surveys (CSUR)}}
  \bibinfo{volume}{55}, \bibinfo{number}{3} (\bibinfo{year}{2022}),
  \bibinfo{pages}{1--44}.
\newblock


\bibitem[Pham et~al\mbox{.}(2020)]%
        {pham2020influence}
\bibfield{author}{\bibinfo{person}{Canh~V Pham}, \bibinfo{person}{Dung~KT Ha},
  \bibinfo{person}{Quang~C Vu}, \bibinfo{person}{Anh~N Su}, {and}
  \bibinfo{person}{Huan~X Hoang}.} \bibinfo{year}{2020}\natexlab{}.
\newblock \showarticletitle{Influence Maximization with Priority in Online
  Social Networks}.
\newblock \bibinfo{journal}{\emph{Algorithms}} \bibinfo{volume}{13},
  \bibinfo{number}{8} (\bibinfo{year}{2020}), \bibinfo{pages}{183}.
\newblock


\bibitem[Pham et~al\mbox{.}(2019)]%
        {pham2019multiple}
\bibfield{author}{\bibinfo{person}{Dung~V Pham}, \bibinfo{person}{Hieu~V
  Duong}, \bibinfo{person}{Canh~V Pham}, \bibinfo{person}{Bui~Q Bao}, {and}
  \bibinfo{person}{Anh~V Nguyen}.} \bibinfo{year}{2019}\natexlab{}.
\newblock \showarticletitle{Multiple Topics Misinformation blocking in Online
  Social Networks}. In \bibinfo{booktitle}{\emph{2019 11th International
  Conference on Knowledge and Systems Engineering (KSE)}}. IEEE,
  \bibinfo{pages}{1--6}.
\newblock


\bibitem[Pitoura et~al\mbox{.}(2021)]%
        {pitoura2021fairness}
\bibfield{author}{\bibinfo{person}{Evaggelia Pitoura}, \bibinfo{person}{Kostas
  Stefanidis}, {and} \bibinfo{person}{Georgia Koutrika}.}
  \bibinfo{year}{2021}\natexlab{}.
\newblock \showarticletitle{Fairness in rankings and recommendations: an
  overview}.
\newblock \bibinfo{journal}{\emph{The VLDB Journal}} (\bibinfo{year}{2021}),
  \bibinfo{pages}{1--28}.
\newblock


\bibitem[Planti{\'e} and Crampes(2013)]%
        {plantie2013survey}
\bibfield{author}{\bibinfo{person}{Michel Planti{\'e}} {and}
  \bibinfo{person}{Michel Crampes}.} \bibinfo{year}{2013}\natexlab{}.
\newblock \showarticletitle{Survey on social community detection}.
\newblock In \bibinfo{booktitle}{\emph{Social media retrieval}}.
  \bibinfo{publisher}{Springer}, \bibinfo{pages}{65--85}.
\newblock


\bibitem[Rahman et~al\mbox{.}(2019)]%
        {rahman2019fairwalk}
\bibfield{author}{\bibinfo{person}{Tahleen Rahman}, \bibinfo{person}{Bartlomiej
  Surma}, \bibinfo{person}{Michael Backes}, {and} \bibinfo{person}{Yang
  Zhang}.} \bibinfo{year}{2019}\natexlab{}.
\newblock \showarticletitle{Fairwalk: Towards fair graph embedding}.
\newblock  (\bibinfo{year}{2019}).
\newblock


\bibitem[Rahmattalabi et~al\mbox{.}(2020)]%
        {rahmattalabi2020fair}
\bibfield{author}{\bibinfo{person}{Aida Rahmattalabi}, \bibinfo{person}{Shahin
  Jabbari}, \bibinfo{person}{Himabindu Lakkaraju}, \bibinfo{person}{Phebe
  Vayanos}, \bibinfo{person}{Max Izenberg}, \bibinfo{person}{Ryan Brown},
  \bibinfo{person}{Eric Rice}, {and} \bibinfo{person}{Milind Tambe}.}
  \bibinfo{year}{2020}\natexlab{}.
\newblock \showarticletitle{Fair influence maximization: A welfare optimization
  approach}.
\newblock \bibinfo{journal}{\emph{arXiv preprint arXiv:2006.07906}}
  (\bibinfo{year}{2020}).
\newblock


\bibitem[Ribeiro et~al\mbox{.}(2018)]%
        {ribeiro2018characterizing}
\bibfield{author}{\bibinfo{person}{Manoel~Horta Ribeiro},
  \bibinfo{person}{Pedro~H Calais}, \bibinfo{person}{Yuri~A Santos},
  \bibinfo{person}{Virg{\'\i}lio~AF Almeida}, {and} \bibinfo{person}{Wagner
  Meira~Jr}.} \bibinfo{year}{2018}\natexlab{}.
\newblock \showarticletitle{Characterizing and detecting hateful users on
  twitter}. In \bibinfo{booktitle}{\emph{Twelfth international AAAI conference
  on web and social media}}.
\newblock


\bibitem[Roberts and Everton(2011)]%
        {roberts2011strategies}
\bibfield{author}{\bibinfo{person}{Nancy Roberts} {and} \bibinfo{person}{Sean~F
  Everton}.} \bibinfo{year}{2011}\natexlab{}.
\newblock \showarticletitle{Strategies for combating dark networks}.
\newblock \bibinfo{journal}{\emph{Journal of Social Structure}}
  (\bibinfo{year}{2011}).
\newblock


\bibitem[Rocha et~al\mbox{.}(2011)]%
        {rocha2011simulated}
\bibfield{author}{\bibinfo{person}{Luis~EC Rocha}, \bibinfo{person}{Fredrik
  Liljeros}, {and} \bibinfo{person}{Petter Holme}.}
  \bibinfo{year}{2011}\natexlab{}.
\newblock \showarticletitle{Simulated epidemics in an empirical spatiotemporal
  network of 50,185 sexual contacts}.
\newblock \bibinfo{journal}{\emph{PLoS computational biology}}
  \bibinfo{volume}{7}, \bibinfo{number}{3} (\bibinfo{year}{2011}),
  \bibinfo{pages}{e1001109}.
\newblock


\bibitem[Rossi and Ahmed(2015)]%
        {rossi2015network}
\bibfield{author}{\bibinfo{person}{Ryan Rossi} {and} \bibinfo{person}{Nesreen
  Ahmed}.} \bibinfo{year}{2015}\natexlab{}.
\newblock \showarticletitle{The network data repository with interactive graph
  analytics and visualization}. In \bibinfo{booktitle}{\emph{Twenty-Ninth AAAI
  Conference on Artificial Intelligence}}.
\newblock


\bibitem[Santosa et~al\mbox{.}(2018)]%
        {santosa2018digital}
\bibfield{author}{\bibinfo{person}{Hedi~Pudjo Santosa}, \bibinfo{person}{Nurul
  Hasfi}, {and} \bibinfo{person}{Triyono Lukmantoro}.}
  \bibinfo{year}{2018}\natexlab{}.
\newblock \showarticletitle{Digital Media Unequality during the 2014th
  Indonesian Presidential Election}. In \bibinfo{booktitle}{\emph{E3S Web of
  Conferences}}, Vol.~\bibinfo{volume}{73}. EDP Sciences,
  \bibinfo{pages}{14006}.
\newblock


\bibitem[Saxena(2022)]%
        {saxena2022evolving}
\bibfield{author}{\bibinfo{person}{Akrati Saxena}.}
  \bibinfo{year}{2022}\natexlab{}.
\newblock \showarticletitle{Evolving Models for Dynamic Weighted Complex
  Networks}.
\newblock In \bibinfo{booktitle}{\emph{Principles of Social Networking}}.
  \bibinfo{publisher}{Springer}, \bibinfo{pages}{177--208}.
\newblock


\bibitem[Saxena et~al\mbox{.}(2021a)]%
        {saxena2021hm}
\bibfield{author}{\bibinfo{person}{Akrati Saxena}, \bibinfo{person}{George
  Fletcher}, {and} \bibinfo{person}{Mykola Pechenizkiy}.}
  \bibinfo{year}{2021}\natexlab{a}.
\newblock \showarticletitle{HM-EIICT: Fairness-aware link prediction in complex
  networks using community information}.
\newblock \bibinfo{journal}{\emph{Journal of Combinatorial Optimization}}
  (\bibinfo{year}{2021}), \bibinfo{pages}{1--18}.
\newblock


\bibitem[Saxena et~al\mbox{.}(2021b)]%
        {saxena2021fair}
\bibfield{author}{\bibinfo{person}{Akrati Saxena}, \bibinfo{person}{George
  Fletcher}, {and} \bibinfo{person}{Mykola Pechenizkiy}.}
  \bibinfo{year}{2021}\natexlab{b}.
\newblock \showarticletitle{How Fair is Fairness-aware Representative
  Ranking?}. In \bibinfo{booktitle}{\emph{Companion Proceedings of the Web
  Conference 2021}}. \bibinfo{pages}{161--165}.
\newblock


\bibitem[Saxena et~al\mbox{.}(2022a)]%
        {saxena2021nodesim}
\bibfield{author}{\bibinfo{person}{Akrati Saxena}, \bibinfo{person}{George
  Fletcher}, {and} \bibinfo{person}{Mykola Pechenizkiy}.}
  \bibinfo{year}{2022}\natexlab{a}.
\newblock \showarticletitle{NodeSim: node similarity based network embedding
  for diverse link prediction}.
\newblock \bibinfo{journal}{\emph{EPJ Data Science}} \bibinfo{volume}{11},
  \bibinfo{number}{1} (\bibinfo{year}{2022}), \bibinfo{pages}{24}.
\newblock


\bibitem[Saxena et~al\mbox{.}(2017a)]%
        {saxena2017fast}
\bibfield{author}{\bibinfo{person}{Akrati Saxena}, \bibinfo{person}{Ralucca
  Gera}, {and} \bibinfo{person}{SRS Iyengar}.}
  \bibinfo{year}{2017}\natexlab{a}.
\newblock \showarticletitle{Fast estimation of closeness centrality ranking}.
  In \bibinfo{booktitle}{\emph{Proceedings of the 2017 IEEE/ACM International
  Conference on Advances in Social Networks Analysis and Mining 2017}}.
  \bibinfo{pages}{80--85}.
\newblock


\bibitem[Saxena et~al\mbox{.}(2017b)]%
        {saxena2017observe}
\bibfield{author}{\bibinfo{person}{Akrati Saxena}, \bibinfo{person}{Ralucca
  Gera}, {and} \bibinfo{person}{SRS Iyengar}.}
  \bibinfo{year}{2017}\natexlab{b}.
\newblock \showarticletitle{Observe locally rank globally}. In
  \bibinfo{booktitle}{\emph{Proceedings of the 2017 IEEE/ACM International
  Conference on Advances in Social Networks Analysis and Mining 2017}}.
  \bibinfo{pages}{139--144}.
\newblock


\bibitem[Saxena et~al\mbox{.}(2018)]%
        {saxena2018estimating}
\bibfield{author}{\bibinfo{person}{Akrati Saxena}, \bibinfo{person}{Ralucca
  Gera}, {and} \bibinfo{person}{SRS Iyengar}.} \bibinfo{year}{2018}\natexlab{}.
\newblock \showarticletitle{Estimating degree rank in complex networks}.
\newblock \bibinfo{journal}{\emph{Social Network Analysis and Mining}}
  \bibinfo{volume}{8}, \bibinfo{number}{1} (\bibinfo{year}{2018}),
  \bibinfo{pages}{1--20}.
\newblock


\bibitem[Saxena et~al\mbox{.}(2019)]%
        {saxena2019heuristic}
\bibfield{author}{\bibinfo{person}{Akrati Saxena}, \bibinfo{person}{Ralucca
  Gera}, {and} \bibinfo{person}{SRS Iyengar}.} \bibinfo{year}{2019}\natexlab{}.
\newblock \showarticletitle{A heuristic approach to estimate nodes’ closeness
  rank using the properties of real world networks}.
\newblock \bibinfo{journal}{\emph{Social Network Analysis and Mining}}
  \bibinfo{volume}{9}, \bibinfo{number}{1} (\bibinfo{year}{2019}),
  \bibinfo{pages}{1--16}.
\newblock


\bibitem[Saxena et~al\mbox{.}(2023)]%
        {saxena2023fairness}
\bibfield{author}{\bibinfo{person}{Akrati Saxena}, \bibinfo{person}{Cristina
  Guti{\'e}rrez~Bierbooms}, {and} \bibinfo{person}{Mykola Pechenizkiy}.}
  \bibinfo{year}{2023}\natexlab{}.
\newblock \showarticletitle{Fairness-aware fake news mitigation using counter
  information propagation}.
\newblock \bibinfo{journal}{\emph{Applied Intelligence}} \bibinfo{volume}{53},
  \bibinfo{number}{22} (\bibinfo{year}{2023}), \bibinfo{pages}{27483--27504}.
\newblock


\bibitem[Saxena et~al\mbox{.}(2020)]%
        {saxena2020mitigating}
\bibfield{author}{\bibinfo{person}{Akrati Saxena}, \bibinfo{person}{Wynne Hsu},
  \bibinfo{person}{Mong~Li Lee}, \bibinfo{person}{Hai Leong~Chieu},
  \bibinfo{person}{Lynette Ng}, {and} \bibinfo{person}{Loo~Nin Teow}.}
  \bibinfo{year}{2020}\natexlab{}.
\newblock \showarticletitle{Mitigating misinformation in online social network
  with top-k debunkers and evolving user opinions}. In
  \bibinfo{booktitle}{\emph{Companion Proceedings of the Web Conference 2020}}.
  \bibinfo{pages}{363--370}.
\newblock


\bibitem[Saxena and Iyengar(2017)]%
        {saxena2017global}
\bibfield{author}{\bibinfo{person}{Akrati Saxena} {and} \bibinfo{person}{SRS
  Iyengar}.} \bibinfo{year}{2017}\natexlab{}.
\newblock \showarticletitle{Global rank estimation}.
\newblock \bibinfo{journal}{\emph{arXiv preprint arXiv:1710.11341}}
  (\bibinfo{year}{2017}).
\newblock


\bibitem[Saxena and Iyengar(2018)]%
        {saxena2018k}
\bibfield{author}{\bibinfo{person}{Akrati Saxena} {and} \bibinfo{person}{SRS
  Iyengar}.} \bibinfo{year}{2018}\natexlab{}.
\newblock \showarticletitle{K-shell rank analysis using local information}. In
  \bibinfo{booktitle}{\emph{International Conference on Computational Social
  Networks}}. Springer, \bibinfo{pages}{198--210}.
\newblock


\bibitem[Saxena and Iyengar(2020)]%
        {saxena2020centrality}
\bibfield{author}{\bibinfo{person}{Akrati Saxena} {and}
  \bibinfo{person}{Sudarshan Iyengar}.} \bibinfo{year}{2020}\natexlab{}.
\newblock \showarticletitle{Centrality measures in complex networks: A survey}.
\newblock \bibinfo{journal}{\emph{arXiv preprint arXiv:2011.07190}}
  (\bibinfo{year}{2020}).
\newblock


\bibitem[Saxena et~al\mbox{.}(2015)]%
        {saxena2015understanding}
\bibfield{author}{\bibinfo{person}{Akrati Saxena}, \bibinfo{person}{SRS
  Iyengar}, {and} \bibinfo{person}{Yayati Gupta}.}
  \bibinfo{year}{2015}\natexlab{}.
\newblock \showarticletitle{Understanding spreading patterns on social networks
  based on network topology}. In \bibinfo{booktitle}{\emph{Proceedings of the
  2015 IEEE/ACM International Conference on Advances in Social Networks
  Analysis and Mining 2015}}. \bibinfo{pages}{1616--1617}.
\newblock


\bibitem[Saxena et~al\mbox{.}(2022b)]%
        {saxena2022fake}
\bibfield{author}{\bibinfo{person}{Akrati Saxena}, \bibinfo{person}{Pratishtha
  Saxena}, {and} \bibinfo{person}{Harita Reddy}.}
  \bibinfo{year}{2022}\natexlab{b}.
\newblock \showarticletitle{Fake News Detection Techniques for Social Media}.
\newblock In \bibinfo{booktitle}{\emph{Principles of Social Networking}}.
  \bibinfo{publisher}{Springer}, \bibinfo{pages}{325--354}.
\newblock


\bibitem[Saxena et~al\mbox{.}(2022c)]%
        {saxena2022fakeprop}
\bibfield{author}{\bibinfo{person}{Akrati Saxena}, \bibinfo{person}{Pratishtha
  Saxena}, {and} \bibinfo{person}{Harita Reddy}.}
  \bibinfo{year}{2022}\natexlab{c}.
\newblock \showarticletitle{Fake News Propagation and Mitigation Techniques: A
  Survey}.
\newblock In \bibinfo{booktitle}{\emph{Principles of Social Networking}}.
  \bibinfo{publisher}{Springer}, \bibinfo{pages}{355--386}.
\newblock


\bibitem[Schwartz and Winship(1980)]%
        {schwartz1980welfare}
\bibfield{author}{\bibinfo{person}{Joseph Schwartz} {and}
  \bibinfo{person}{Christopher Winship}.} \bibinfo{year}{1980}\natexlab{}.
\newblock \showarticletitle{The welfare approach to measuring inequality}.
\newblock \bibinfo{journal}{\emph{Sociological methodology}}
  \bibinfo{volume}{11} (\bibinfo{year}{1980}), \bibinfo{pages}{1--36}.
\newblock


\bibitem[Seierstad and Opsahl(2011)]%
        {seierstad2011few}
\bibfield{author}{\bibinfo{person}{Cathrine Seierstad} {and}
  \bibinfo{person}{Tore Opsahl}.} \bibinfo{year}{2011}\natexlab{}.
\newblock \showarticletitle{For the few not the many? The effects of
  affirmative action on presence, prominence, and social capital of women
  directors in Norway}.
\newblock \bibinfo{journal}{\emph{Scandinavian journal of management}}
  \bibinfo{volume}{27}, \bibinfo{number}{1} (\bibinfo{year}{2011}),
  \bibinfo{pages}{44--54}.
\newblock


\bibitem[Shetty and Adibi(2004)]%
        {shetty2004enron}
\bibfield{author}{\bibinfo{person}{Jitesh Shetty} {and} \bibinfo{person}{Jafar
  Adibi}.} \bibinfo{year}{2004}\natexlab{}.
\newblock \showarticletitle{The Enron email dataset database schema and brief
  statistical report}.
\newblock \bibinfo{journal}{\emph{Information sciences institute technical
  report, University of Southern California}} \bibinfo{volume}{4},
  \bibinfo{number}{1} (\bibinfo{year}{2004}), \bibinfo{pages}{120--128}.
\newblock


\bibitem[Singh and Joachims(2018)]%
        {singh2018fairness}
\bibfield{author}{\bibinfo{person}{Ashudeep Singh} {and}
  \bibinfo{person}{Thorsten Joachims}.} \bibinfo{year}{2018}\natexlab{}.
\newblock \showarticletitle{Fairness of exposure in rankings}. In
  \bibinfo{booktitle}{\emph{Proceedings of the 24th ACM SIGKDD International
  Conference on Knowledge Discovery \& Data Mining}}.
  \bibinfo{pages}{2219--2228}.
\newblock


\bibitem[Snijders et~al\mbox{.}(2010)]%
        {snijders2010introduction}
\bibfield{author}{\bibinfo{person}{Tom~AB Snijders}, \bibinfo{person}{Gerhard~G
  Van~de Bunt}, {and} \bibinfo{person}{Christian~EG Steglich}.}
  \bibinfo{year}{2010}\natexlab{}.
\newblock \showarticletitle{Introduction to stochastic actor-based models for
  network dynamics}.
\newblock \bibinfo{journal}{\emph{Social networks}} \bibinfo{volume}{32},
  \bibinfo{number}{1} (\bibinfo{year}{2010}), \bibinfo{pages}{44--60}.
\newblock


\bibitem[Spinelli et~al\mbox{.}(2021)]%
        {spinelli2021biased}
\bibfield{author}{\bibinfo{person}{Indro Spinelli}, \bibinfo{person}{Simone
  Scardapane}, \bibinfo{person}{Amir Hussain}, {and} \bibinfo{person}{Aurelio
  Uncini}.} \bibinfo{year}{2021}\natexlab{}.
\newblock \showarticletitle{Biased Edge Dropout for Enhancing Fairness in Graph
  Representation Learning}.
\newblock \bibinfo{journal}{\emph{arXiv preprint arXiv:2104.14210}}
  (\bibinfo{year}{2021}).
\newblock


\bibitem[Stein et~al\mbox{.}(2017)]%
        {stein2017heuristic}
\bibfield{author}{\bibinfo{person}{Sebastian Stein}, \bibinfo{person}{Soheil
  Eshghi}, \bibinfo{person}{Setareh Maghsudi}, \bibinfo{person}{Leandros
  Tassiulas}, \bibinfo{person}{Rachel~KE Bellamy}, {and}
  \bibinfo{person}{Nicholas~R Jennings}.} \bibinfo{year}{2017}\natexlab{}.
\newblock \showarticletitle{Heuristic algorithms for influence maximization in
  partially observable social networks}. In \bibinfo{booktitle}{\emph{SocInf@
  IJCAI}}.
\newblock


\bibitem[Stoica(2018)]%
        {stoica2018homophily}
\bibfield{author}{\bibinfo{person}{Adelina-Alexandra Stoica}.}
  \bibinfo{year}{2018}\natexlab{}.
\newblock \showarticletitle{Homophily in co-autorship networks}.
\newblock \bibinfo{journal}{\emph{Int. Rev. Soc. Res}} \bibinfo{volume}{8},
  \bibinfo{number}{2} (\bibinfo{year}{2018}), \bibinfo{pages}{119--128}.
\newblock


\bibitem[Stoica and Chaintreau(2019)]%
        {stoica2019fairness}
\bibfield{author}{\bibinfo{person}{Ana-Andreea Stoica} {and}
  \bibinfo{person}{Augustin Chaintreau}.} \bibinfo{year}{2019}\natexlab{}.
\newblock \showarticletitle{Fairness in social influence maximization}. In
  \bibinfo{booktitle}{\emph{Companion Proceedings of The 2019 World Wide Web
  Conference}}. \bibinfo{pages}{569--574}.
\newblock


\bibitem[Stoica et~al\mbox{.}(2020)]%
        {stoica2020seeding}
\bibfield{author}{\bibinfo{person}{Ana-Andreea Stoica},
  \bibinfo{person}{Jessy~Xinyi Han}, {and} \bibinfo{person}{Augustin
  Chaintreau}.} \bibinfo{year}{2020}\natexlab{}.
\newblock \showarticletitle{Seeding network influence in biased networks and
  the benefits of diversity}. In \bibinfo{booktitle}{\emph{Proceedings of The
  Web Conference 2020}}. \bibinfo{pages}{2089--2098}.
\newblock


\bibitem[Stoica et~al\mbox{.}(2018)]%
        {stoica2018algorithmic}
\bibfield{author}{\bibinfo{person}{Ana-Andreea Stoica},
  \bibinfo{person}{Christopher Riederer}, {and} \bibinfo{person}{Augustin
  Chaintreau}.} \bibinfo{year}{2018}\natexlab{}.
\newblock \showarticletitle{Algorithmic Glass Ceiling in Social Networks: The
  effects of social recommendations on network diversity}. In
  \bibinfo{booktitle}{\emph{Proceedings of the 2018 World Wide Web
  Conference}}. \bibinfo{pages}{923--932}.
\newblock


\bibitem[Sumith et~al\mbox{.}(2018)]%
        {sumith2018influence}
\bibfield{author}{\bibinfo{person}{N Sumith}, \bibinfo{person}{B Annappa},
  {and} \bibinfo{person}{Swapan Bhattacharya}.}
  \bibinfo{year}{2018}\natexlab{}.
\newblock \showarticletitle{Influence maximization in large social networks:
  Heuristics, models and parameters}.
\newblock \bibinfo{journal}{\emph{Future Generation Computer Systems}}
  \bibinfo{volume}{89} (\bibinfo{year}{2018}), \bibinfo{pages}{777--790}.
\newblock


\bibitem[Takac and Zabovsky(2012)]%
        {takac2012data}
\bibfield{author}{\bibinfo{person}{Lubos Takac} {and} \bibinfo{person}{Michal
  Zabovsky}.} \bibinfo{year}{2012}\natexlab{}.
\newblock \showarticletitle{Data analysis in public social networks}. In
  \bibinfo{booktitle}{\emph{International scientific conference and
  international workshop present day trends of innovations}},
  Vol.~\bibinfo{volume}{1}.
\newblock


\bibitem[Tang et~al\mbox{.}(2015)]%
        {tang2015influence}
\bibfield{author}{\bibinfo{person}{Youze Tang}, \bibinfo{person}{Yanchen Shi},
  {and} \bibinfo{person}{Xiaokui Xiao}.} \bibinfo{year}{2015}\natexlab{}.
\newblock \showarticletitle{Influence maximization in near-linear time: A
  martingale approach}. In \bibinfo{booktitle}{\emph{Proceedings of the 2015
  ACM SIGMOD international conference on management of data}}.
  \bibinfo{pages}{1539--1554}.
\newblock


\bibitem[Tang et~al\mbox{.}(2014)]%
        {tang2014influence}
\bibfield{author}{\bibinfo{person}{Youze Tang}, \bibinfo{person}{Xiaokui Xiao},
  {and} \bibinfo{person}{Yanchen Shi}.} \bibinfo{year}{2014}\natexlab{}.
\newblock \showarticletitle{Influence maximization: Near-optimal time
  complexity meets practical efficiency}. In
  \bibinfo{booktitle}{\emph{Proceedings of the 2014 ACM SIGMOD international
  conference on Management of data}}. \bibinfo{pages}{75--86}.
\newblock


\bibitem[Tarbush and Teytelboym(2012)]%
        {tarbush2012homophily}
\bibfield{author}{\bibinfo{person}{Bassel Tarbush} {and}
  \bibinfo{person}{Alexander Teytelboym}.} \bibinfo{year}{2012}\natexlab{}.
\newblock \showarticletitle{Homophily in online social networks}. In
  \bibinfo{booktitle}{\emph{International Workshop on Internet and Network
  Economics}}. Springer, \bibinfo{pages}{512--518}.
\newblock


\bibitem[Tassier and Menczer(2008)]%
        {tassier2008social}
\bibfield{author}{\bibinfo{person}{Troy Tassier} {and} \bibinfo{person}{Filippo
  Menczer}.} \bibinfo{year}{2008}\natexlab{}.
\newblock \showarticletitle{Social network structure, segregation, and equality
  in a labor market with referral hiring}.
\newblock \bibinfo{journal}{\emph{Journal of Economic Behavior \&
  Organization}} \bibinfo{volume}{66}, \bibinfo{number}{3-4}
  (\bibinfo{year}{2008}), \bibinfo{pages}{514--528}.
\newblock


\bibitem[Teng et~al\mbox{.}(2020)]%
        {teng2020influencing}
\bibfield{author}{\bibinfo{person}{Arwen Teng}, \bibinfo{person}{Ting-Wei Li},
  \bibinfo{person}{Yu-chi Liao}, \bibinfo{person}{Hsi-Wen Chen},
  \bibinfo{person}{Yvonne-Anne Pignolet}, \bibinfo{person}{De-Nian Yang}, {and}
  \bibinfo{person}{Lydia~Y Chen}.} \bibinfo{year}{2020}\natexlab{}.
\newblock \showarticletitle{On Influencing the Influential: Disparity Seeding}.
\newblock \bibinfo{journal}{\emph{arXiv preprint arXiv:2011.08946}}
  (\bibinfo{year}{2020}).
\newblock


\bibitem[Teng et~al\mbox{.}(2021)]%
        {teng2021influencing}
\bibfield{author}{\bibinfo{person}{Ya-Wen Teng}, \bibinfo{person}{Hsi-Wen
  Chen}, \bibinfo{person}{De-Nian Yang}, \bibinfo{person}{Yvonne-Anne
  Pignolet}, \bibinfo{person}{Ting-Wei Li}, {and} \bibinfo{person}{Lydia
  Chen}.} \bibinfo{year}{2021}\natexlab{}.
\newblock \showarticletitle{On influencing the influential: disparity seeding}.
  In \bibinfo{booktitle}{\emph{Proceedings of the 30th ACM International
  Conference on Information \& Knowledge Management}}.
  \bibinfo{pages}{1804--1813}.
\newblock


\bibitem[Tong et~al\mbox{.}(2016)]%
        {tong2016adaptive}
\bibfield{author}{\bibinfo{person}{Guangmo Tong}, \bibinfo{person}{Weili Wu},
  \bibinfo{person}{Shaojie Tang}, {and} \bibinfo{person}{Ding-Zhu Du}.}
  \bibinfo{year}{2016}\natexlab{}.
\newblock \showarticletitle{Adaptive influence maximization in dynamic social
  networks}.
\newblock \bibinfo{journal}{\emph{IEEE/ACM Transactions on Networking}}
  \bibinfo{volume}{25}, \bibinfo{number}{1} (\bibinfo{year}{2016}),
  \bibinfo{pages}{112--125}.
\newblock


\bibitem[Tran et~al\mbox{.}(2021)]%
        {tran2021community}
\bibfield{author}{\bibinfo{person}{Cong Tran}, \bibinfo{person}{Won-Yong Shin},
  {and} \bibinfo{person}{Andreas Spitz}.} \bibinfo{year}{2021}\natexlab{}.
\newblock \showarticletitle{Community detection in partially observable social
  networks}.
\newblock \bibinfo{journal}{\emph{ACM Transactions on Knowledge Discovery from
  Data (TKDD)}} \bibinfo{volume}{16}, \bibinfo{number}{2}
  (\bibinfo{year}{2021}), \bibinfo{pages}{1--24}.
\newblock


\bibitem[Tsang et~al\mbox{.}(2019)]%
        {tsang2019group}
\bibfield{author}{\bibinfo{person}{Alan Tsang}, \bibinfo{person}{Bryan Wilder},
  \bibinfo{person}{Eric Rice}, \bibinfo{person}{Milind Tambe}, {and}
  \bibinfo{person}{Yair Zick}.} \bibinfo{year}{2019}\natexlab{}.
\newblock \showarticletitle{Group-fairness in influence maximization}.
\newblock \bibinfo{journal}{\emph{arXiv preprint arXiv:1903.00967}}
  (\bibinfo{year}{2019}).
\newblock


\bibitem[Tsioutsiouliklis et~al\mbox{.}(2022)]%
        {tsioutsiouliklis2022link}
\bibfield{author}{\bibinfo{person}{Sotiris Tsioutsiouliklis},
  \bibinfo{person}{Evaggelia Pitoura}, \bibinfo{person}{Konstantinos
  Semertzidis}, {and} \bibinfo{person}{Panayiotis Tsaparas}.}
  \bibinfo{year}{2022}\natexlab{}.
\newblock \showarticletitle{Link Recommendations for PageRank Fairness}. In
  \bibinfo{booktitle}{\emph{Proceedings of the ACM Web Conference 2022}}.
  \bibinfo{pages}{3541--3551}.
\newblock


\bibitem[Tsioutsiouliklis et~al\mbox{.}(2021)]%
        {tsioutsiouliklis2021fairness}
\bibfield{author}{\bibinfo{person}{Sotiris Tsioutsiouliklis},
  \bibinfo{person}{Evaggelia Pitoura}, \bibinfo{person}{Panayiotis Tsaparas},
  \bibinfo{person}{Ilias Kleftakis}, {and} \bibinfo{person}{Nikos Mamoulis}.}
  \bibinfo{year}{2021}\natexlab{}.
\newblock \showarticletitle{Fairness-Aware PageRank}. In
  \bibinfo{booktitle}{\emph{Proceedings of the Web Conference 2021}}.
  \bibinfo{pages}{3815--3826}.
\newblock


\bibitem[ur~Rasheed et~al\mbox{.}(2018)]%
        {ur2018detecting}
\bibfield{author}{\bibinfo{person}{Haroon ur Rasheed},
  \bibinfo{person}{Farhan~Hassan Khan}, \bibinfo{person}{Saba Bashir}, {and}
  \bibinfo{person}{Irsa Fatima}.} \bibinfo{year}{2018}\natexlab{}.
\newblock \showarticletitle{Detecting suspicious discussion on online forums
  using data mining}. In \bibinfo{booktitle}{\emph{International Conference on
  Intelligent Technologies and Applications}}. Springer,
  \bibinfo{pages}{262--273}.
\newblock


\bibitem[Varshney et~al\mbox{.}(2014)]%
        {varshney2014modeling}
\bibfield{author}{\bibinfo{person}{Devesh Varshney}, \bibinfo{person}{Sandeep
  Kumar}, {and} \bibinfo{person}{Vineet Gupta}.}
  \bibinfo{year}{2014}\natexlab{}.
\newblock \showarticletitle{Modeling information diffusion in social networks
  using latent topic information}. In \bibinfo{booktitle}{\emph{Intelligent
  Computing Theory: 10th International Conference, ICIC 2014, Taiyuan, China,
  August 3-6, 2014. Proceedings 10}}. Springer, \bibinfo{pages}{137--148}.
\newblock


\bibitem[Varshney et~al\mbox{.}(2017)]%
        {varshney2017predicting}
\bibfield{author}{\bibinfo{person}{Devesh Varshney}, \bibinfo{person}{Sandeep
  Kumar}, {and} \bibinfo{person}{Vineet Gupta}.}
  \bibinfo{year}{2017}\natexlab{}.
\newblock \showarticletitle{Predicting information diffusion probabilities in
  social networks: A Bayesian networks based approach}.
\newblock \bibinfo{journal}{\emph{Knowledge-Based Systems}}
  \bibinfo{volume}{133} (\bibinfo{year}{2017}), \bibinfo{pages}{66--76}.
\newblock


\bibitem[Venkatasubramanian et~al\mbox{.}(2021)]%
        {venkatasubramanian2021fairness}
\bibfield{author}{\bibinfo{person}{Suresh Venkatasubramanian},
  \bibinfo{person}{Carlos Scheidegger}, \bibinfo{person}{Sorelle Friedler},
  {and} \bibinfo{person}{Aaron Clauset}.} \bibinfo{year}{2021}\natexlab{}.
\newblock \showarticletitle{Fairness in Networks, a tutorial}.
\newblock  (\bibinfo{year}{2021}).
\newblock


\bibitem[Villa et~al\mbox{.}(2021)]%
        {villa2021echo}
\bibfield{author}{\bibinfo{person}{Giacomo Villa}, \bibinfo{person}{Gabriella
  Pasi}, {and} \bibinfo{person}{Marco Viviani}.}
  \bibinfo{year}{2021}\natexlab{}.
\newblock \showarticletitle{Echo chamber detection and analysis}.
\newblock \bibinfo{journal}{\emph{Social Network Analysis and Mining}}
  \bibinfo{volume}{11}, \bibinfo{number}{1} (\bibinfo{year}{2021}),
  \bibinfo{pages}{1--17}.
\newblock


\bibitem[Wang et~al\mbox{.}(2021)]%
        {wang2021information}
\bibfield{author}{\bibinfo{person}{Xindi Wang}, \bibinfo{person}{Onur Varol},
  {and} \bibinfo{person}{Tina Eliassi-Rad}.} \bibinfo{year}{2021}\natexlab{}.
\newblock \showarticletitle{Information Access Equality on Network Generative
  Models}.
\newblock \bibinfo{journal}{\emph{arXiv preprint arXiv:2107.02263}}
  (\bibinfo{year}{2021}).
\newblock


\bibitem[Wani and Jabin(2018)]%
        {wani2018mutual}
\bibfield{author}{\bibinfo{person}{Mudasir~Ahmad Wani} {and}
  \bibinfo{person}{Suraiya Jabin}.} \bibinfo{year}{2018}\natexlab{}.
\newblock \showarticletitle{Mutual clustering coefficient-based suspicious-link
  detection approach for online social networks}.
\newblock \bibinfo{journal}{\emph{Journal of King Saud University-Computer and
  Information Sciences}} (\bibinfo{year}{2018}).
\newblock


\bibitem[Weerts et~al\mbox{.}(2023a)]%
        {weerts2023can}
\bibfield{author}{\bibinfo{person}{Hilde Weerts}, \bibinfo{person}{Florian
  Pfisterer}, \bibinfo{person}{Matthias Feurer}, \bibinfo{person}{Katharina
  Eggensperger}, \bibinfo{person}{Edward Bergman}, \bibinfo{person}{Noor Awad},
  \bibinfo{person}{Joaquin Vanschoren}, \bibinfo{person}{Mykola Pechenizkiy},
  \bibinfo{person}{Bernd Bischl}, {and} \bibinfo{person}{Frank Hutter}.}
  \bibinfo{year}{2023}\natexlab{a}.
\newblock \showarticletitle{Can Fairness be Automated? Guidelines and
  Opportunities for Fairness-aware AutoML}.
\newblock \bibinfo{journal}{\emph{arXiv preprint arXiv:2303.08485}}
  (\bibinfo{year}{2023}).
\newblock


\bibitem[Weerts et~al\mbox{.}(2022)]%
        {weerts2022does}
\bibfield{author}{\bibinfo{person}{Hilde Weerts}, \bibinfo{person}{Lamb{\`e}r
  Royakkers}, {and} \bibinfo{person}{Mykola Pechenizkiy}.}
  \bibinfo{year}{2022}\natexlab{}.
\newblock \showarticletitle{Does the End Justify the Means? On the Moral
  Justification of Fairness-Aware Machine Learning}.
\newblock \bibinfo{journal}{\emph{arXiv preprint arXiv:2202.08536}}
  (\bibinfo{year}{2022}).
\newblock


\bibitem[Weerts et~al\mbox{.}(2023b)]%
        {weerts2023algorithmic}
\bibfield{author}{\bibinfo{person}{Hilde Weerts}, \bibinfo{person}{Rapha{\"e}le
  Xenidis}, \bibinfo{person}{Fabien Tarissan}, \bibinfo{person}{Henrik~Palmer
  Olsen}, {and} \bibinfo{person}{Mykola Pechenizkiy}.}
  \bibinfo{year}{2023}\natexlab{b}.
\newblock \showarticletitle{Algorithmic unfairness through the lens of EU
  non-discrimination law: Or why the law is not a decision tree}. In
  \bibinfo{booktitle}{\emph{Proceedings of the 2023 ACM Conference on Fairness,
  Accountability, and Transparency}}. \bibinfo{pages}{805--816}.
\newblock


\bibitem[Wehmuth and Ziviani(2013)]%
        {wehmuth2013daccer}
\bibfield{author}{\bibinfo{person}{Klaus Wehmuth} {and} \bibinfo{person}{Artur
  Ziviani}.} \bibinfo{year}{2013}\natexlab{}.
\newblock \showarticletitle{Daccer: Distributed assessment of the closeness
  centrality ranking in complex networks}.
\newblock \bibinfo{journal}{\emph{Computer Networks}} \bibinfo{volume}{57},
  \bibinfo{number}{13} (\bibinfo{year}{2013}), \bibinfo{pages}{2536--2548}.
\newblock


\bibitem[Wilder et~al\mbox{.}(2020)]%
        {wilder2020clinical}
\bibfield{author}{\bibinfo{person}{Bryan Wilder}, \bibinfo{person}{Laura
  Onasch-Vera}, \bibinfo{person}{Graham Diguiseppi}, \bibinfo{person}{Robin
  Petering}, \bibinfo{person}{Chyna Hill}, \bibinfo{person}{Amulya Yadav},
  \bibinfo{person}{Eric Rice}, {and} \bibinfo{person}{Milind Tambe}.}
  \bibinfo{year}{2020}\natexlab{}.
\newblock \showarticletitle{Clinical trial of an AI-augmented intervention for
  HIV prevention in youth experiencing homelessness}.
\newblock \bibinfo{journal}{\emph{arXiv preprint arXiv:2009.09559}}
  (\bibinfo{year}{2020}).
\newblock


\bibitem[Wu and Huberman(2004)]%
        {wu2004social}
\bibfield{author}{\bibinfo{person}{Fang Wu} {and} \bibinfo{person}{Bernardo~A
  Huberman}.} \bibinfo{year}{2004}\natexlab{}.
\newblock \showarticletitle{Social structure and opinion formation}.
\newblock \bibinfo{journal}{\emph{arXiv preprint cond-mat/0407252}}
  (\bibinfo{year}{2004}).
\newblock


\bibitem[Wu et~al\mbox{.}(2016)]%
        {wu2016mining}
\bibfield{author}{\bibinfo{person}{Liang Wu}, \bibinfo{person}{Fred
  Morstatter}, \bibinfo{person}{Xia Hu}, {and} \bibinfo{person}{Huan Liu}.}
  \bibinfo{year}{2016}\natexlab{}.
\newblock \showarticletitle{Mining misinformation in social media}.
\newblock \bibinfo{journal}{\emph{Big Data in Complex and Social Networks}}
  (\bibinfo{year}{2016}), \bibinfo{pages}{123--152}.
\newblock


\bibitem[Wu and Pan(2017)]%
        {wu2017scalable}
\bibfield{author}{\bibinfo{person}{Peng Wu} {and} \bibinfo{person}{Li Pan}.}
  \bibinfo{year}{2017}\natexlab{}.
\newblock \showarticletitle{Scalable influence blocking maximization in social
  networks under competitive independent cascade models}.
\newblock \bibinfo{journal}{\emph{Computer Networks}}  \bibinfo{volume}{123}
  (\bibinfo{year}{2017}), \bibinfo{pages}{38--50}.
\newblock


\bibitem[Yadav et~al\mbox{.}(2018)]%
        {yadav2018bridging}
\bibfield{author}{\bibinfo{person}{Amulya Yadav}, \bibinfo{person}{Bryan
  Wilder}, \bibinfo{person}{Eric Rice}, \bibinfo{person}{Robin Petering},
  \bibinfo{person}{Jaih Craddock}, \bibinfo{person}{Amanda Yoshioka-Maxwell},
  \bibinfo{person}{Mary Hemler}, \bibinfo{person}{Laura Onasch-Vera},
  \bibinfo{person}{Milind Tambe}, {and} \bibinfo{person}{Darlene Woo}.}
  \bibinfo{year}{2018}\natexlab{}.
\newblock \showarticletitle{Bridging the Gap Between Theory and Practice in
  Influence Maximization: Raising Awareness about HIV among Homeless Youth.}.
  In \bibinfo{booktitle}{\emph{IJCAI}}. \bibinfo{pages}{5399--5403}.
\newblock


\bibitem[Yang et~al\mbox{.}(2013)]%
        {yang2013community}
\bibfield{author}{\bibinfo{person}{Jaewon Yang}, \bibinfo{person}{Julian
  McAuley}, {and} \bibinfo{person}{Jure Leskovec}.}
  \bibinfo{year}{2013}\natexlab{}.
\newblock \showarticletitle{Community detection in networks with node
  attributes}. In \bibinfo{booktitle}{\emph{2013 IEEE 13th international
  conference on data mining}}. IEEE, \bibinfo{pages}{1151--1156}.
\newblock


\bibitem[Yen et~al\mbox{.}(2013)]%
        {yen2013efficient}
\bibfield{author}{\bibinfo{person}{Chia-Chen Yen}, \bibinfo{person}{Mi-Yen
  Yeh}, {and} \bibinfo{person}{Ming-Syan Chen}.}
  \bibinfo{year}{2013}\natexlab{}.
\newblock \showarticletitle{An efficient approach to updating closeness
  centrality and average path length in dynamic networks}. In
  \bibinfo{booktitle}{\emph{2013 IEEE 13th International Conference on Data
  Mining}}. IEEE, \bibinfo{pages}{867--876}.
\newblock


\bibitem[Zhang et~al\mbox{.}(2022)]%
        {zhang2022fairness}
\bibfield{author}{\bibinfo{person}{Wenbin Zhang}, \bibinfo{person}{Jeremy~C
  Weiss}, \bibinfo{person}{Shuigeng Zhou}, {and} \bibinfo{person}{Toby Walsh}.}
  \bibinfo{year}{2022}\natexlab{}.
\newblock \showarticletitle{Fairness amidst non-iid graph data: A literature
  review}.
\newblock \bibinfo{journal}{\emph{arXiv preprint arXiv:2202.07170}}
  (\bibinfo{year}{2022}).
\newblock


\bibitem[Zhou et~al\mbox{.}(2008)]%
        {zhou2008brief}
\bibfield{author}{\bibinfo{person}{Bin Zhou}, \bibinfo{person}{Jian Pei}, {and}
  \bibinfo{person}{WoShun Luk}.} \bibinfo{year}{2008}\natexlab{}.
\newblock \showarticletitle{A brief survey on anonymization techniques for
  privacy preserving publishing of social network data}.
\newblock \bibinfo{journal}{\emph{ACM Sigkdd Explorations Newsletter}}
  \bibinfo{volume}{10}, \bibinfo{number}{2} (\bibinfo{year}{2008}),
  \bibinfo{pages}{12--22}.
\newblock


\bibitem[Zhou et~al\mbox{.}(2009)]%
        {zhou2009predicting}
\bibfield{author}{\bibinfo{person}{Tao Zhou}, \bibinfo{person}{Linyuan L{\"u}},
  {and} \bibinfo{person}{Yi-Cheng Zhang}.} \bibinfo{year}{2009}\natexlab{}.
\newblock \showarticletitle{Predicting missing links via local information}.
\newblock \bibinfo{journal}{\emph{The European Physical Journal B}}
  \bibinfo{volume}{71}, \bibinfo{number}{4} (\bibinfo{year}{2009}),
  \bibinfo{pages}{623--630}.
\newblock


\bibitem[Zhu et~al\mbox{.}(2014)]%
        {zhu2014maximizing}
\bibfield{author}{\bibinfo{person}{Tian Zhu}, \bibinfo{person}{Bai Wang},
  \bibinfo{person}{Bin Wu}, {and} \bibinfo{person}{Chuanxi Zhu}.}
  \bibinfo{year}{2014}\natexlab{}.
\newblock \showarticletitle{Maximizing the spread of influence ranking in
  social networks}.
\newblock \bibinfo{journal}{\emph{Information Sciences}}  \bibinfo{volume}{278}
  (\bibinfo{year}{2014}), \bibinfo{pages}{535--544}.
\newblock


\end{thebibliography}

\appendix

\section{Abbreviations}\label{appendixabbrv}
In Table \ref{abbrvtable}, we list all abbreviations used in the paper.

\begin{table}[htbp]
\caption{Abbreviations used in the paper.}
\label{abbrvtable}
\begin{tabular}{ll}
\hline
\textbf{Abbreviation} & \textbf{Explanation} \\ \hline
CD & Community Detection \\  
CR & Centrality Ranking \\
FairSNA & Fairness-aware Social Network Analysis \\
IBM  & Influence Blocking Maximization  \\
IC & Independent Cascade \\
IM & Influence Maximization  \\
LP & Link Prediction \\
ML & Machine Learning \\
NLP & Natural Language Processing \\
OSNs & Online Social Networks \\
SNA & Social Network Analysis \\
 \hline
\end{tabular}
\end{table}

\section{Fairness Evaluation Metrics}\label{metrics}

Here, we summarize metrics used to evaluate fairness with their applications.

\begin{enumerate}[noitemsep,topsep=0pt]
    \item Price of Fairness (PoF): It is the ratio of the optimal solution without a fairness constraint to the best achievable solution under the fairness constraint. For example, in Influence maximization, the PoF is the ratio of maximum outreach for any choice of $k$ seed nodes to the outreach achieved under a given fairness constraint. It can be computed as,
    
    \begin{equation*}
     PoF=\frac{I_{opt}}{I_{fairness}}
    \end{equation*}
    
    \item Utility Gap: It computes the gap in utilities of communities, or mainly, the gap between the communities with the highest and lowest utilities. In influence maximization, it can measure the utility gap between the communities having the highest and lowest outreach. 
    
    \item Kullback–Leibler (K-L) Divergence \cite{kullback1951information}: It is used to compare two distributions. In influence maximization, it can be used to compare the achieved outreach with the expected outreach ratio of different communities. 
    
    \item Earth Mover Distance (EMD): The EMD is used to compare two distributions, and it also considers the distance between the values that is not considered in the K-L divergence \cite{wang2021information}. For example, the divergence between [1, 1, 1, ...] and [2, 2, 2, ...] is equal to the divergence between [1, 1, 1, ...] and [10, 10, 10, ...], though they have different EMD values.
\end{enumerate}

The other evaluation metrics include Power inequality \cite{avin2015homophily, wang2021information}, Moment Glass Ceiling \cite{avin2015homophily, hymowitz1986women, wang2021information}, and Modularity Reduction \cite{masrour2020bursting}. 

\section{Synthetic Network Generating Models}\label{appendixsynmodel}

Following are the synthetic models to generate homophilic networks with minor and major communities that have been used in studying fairSNA.

\begin{enumerate}
    \item \textbf{Homophily BA Model} \cite{karimi2018homophily, lee2019homophily}: In this model, the links are formed using two rules, preferential attachment and homophily. The model uses the following parameters, (i) $m$, i.e., the fraction of minority nodes, (ii) $l$, i.e., the number of edges formed by each new node, and (iii) $H$, i.e., the homophily matrix where the entry $H_{g_ug_v}$ is the probability to connect two nodes $u$ and $v$ belonging to group $g_u$ and $g_v$, respectively. The network evolves as follows, at each time step,
    \begin{enumerate}
        \item A new node $v$ is added to the network. It belongs to the minority group with probability $m$ and to the majority with $1-m$. The group of node $v$ is denoted as $g_v$.
        \item Node $v$ makes $l$ connections, and the probability to connect with a node $u$ is defined as,
        \begin{equation*}
            \Pi_u=\frac{H_{g_vg_u}d(u)^{\alpha}}{\sum_w H_{g_vg_w}d(w)^{\alpha}}
        \end{equation*}
        where $d(u)$ is the degree of node $u$.
    \end{enumerate}
    If $\alpha=0$, the network has no preferential attachment and is referred to as \textbf{Random Homophily}.
    
    \item \textbf{Diversified Homophily BA:} Wang et al. \cite{wang2021information} proposed the Diversified Homophily BA model to promote inter-community edges while maintaining some degree of homophily. The model requires the following parameters: (i) $m$, the fraction of minority nodes, (ii) $l$, the total number of edges for each new node, (iii) $H$, the homophily matrix that contains the probability of inter and intra-group connections, (iv) $\alpha$, preferential attachment strength, (v) $l_d$, the number of diversified edges for each node, and (vi) $p_d$, the diversification probability. The network evolves using the following steps at each time step,
    \begin{enumerate}
        \item A new node $v$ is added to the network. It belongs to the minority group with probability $m$ and to the majority with $1-m$. The group of node $v$ is denoted as $g_v$.
        \item The node $v$ is connected with $l-l_d$ nodes using Homophily BA attachment rule where the probability to connect with a node $u$ is defined as, $\Pi_u=\frac{H_{g_vg_u}d(u)^{\alpha}}{\sum_u H_{g_vg_u}d(u)^{\alpha}}$. The nodes connected at this step are referred to as $S_v$.
        \item In this step, the node $v$ will make $l_d$ diversified connections. The probability of connecting two nodes $v$ and $w$ is computed as,
        \begin{equation*}
            p_{vw}=\left\{\begin{matrix}
            p_d, & g_v \neq g_w\\ 
            1-p_d, & g_v=g_w.
            \end{matrix}\right.
        \end{equation*}
        
        For node $u \in S_v$, create a set of their neighbors denoted by $N_{S_v}$. Next, the node will make $l_d$ diversified connections by connecting to $k \in N_{S_v}$ with the probability $\Pi_{vw} \propto p_{vw}\frac{1}{|d(w)-d(u)|}$, where $d(u)$ is the degree of node $u$ and $u \in S_v$. The main idea behind this is that the nodes belonging to opposite groups should be connected by maintaining a similar degree as of existing neighbors. When $\alpha=0$, it will remove the preferential attachment, and the generated network will be called \textbf{Diversified Homophily}.
    \end{enumerate}
    
    \item \textbf{Directed Homophily Network:} Anwar et al. \cite{anwar2021balanced} proposed a model to generate directed homophilic networks based on the network evolution models proposed by Bollobás et al. \cite{bollobas2003directed} and Karimi et al. \cite{karimi2018homophily}. In the evolution process, at each time step, a directed edge is added to the network as follows:
    
        \begin{enumerate}
            \item With probability $\alpha$, a new node $v$ is added, and it is connected to an existing node $w$. The node $v$ is assigned to the majority category with probability $p_M$, and to the minority category with probability $(1- p_M)$. The node $w$ is chosen with the probability proportional to $h(v,w)d_{in}(w) +\delta_{in}$, and $h(v,w)=h$ if $v$ and $w$ belong to the same category, otherwise $h(v,w)= (1-h)$, where $h$ is a variable and its value depends on the required homophily in the network, $d_{in}(w)$ is the in-degree of node $w$, and $\delta_{in}$ is a constant.
            
            \item With probability $\beta$, an edge is added from an existing node $v$ to an existing node $w$. Node $v$ is chosen from all existing nodes with the probability proportional to $d_{out}(v) + \delta_{out}$, where $d_{out}(v)$ is the out-degree of node $v$ and $\delta_{out}$ is a constant, and node $w$ is chosen from all existing nodes with probability proportional to $h(v,w)d_{in}(w) + \delta_{in}$.
            
            \item With probability $\gamma$, a new node $w$ is added and is connected with an edge from an existing node $v$. The node $w$ is assigned to the majority category with probability $p_M$, otherwise to the minority category. Node $v$ is chosen from existing nodes with probability proportional to $h(v,w)d_{out} (v) + \delta_{out}$.
        \end{enumerate}
        
        \item \textbf{Organic Growth Model:} Stoica et al. \cite{stoica2018algorithmic} proposed this model to study the glass ceiling effect in the network. The proposed model assumes that each node belongs to one of the two communities, called minor or major communities. As the network grows, one node and one directed edge is added at each time step using the following steps.
        \begin{enumerate}
            \item Minority-majority partition: a new node $u$ joins the network and belongs to the minor community with probability $r$ and major community with probability $1 - r$ (for $0 \leq r \leq 1/2$).
            \item Randomness: With probability $\eta$, the new node $u$ connects with a randomly chosen existing node $v$.
            \item Preferential attachment: with probability $1- \eta$, the node $u$ chooses an existing node uniformly at random and copies one of its edges. In simple words, the probability of choosing another endpoint is directly proportional to its degree, as in preferential attachment law.
            \item Homophily: if the newly added node belongs to a different group than the chosen node for the connection, the connection is created with probability $\rho$, and this process will be repeated until an edge is formed. $\rho$ is the homophily factor ($0 \leq \rho \leq 1$) and controls that a node that is less similar is less probable to be chosen for making the connection.
        \end{enumerate}
        The model is referred to as ``organic growth'' of the network, as the newly added nodes form new connections without the influence of any external forces.
        
\end{enumerate}

\end{document}